\documentclass[qps,rmp,10pt,twocolumn,superscriptaddress,nofootinbib,floatfix,showpacs]{revtex4}
\usepackage{graphics, graphicx, amsmath, amssymb}
\usepackage[usenames,dvipsnames]{color}
\definecolor{Red}{rgb}{1,0,0}

\def\bra#1{\mathinner{\langle{#1}|}}
\def\ket#1{\mathinner{|{#1}\rangle}}
\def\braket#1{\mathinner{\langle{#1}\rangle}}
\def\ketbra#1#2{{\ket{#1}}{\bra{#2}}}

{\catcode`\|=\active
 \gdef\Braket#1{\begingroup
\mathcode`\|32768\let|\BraVert\left<{#1}\right>\endgroup}}
\def\BraVert{\egroup\,\mid\,\bgroup}

\newcommand{\tr}{\mathrm{tr}}
\newcommand{\thm}{\mathrm{th}}
\newcommand{\Sy}{\mathcal{S}}
\newcommand{\E}{\mathcal{E}}
\newcommand{\SE}{\mathcal{SE}}

\begin{document}

\title{The classical-quantum boundary for correlations: discord and related measures}
\date{\today}

\author{Kavan Modi}
\email{kavan@quantumlah.org}
\affiliation{Department of Physics, University of Oxford, Clarendon Laboratory, Oxford UK}
\affiliation{Centre for Quantum Technologies, National University of Singapore, Singapore}

\author{Aharon Brodutch}
%\email{cap.fwiffo@gmail.com}
\affiliation{Department of Physics \& Astronomy, Faculty of Science Macquarie University, Sydney, Australia}
\affiliation{Institute for Quantum Computing and Department of Physics and Astronomy, University of Waterloo, Canada}

\author{Hugo Cable}
%\email{cqthvc@nus.edu.sg}
\affiliation{Centre for Quantum Technologies, National University of Singapore, Singapore}

\author{Tomasz Paterek}
%\email{tomasz.paterek@nus.edu.sg}
\affiliation{Division of Physics and Applied Physics, School of Physical and Mathematical Sciences, Nanyang Technological University, Singapore}
\affiliation{Centre for Quantum Technologies, National University of Singapore, Singapore}

\author{Vlatko Vedral}
%\email{phyvv@nus.edu.sg}
\affiliation{Department of Physics, University of Oxford, Clarendon Laboratory, Oxford UK}
\affiliation{Centre for Quantum Technologies, National University of Singapore, Singapore}
\affiliation{Department of Physics, National University of Singapore, Singapore}
	
\begin{abstract}
One of the best signatures of nonclassicality in a quantum system is the existence of correlations that have no classical counterpart. Different methods for quantifying the quantum and classical parts of correlations are amongst the more actively-studied topics of quantum information theory over the past decade. Entanglement is the most prominent of these correlations, but in many cases unentangled states exhibit nonclassical behavior too. Thus distinguishing quantum correlations other than entanglement provides a better division between the quantum and classical worlds, especially when considering mixed states. Here we review different notions of classical and quantum correlations quantified by quantum discord and other related measures. In the first half, we review the mathematical properties of the measures of quantum correlations, relate them to each other, and discuss the classical-quantum division that is common among them. In the second half, we show that the measures identify and quantify the deviation from classicality in various quantum-information-processing tasks, quantum thermodynamics, open-system dynamics, and many-body physics. We show that in many cases quantum correlations indicate an advantage of quantum methods over classical ones.
\end{abstract}


\maketitle
\tableofcontents

%**************************************************************
%**************************************************************
\section{Introduction}
%**************************************************************
%**************************************************************

In the early days of quantum information, entanglement was viewed as the main feature that gives quantum computers an advantage over their classical counterparts. Superpositions without entanglement were seen as insufficient, especially given the fact that the concept of superposition exists in the classical physics of waves, as it does in the classical theory of electromagnetism, for instance. The view that entanglement is crucial is also supported by foundational considerations, for it is known that Bell's inequalities cannot be violated by either classical or quantum superpositions, and require genuine entanglement to exceed the classically-determined limit for correlations. \cite{Naturwissenschaften.23.844} captured all this in his highly-influential ``cat paper", saying entanglement is ``not just one of many traits, but the characteristic trait of quantum physics". However, this straightforward and simple view about the efficiency of quantum-information processing changed dramatically about ten years ago, when several developments took place.

First \cite{PhysRevLett.81.5672} showed that quantum computation in which only one qubit is not in a maximally-mixed state, while the rest are, can achieve an exponential improvement in efficiency over classical computers for a limited set of tasks (see Sec.~\ref{Sec:discinQC}). This started to throw doubt on entanglement being responsible for all quantum speedups, since a computer register which is so mixed as to have only one nonmaximally-mixed qubit is unlikely to be entangled. The Knill-Laflamme model is experimentally motivated by (liquid-state) nuclear-magnetic-resonance (NMR) information processing, at room temperature, and is therefore important for resolving the question of whether NMR can provide a genuine implementation of a quantum computer.

Another development came in 2001 while analyzing different measures of information in quantum theory. \cite{arXiv:quant-ph/0011039, JPhysA.34.6899, arXiv:quant-ph/0105072} concluded that entanglement does not account for all non-classical correlations and that even separable states usually contain correlations that are not entirely classical. These correlations are aptly named quantum discord. Soon after its inception \cite{arXiv:quant-ph/0110029} gave an intuitive argument that quantum discord may be connected to the performance of certain quantum computers. In a seminal paper \cite{arXiv:0709.0548} put this on a firm quantitative basis . They calculated discord in the Knill-Laflamme algorithm and showed that it scales with the quantum efficiency, unlike entanglement which remains vanishingly small throughout the computation (see \cite{Nature.474.24} for a popular account). This triggered a flurry of activity in applying discord to many different protocols and problems in quantum information.

About the same time another form of quantum correlations different from entanglement emerged in an information-theoretic approach to thermodynamics: \cite{arXiv:quant-ph/0112074} showed that the advantage of using nonlocal operation to extract work from a correlated system coupled to a heat bath is related to entanglement only in the case of pure states. In the general case, the advantage is related to more general forms of quantum correlations. This work was followed by a series of results which we review in Sec.~\ref{SEC_DEFICIT} and \ref{maxwell}. Other results linking discord to various areas of physics involved open systems, which provided Zurek's original motivation for introducing quantum discord, \cite{arXiv:quant-ph/0011039}. While Zurek's main interest was decoherence (see Sec.~\ref{Einselection}), \cite{arXiv:quant-ph/0703022} linked discord to open-system dynamics and their description via dynamical maps (see Sec.~\ref{dynamics}). At the same time, \cite{arXiv:0809.1723} studied the relation between discord and quantum phase transitions opening the way for further studies on discord in many-body systems (see Sec.~\ref{mbp}).

Nowadays,\footnote{The application of entropic measures to quantum correlations dates back to the works of Everett, Lindblad, and Holevo \cite{Coll.Everett, CommMathPhys.33.305, ProblPeredachiInf.9.31, ProblInfTransm.9.110}.} there are many ways of understanding the gap in correlations, that is to say the fact that classical correlations and entanglement do not exhaust all possible correlations in quantum systems. The widely-used measures of quantum correlations are quantum discord, quantum deficit, measurement-induced disturbance, and relative entropy of discord. In the first half of this review we introduce these different measures and show the fundamental differences and similarities between them. In the second half of the review we identify and discuss the major directions of research that make use of measures of quantum correlations. They are: quantum information, quantum algorithms, quantum thermodynamics, dynamics of open systems, and many-body physics.

%**************************************************************
%**************************************************************
\section{Different measures of quantum correlations}\label{SEC_MEASURES}
%**************************************************************
%**************************************************************

Quantum systems can be correlated in ways inaccessible to classical objects and the existence of non-classical correlations in a system can be seen as a signature that subsystems are genuinely quantum. Various notions of classicality exist and give rise to the hierarchy of states and correlations considered to be genuinely quantum, \cite{arXiv:1107.3428}. It is not our aim to discuss all notions of classicality present in the literature, rather we focus on some of those directly related to correlations. For example, one may regard as classical the local-realistic worldview put forward in the famous Einstein-Podolsky-Rosen (EPR) paper, \cite{PhysRev.47.777}. Using modern language this is the world in which the results of experiments can be calculated by local algorithms supplied with data transmitted no faster than the speed of light. Bell showed that correlations between outcomes of such local programs are bounded, and there exists entangled quantum states with correlations violating this bound \cite{Physics.1.195}. Interestingly, Werner proved that there are entangled quantum states that generate outcomes perfectly in agreement with a local-realistic view \cite{PhysRevA.40.2477}. Therefore according to local realism even correlations generated by some entangled states are classical.

Clearly one can object to the notion that local realism is all there is to the classical world. The set of states admitting a local-realistic model is reduced if another notion of classicality is introduced. One may regard as classical those states which can be prepared with the help of local operations and classical communication (LOCC). According to this notion, the set of classical states is exactly the set of separable (not entangled) quantum states, \cite{RevModPhys.81.865}, and quantum correlations correspond exactly to entanglement. However, one may object to this notion of classicality too, having in mind the nature of the operations allowed in the framework of LOCC. For example, local operations here allow for the preparation of indistinguishable pure quantum states, whereas it is impossible to prepare pure-indistinguishable states of a classical bit: a classical bit about which we have full knowledge (in a pure state) can either be in state `0' or in state `1', i.e. in one of two fully-distinguishable states. General quantum states which satisfy this final classicality constraint form a subset of the separable quantum states, and accordingly define some separable states as quantum correlated. This is the spirit of the present section, in which certain notions and measures of classicality are discussed, according to which the classical states form a subset of the separable states.

%**************************************************************
\subsection{Quantum discord}\label{SEC_DISCORD}
%**************************************************************

The notion of classicality related to quantum discord revolves around information theory, \cite{arXiv:quant-ph/0105072, arXiv:quant-ph/0011039, JPhysA.34.6899}. Two systems are correlated if together they contain more information than taken separately. If we measure the lack of information by entropy, this definition of correlations is captured by the mutual information
\begin{gather}\label{MI}
I(A:B) \equiv S(A) + S(B) - S(AB),
\end{gather}
where $S(X)$ is the Shannon entropy $S(X) = - \sum_x p_x \log p_x$ if $X$ is a classical variable with values $x$ occurring with probability $p_x$, or $S(X)$ is the von Neumann entropy $S(X) = - \tr(\rho_X \log \rho_X)$ if $\rho_X$ is a quantum state of system $X$ (all logarithms are base two). For classical variables, Bayes' rule defines a conditional probability as $p_{x|y} = p_{xy}/p_y$. This implies an equivalent form for the classical mutual information
\begin{gather}\label{CL_CORR}
J_{\mathrm{cl}}(B|A) = S(B) - S(B|A),
\end{gather}
where the conditional entropy $S(B|A) = \sum_a p_a S(B|a)$ is the average of entropies $S(B|a) = - \sum_b p_{b|a} \log p_{b|a}$. The classical correlations can therefore be interpreted as information gain about one subsystem as a result of a measurement on the other.

In contradistinction to the classical case, in the quantum analog there are many different measurements that can be performed on a system, and measurements generally disturb the quantum state. A measurement on subsystem $A$ is described by a positive-operator-valued measure (POVM) with elements $E_a = M_a^{\dagger} M_a$, where $M_{a}$ is the measurement operator and $a$ is the classical outcome. The initial state $\rho_{AB}$ is transformed under the measurement (with unknown result) to
\begin{gather}\label{Postmesstate}
\rho_{AB} \to \rho'_{AB} = \sum_{a} M_{a} \rho_{AB} M^\dag_{a},
\end{gather}
where party $A$ observes outcome $a$ with probability $p_a = \tr (E_a \rho_{AB})$ and $B$ has the conditional state $\rho_{B|a} = \tr_A (E_a \rho_{AB}) / p_a$. This allows us to define a classical-quantum version of the conditional entropy, $S(B|\{E_a\}) \equiv \sum_a p_a S(\rho_{B|a})$, and introduce classical correlations of the state $\rho_{AB}$ in analogy with Eq.~\eqref{CL_CORR}, \cite{JPhysA.34.6899}:
\begin{gather}\label{HV_CL_CORR}
J(B|\{E_a\}) \equiv S(B) - S(B|\{E_a\}).
\end{gather}
To quantify the classical correlations of the state independently of a measurement $J(B|\{E_a\})$ is maximized over all measurements,
\begin{gather}
J(B|A) \equiv \max_{\{E_a\}} J(B|\{E_a\}).
\label{JBA}
\end{gather}

When the measurement is carried out by a set of rank-one orthogonal projections $\{\Pi_a\}$, the state on the right hand side of Eq.~\eqref{Postmesstate} has the form
\begin{gather}\label{CLASSICAL_QUANTUM_STATES}
\chi_{aB} = \sum_a p_a \Pi_a \otimes \rho_{B|a},
\end{gather}
which involves only fully-distinguishable states for $A$ and some indistinguishable states for $B$. Such states are often called classical-quantum (CQ) states, or quantum-classical (QC) when one exchanges the roles of $A$ and $B$. Note that for a CQ state there exists a von Neumann measurement of $A$ which does not perturb the state.

The quantum discord of a state $\rho_{AB}$ under a measurement $\{E_a\}$ is defined as a difference between total correlations, as given by the quantum mutual information in Eq.~\eqref{MI}, and the classical correlations Eq.~\eqref{HV_CL_CORR}, \cite{arXiv:quant-ph/0105072}:
\begin{align}
D(B|A) \equiv & I(A:B) - J(B|A) \nonumber\\
 =& \min_{\{E_a\}} \sum_a p_a S(\rho_{B|a}) + S(A) - S(AB).\label{DISCORD}
\end{align}
Note that the minimization here is equivalent to maximization in Eq.~\eqref{JBA}. This is just a difference between two classically-equivalent versions of conditional entropy $D(B|A) = \min_{\{E_a\}} S(B|\{E_a\}) - S(B|A)$, where $S(B|A) = S(AB) - S(A)$ is the usual conditional entropy, \cite{Book.Nielsen.Chuang}. This equivalence holds for rank-one POVM measurements which in classical theory correspond to questions about a value of a classical random variable. One could imagine classical coarse-grained measurements with outcomes which merge several values of the random variable (a given value may be present in several coarse-grained outcomes). Such measurements can be regarded as higher-rank classical POVMs, and conditional entropy under such measurements can be strictly bigger than the usual conditional entropy as the POVMs do not provide as much information as fine-grained measurements. This would not be satisfactory as the correlations are obtained with classical measurements on a classical random variable. In agreement with this conceptual point, it turns out that rank-one POVM measurements minimize the discord (see Sec.~\ref{POVM}).

%**************************************************************
\subsubsection{Properties of discord}

Quantum discord has the following properties: (a) It is not symmetric, i.e. in general $D(B|A) \ne D(A|B)$, which may be expected because conditional entropy is not symmetric. This can be interpreted in terms of the probability of confusing certain quantum states, see Sec.~\ref{SEC_REL_ENT_INTERPRETATION}; (b) Discord is nonnegative, $D \ge 0$, which is a direct consequence of the concavity of conditional entropy \cite{RevModPhys.50.221}; (c) Discord is invariant under local unitary transformations, i.e. it is the same for state $\rho_{AB}$ and state $(U_A \otimes U_B) \rho_{AB} (U_A \otimes U_B)^\dagger$. This follows from the fact that discord is defined via entropies, and the value obtained for measurement $\{E_a\}$ on the state $\rho_{AB}$ can also be achieved with measurement $\{U_A E_a U_A^\dagger\}$ on the transformed state. Note that discord is not contractive under general local operations, and therefore should not be regarded as a strict measure of correlations satisfying postulates of \cite{PhysRevA.83.012312}. However, $J(B|A)$ is contractive under general local operations. (d) Discord $D(B|A)$ vanishes if and only if the state is classical-quantum, \cite{arXiv:quant-ph/0105072, arXiv:0807.4490}. (e) discord is bounded from above as $D(B|A) \le S(A)$, while $J(B|A) \le \min \{S(A),S(B)\}$ \cite{PhysRevA.84.042124}.

%**************************************************************
\subsubsection{Thermal discord}\label{SEC_THM_DISCORD}

\cite{arXiv:quant-ph/0202123} present a slightly-different version of quantum discord:
\begin{gather}\label{thermal}
\tilde D_{\thm}(B|A) = \min_{\{ \Pi_a \}} [S(A') + S(B|\{ \Pi_a \})] - S(AB).
\end{gather}
Here the entropy $S(A')$ is the entropy of outcomes of $A$ after the orthogonal measurement $\{ \Pi_a \}$. \footnote{In general, tilde above a quantity means that it is optimized over rank-one orthogonal projective measurements.} The term being minimized above is exactly the entropy of the state $\chi_{aB}$ in Eq.~\eqref{CLASSICAL_QUANTUM_STATES}. Compared to the discord of Eq.~\eqref{DISCORD}, $D_{\thm}$ involves minimization of a sum of local entropy and the conditional entropy and therefore additionally takes into account the entropic cost of performing local measurements. This is relevant when discussing thermodynamics of correlated systems (see Sec.~\ref{SEC_PHYS_INTERPRETATION}), we therefore call $D_{\thm}$ the \emph{thermal discord}.

We note that this quantity may be further optimized by considering rank-one POVM measurements:
\begin{gather}\label{thermalPOVM}
D_{\thm}(B|A) \equiv \min_{\{ E_a \}} [S(\{p_a\}) + S(B|\{ E_a \})] - S(AB).
\end{gather}
However, in much of the literature only rank-one orthogonal projections are used when dealing with it. See Sec.~\ref{POVM} for a discussion on projective measurements versus POVM for thermal discord.

%**************************************************************
\subsubsection{Measurement-dependent discords}\label{SEC_MD_DISCORDS}

The two discords above are defined to be independent of external constraints by requiring the optimization over all measurements. There are however some circumstances where only a particular measurement (or a set of measurements) is relevant, for example when considering a particular measuring device. One can define the \emph{measurement-dependent discord} as:
\begin{gather}
D(B|\{E_a\})=I(A:B)-J(B|\{E_a\}),
\end{gather}
with fixed $\{E_a\}$. Note that since it involves no optimization it is not a particularly good measure of correlations \cite{arXiv:1108.3649}. In general $D(B|\{E_a\}) \le D_\thm(B|\{E_a\})$ for the same measurement with equality if $\rho_A= \sum_a M_a \rho_A M_a^\dag$ \cite{arXiv:1002.4913}, i.e., $\{M_a\}=\{\Pi_a^{\rm Eig}\}$, where $\rho_A=\sum_a p_a\Pi_a^{\rm Eig}$. The quantity $D(B|\{\Pi^{\rm Eig}_a\})$ is related to measurement-induced disturbance in Sec.~\ref{SEC_MID}.

%**************************************************************
\subsection{Quantum deficit}\label{SEC_DEFICIT}
%**************************************************************

This measure of quantum correlations originates in questions regarding work extraction from quantum systems coupled to a heat bath, \cite{arXiv:quant-ph/0112074, PhysRevLett.90.100402}. Their operational approach links quantum-correlations theory and quantum thermodynamics. \cite{arXiv:quant-ph/0202123} uses a similar approach to justify a physical interpretation of the optimized thermal discord in terms of Maxwell's demon (see Sec.~\ref{maxwell}). The corresponding notion of classicality is in the spirit of LOCC \cite{PhysRevLett.90.100402}. A system is classical, (only has classical correlations), if the same amount of work that can be extracted from the total system as one, $W_t$, can also be extracted from the subsystems after suitable LOCC operations, $W_l$. This motivates the definition of \emph{quantum deficit} as a work deficit
\begin{gather}
\Delta \equiv W_t - W_l.
\end{gather}
\cite{arXiv:quant-ph/0112074,Book.VN} relate the work extractable from the total system described by density operator $\rho_{AB}$ to its entropy:
\begin{gather}
W_t = \log d_{AB} - S(\rho_{AB}),
\label{DEFICIT_WT}
\end{gather}
where $d_{AB}$ is the dimension of the Hilbert space $\mathcal{H}_{AB}$, and we set the units such that work is measured in bits, i.e. the Boltzmann constant times the temperature of the bath are set to $k_B T=1$. In essence, the purer the state the more work can be extracted from it. In keeping with this, a subclass of LOCC operations should be allowed for the process of extracting work from the subsystems as adding ancillary systems in pure states, allowed in LOCC, would artificially increase the amount of extractable work. In order to make statements about the system of interest alone, it is considered closed and the only non-unitary operation allowed is sending a subsystem down the dephasing channel, which models classical communication. The resulting subclass of LOCC operations is called \emph{closed LOCC} (CLOCC). CLOCC does not allow a change in total number of particles and contains the following members: (a) local unitary operations, (b) sending a system down a dephasing channel (which can also be applied locally). The action of a dephasing channel on a state $\rho$ is to remove all of its off-diagonal elements in a specified basis $\rho \to \sum_j \Pi_j \rho \Pi_j$, where the projectors $\Pi_j$ define the basis of dephasing. It is allowed to change the basis of the dephasing channel from one use of it to another. Note that dephasing is equivalent to a local projective measurement with an unknown result, and general POVMs are not allowed within CLOCC paradigm (see Sec.~\ref{SEC_POVM_DEFICIT} for a general treatment).

Since entropy is a measure of ignorance, Eq.~\eqref{DEFICIT_WT} establishes an equivalence between the extractable work and information in a state. These two notions are interchangeable and the quantum deficit $\Delta = \mathsf{I}_t - \mathsf{I}_l$, is the difference between information contained in the whole system $\mathsf{I}_t \equiv W_t$ and the \emph{localizable information} $\mathsf{I}_l \equiv W_l$ \cite{arXiv:quant-ph/0410090, PhysRevA.67.062104}.

In the process of information localization, the initial state $\rho_{AB}$ is transformed via CLOCC operations to a state $\rho'_{AB}$. Since CLOCC operations keep the number of particles constant, for the final state the particles are just relocated, i.e. $d_{AB}' = d_{AB}$. By definition, the work that can be extracted locally from the subsystems is the sum of work extracted from subsystem $A$ in the state $\rho_A'$ and work extracted from subsystem $B$ in the state $\rho_{B}'$:
\begin{gather}
W_l = [\log d_A' - S(\rho'_A)] + [\log d_B' - S(\rho'_B)].
\end{gather}
In this way the following expression is obtained for the quantum deficit:
\begin{gather}
\Delta = \min [S(\rho_A') + S(\rho_B')] - S(\rho_{AB}),
\label{DEFICIT_DEL_MIN}
\end{gather}
where the minimum is over CLOCC operations. This is also called two-way deficit. For pure states this quantity measures entanglement \cite{PhysRevLett.90.100402}. In general, the tools from entanglement theory can be adopted here in order to distill local pure states and obtain bounds for the quantum deficit \cite{PhysRevLett.90.100402}. Similar techniques show that, for Werner and isotropic states, the deficit is lower-bounded by the (regularized) relative entropy of entanglement, in agreement with the intuition that it captures more quantum correlations than entanglement \cite{arXiv:quant-ph/0405149}.

%**************************************************************
\subsubsection{Zero-way deficit}

Various simpler forms of deficit differ in the type of communication allowed between parts $A$ and $B$. There are: zero-way, one-way, and two-way quantum deficits. For the \emph{zero-way deficit}, $A$ and $B$ are required to first fully dephase their local states before communicating, and only then to send the resulting states to use the obtained classical correlations. Therefore, the minimization in Eq.~\eqref{DEFICIT_DEL_MIN} is now over local dephasings. Due to subadditivity $S(\rho_{AB}') \le S(\rho_A') + S(\rho_B')$, it is best to transmit one of the subsystems to the other party and the zero-way deficit reads:
\begin{gather}
\tilde \Delta^{\emptyset} = \min_{\{\Pi_a \otimes \Pi_b\}} S(\rho_{AB}') - S(\rho_{AB}).
\label{DEFICIT_ZERO-WAY}
\end{gather}
Now the state after the dephasing is
\begin{gather}\label{STATE_AFTER_DEPHASING}
\rho_{AB} \to \rho'_{AB} = \sum_{a,b} \Pi_a \otimes \Pi_b \rho_{AB} \Pi_a \otimes \Pi_b,
\end{gather}
and is therefore of a general form
\begin{gather}\label{CLASSICAL_CLASSICAL_STATES}
\chi_{ab} = \sum_{a,b} p_{ab} \Pi_a \otimes \Pi_b.
\end{gather}
Such a state is called a classical-classical state or CC state \cite{arXiv:quant-ph/0112074, arXiv:0707.0848}.\footnote{When it does not lead to confusion we use the generic name classical states to denote CQ, QC or CC states.} Since all projectors in this decomposition correspond to fully distinguishable states, the probability $p_{ab}$ can be regarded as a classical joint probability of random variables $a$ and $b$.

Zero-way deficit equals minimal relative entropy between $\rho_{AB}$ and a state belonging to the set of CC states \cite{arXiv:quant-ph/0410090}.

%**************************************************************
\subsubsection{One-way deficit}

In the \emph{one-way deficit} $A$ can communicate (via a dephasing channel) to $B$. In this way they produce a state in the from of Eq.~\eqref{CLASSICAL_QUANTUM_STATES}. If another state is obtained, $A$ can always dephase $\rho_A'$ in its eigenbasis without changing local entropy and bring it to the zero-discord state. Therefore, one-way deficit reads:
\begin{gather}\label{one-way-deficit}
\tilde \Delta^{\to} = \min_{\{\Pi_a\}} S(\rho'_{AB}) - S(\rho_{AB}).
\end{gather}
This quantity is equal to the thermal discord $\tilde D_{\thm}$ \cite{arXiv:quant-ph/0202123}. It is also given by the relative entropy to the set of CQ states \cite{arXiv:quant-ph/0410090}.

%**************************************************************
\subsubsection{Two-way deficit}

We finish this brief survey of different types of deficit by noting that the two-way quantum deficit, $\Delta$, can be strictly smaller than the one-way deficit $\tilde \Delta^{\to}$ \cite{arXiv:quant-ph/0410090}. Let $\rho_{\to} \equiv \sum_a p_a \Pi_a \otimes \rho_{B|a}$ and $\rho_{\leftarrow} \equiv \sum_b p_b \rho_{A|b} \otimes \Pi_b$, and consider the mixture
\begin{gather}\label{twoway}
\rho \equiv p_{\to} \rho_{\to} \otimes \Pi_{a'=0} + p_{\leftarrow} \rho_{\leftarrow} \otimes \Pi_{a'=1}.
\end{gather}
Here $A$ holds the additional system $A'$. This state has vanishing $\Delta$ because once $A'$ is measured $A$ and $B$ can use suitable one-way communication to localize all its information. However, one-way deficit is strictly positive, e.g. $\tilde \Delta^{\to} > 0$, because with probability $p_{\leftarrow}$ the state $\rho_{\leftarrow}$ has positive $\tilde \Delta^{\to}$. If $A$ observes $A'$ to be in `0' then $AB$ have a CQ state and if she observes `1' then they have a QC state.

%**************************************************************
\subsubsection{Classical deficit}\label{SEC_CLASSICAL_DEFICIT}

In a similar manner \emph{classical deficit} is introduced as:
\begin{gather}
\Delta_{cl} \equiv \mathsf{I}_l - \mathsf{I}_{LO},
\end{gather}
which captures how much more information can be obtained from $\rho_{AB}$ by exploiting classical correlations with the help of dephasing channel. Here $\mathsf{I}_{LO} \equiv \log d_{AB} - S(\rho_A) - S(\rho_B)$ stands for local information of the initial state. Surprisingly, the one-way version of this quantity $\Delta_{cl}^{\to} \equiv \mathsf{I}_l^{\to} - \mathsf{I}_{LO}$, although similar to classical correlations Eq.~\eqref{HV_CL_CORR}, is not a proper measure because it can increase under local operations \cite{arXiv:quant-ph/0403167}.

%**************************************************************
\subsection{Distillable common randomness}
%**************************************************************

What is the amount of classical correlations in a bipartite quantum state? \cite{arXiv:quant-ph/0304196} take yet another information-theoretic approach to this question.

A natural amount of classical correlations is present in a perfectly-correlated pair of classical binary variables each having a full bit of entropy:
\begin{gather}
\rho_{\mathrm{cr}} = \frac{1}{2} \ket{00} \bra{00} + \frac{1}{2} \ket{11} \bra{11}.
\end{gather}
This state is referred to as common randomness. Roughly speaking, the measure of classical correlations we describe below gives the number of states $\rho_{\mathrm{cr}}$ into which the initial state can be converted asymptotically. More rigorously, one first considers many copies of a bipartite state $\rho_{AB}$, and a fixed amount $m$ of classical communication from $A$ to $B$ in order to optimize the amount of common randomness $C(m)$ per copy.
The \emph{one-way distillable common randomness} is defined as
\begin{gather}
D_{cr}^{\to} \equiv \lim_{m \to \infty}[C(m) - m],
\end{gather}
and therefore quantifies obtainable common randomness in excess of the invested classical communication.

It turns out that this quantity is equivalent to a regularized version of the classical correlations $J(B|A)$ of Sec.~\ref{SEC_DISCORD}, giving it operational meaning \cite{arXiv:quant-ph/0304196} (see Sec.~\ref{SEC_REGULARIZATION}). It is also equal to the regularized version of one-way classical deficit \cite{PhysRevA.71.062303}. Note the subtlety here: It follows that the regularized one-way classical deficit is equivalent to the regularized classical correlations $J(B|A)$ but, as mentioned in Sec.~\ref{SEC_CLASSICAL_DEFICIT}, the equivalence no longer holds in a single-copy scenario where the one-way classical deficit is not monotonic (may increase) under local operations. Therefore, regularization here regains monotonicity of the one-way classical deficit \cite{arXiv:quant-ph/0403167}.

%**************************************************************
\subsection{Measurement-induced disturbance}\label{SEC_MID}
%**************************************************************

At the heart of classical physics lies realism: in principle, measurements can reveal properties of a classical system without modifying the system. \cite{PhysRevA.77.022301} formalize this notion of classicality with the \emph{measurement-induced disturbance} (MID). Mathematically the same quantity as that given in Eq.~\eqref{MID_SIMP} is also introduced in \cite{PhysRevA.66.022104} under the name deficit, though not to be confused with the deficit from Sec.~\ref{SEC_DEFICIT}.

When a bipartite state is measured by local projective measurements, the post-measurement state is given in Eq.~\eqref{STATE_AFTER_DEPHASING}. A state $\rho_{AB}$ is classical if there exists local measurements which do not perturb it, i.e. $\rho_{AB} = \rho'_{AB}$. MID is defined as the difference:
\begin{gather}\label{MID}
M = I(\rho_{AB}) - I(\rho'_{AB}),
\end{gather}
where $I(\cdot)$ denotes quantum mutual information, and $\rho_{AB}'$ is given by Eq.~\eqref{CLASSICAL_CLASSICAL_STATES} with local measurements induced by the spectral decomposition of the reduced states $\rho_A = \sum_a p_a \Pi_a$ and $\rho_B = \sum_b p_b \Pi_b$. Since the reduced states are not affected by this measurement, MID is just the entropic cost of a measurement in this basis:
\begin{gather}\label{MID_SIMP}
M = S(\rho'_{AB}) - S(\rho_{AB}).
\end{gather}

On the other hand, depending on the context, we can choose any local measurements to define a fixed-measurement measure similar to MID. Due to the concavity of von Neumann entropy, this is a nonnegative quantity with the advantage that it is simple to calculate. An asymmetric version of MID relates to the work-deficit within Maxwell's demon paradigm \cite{arXiv:1002.4913}, see Sec.~\ref{SEC_MD_DISCORDS}. Another asymmetric measure in the spirit of MID is the so-called \emph{measurement-induced nonlocality} quantifying a change of the whole system under a measurement on its subsystem only \cite{PhysRevLett.106.120401, IJQI.9.1587}.

%**************************************************************
\subsection{Symmetric discord}
%**************************************************************

The fact that MID does not involve any optimization has been criticized as this results in overestimation of the amount of nonclassical correlations. Moreover, for states whose reduced operators have degenerated spectrum, $M$ is not uniquely defined. This leads to positive (even maximal) values of $M$ even for classical states, e.g. if their reduced operators are completely mixed \cite{arXiv:1008.4136}. In this case MID is also discontinuous \cite{arXiv:0905.2123, arXiv:1108.3649}. To circumvent these problems, an optimized version of MID is proposed
\begin{gather}\label{AMID}
D_S \equiv I(\rho_{AB}) - \max_{\{E_a \otimes E_b\}} I(\rho_{AB}'),
\end{gather}
where optimization is over general local measurements. We call this optimized quantity \emph{symmetric discord}, but it is also known as WPM discord, after the authors of \cite{arXiv:0905.2123}, and ameliorated MID.

\cite{arXiv:0707.0848, arXiv:0905.2123} introduce and study symmetric discord with POVMs, while \cite{arXiv:1008.4136} study it with with projective measurements.  \cite{arXiv:0905.2123, arXiv:1105.2993, PhysRevA.82.052122} show that for one-- or two--sided measurements the classical part of correlations, i.e. $\max I(\rho_{AB}')$, is greater than $\min[S(\rho_A),S(\rho_B)]$. \cite{PhysRevA.82.052122} conjecture the same bound for quantum discord, and \cite{arXiv:1105.2993, PhysRevA.84.042124} show the conjecture to be true. \cite{arXiv:1004.2082}, \cite{arXiv:1012.4302} and \cite{JStatPhys.136.165} also discuss symmetric discord. The latter authors trace its origins as far back as \cite{CommMathPhys.33.305, LectNotePhys.378.71}. See \cite{Phys.Rev.Lett.103230502} for an argument for symmetric classical correlations.

%**************************************************************
\subsection{Relative entropy of discord and dissonance}\label{SEC_DISSONANCE}
%**************************************************************

%**************************************************************
\begin{figure}[t]
\resizebox{8 cm}{3.43 cm}{\includegraphics{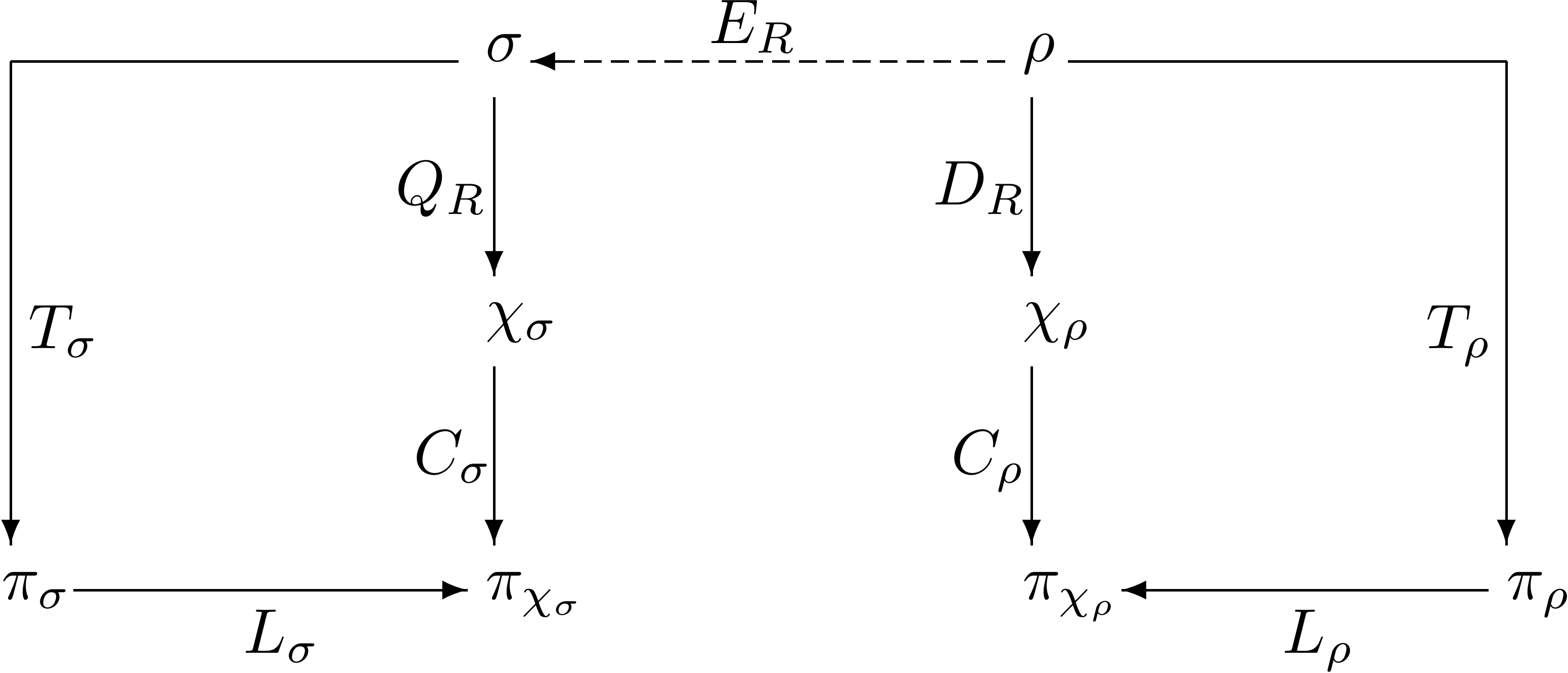}}
\caption{\label{ALLSTATES} \emph{Relative entropy of discord and dissonance.} The diagram shows the relationships between various states used in constructing correlations measures based on relative entropy. An arrow $X \to Y$ indicates that $Y$ is the closest state to $X$ as measured by the relative entropy $S(X\|Y)$. The relevant states $Y$ belong to different subsets as follows: $\rho\in\mathcal{E}$ (the set of entangled states), $\sigma\in\mathcal{S}$ (the set of separable states), $\chi\in\mathcal{C}$ (the set of classical states), and $\pi\in\mathcal{P}$ (the set of product states). The resulting measures are relative entropy of entanglement $E_R$, relative entropy of discord $D_R$, relative entropy of dissonance $Q_R$, total mutual information $T_\rho$ and $T_\sigma$, classical correlations $C_\sigma$ and $C_\rho$, and the local-entropic cost of a dephasing channel $L_\rho$ and $L_\sigma$. All relative entropies, except for $E_R$ (dotted line), reduce to the differences in entropies of $Y$ and $X$, $S(X\|Y)=S(Y)-S(X)$, leading to additivity relations across closed paths. This figure is reproduced from \cite{arXiv:0911.5417}.}
\end{figure}

The Kullback-Leibler divergence or relative entropy is a frequently-used tool to distinguish two probability distributions or density operators. It resembles a distance measure, however it is not symmetric. \cite{arXiv:0911.5417} laid out a unifying approach to various correlations based on the idea that a distance from a given state to the closest state without the desired property (e.g., entanglement or discord) is a measure of that property. For example, the shortest distance to the set of separable states (belonging to set $\mathcal{S}$) is a meaningful measure of entanglement called relative entropy of entanglement \cite{PhysRevLett.78.2275, PhysRevA.57.1619, PhysRevA.67.012313}. Similarly, one defines the shortest distances to the set of classical states (states belonging to set $\mathcal{C}$) or product states (states belonging to set $\mathcal{P}$). If all the distances are measured with relative entropy $S(X\|Y) \equiv -\tr(X \log Y) - S(X)$, the resulting measures are \cite{arXiv:0911.5417}:
\begin{align}
E_R = & \min_{\sigma \in \mathcal{S}} S(\rho\|\sigma) \quad \textrm{(relative entropy of entanglement)}, \nonumber \\
D_R = & \min_{\chi \in \mathcal{C}} S(\rho\|\chi) \quad \textrm{(relative entropy of discord)}, \nonumber \\
Q _R = & \min_{\chi \in \mathcal{C}} S(\sigma\|\chi) \quad \textrm{(relative entropy of dissonance)}. \nonumber
\end{align}
The state $\rho$ in these expressions belongs to the set of entangled states $\mathcal{E}$, $\sigma$ is in the set of separable states $\mathcal{S}$, $\chi$ is in the set of classical states $\mathcal{C}$, and $\pi$ is in the set of product states $\mathcal{P}$. In this way quantum dissonance is defined as nonclassical correlations which exclude entanglement. Interestingly, relative entropy of dissonance is not present in pure bipartite states, but can appear in pure multipartite states. For example the $\ket{W}$ state of three qubits $\ket{W} = \frac{1}{\sqrt{3}}(\ket{100} + \ket{010}+\ket{001})$ admits almost $1$ bit of dissonance \cite{arXiv:0911.5417}. It is conjectured that $N$-partite $\ket{W}$ states contain $\log N$ bits of relative entropy of discord and it is unknown which part of it is the dissonance \cite{arXiv:0909.4443}. An advantage of using distance-like measures is that everything can be defined for multipartite states, see Sec.~\ref{multipartite}. It also turns out that $D_R$ and $Q_R$ are optimized by an orthogonal projective measurement \cite{arXiv:0911.5417}.

Various relations between these measures are presented on the diagram in Fig.~\ref{ALLSTATES}. It turns out that most of the quantities are given by the entropic cost (difference of entropies) of performing operations bringing the initial state to the closest state without the desired property. In particular, if the set of classical states is considered to be the set of states of Eq.~\eqref{CLASSICAL_CLASSICAL_STATES}, the relative entropy of discord is just zero-way quantum deficit $D_R = \tilde \Delta^{\emptyset}$ \cite{arXiv:quant-ph/0410090}. If the set of classical states is considered as the set of classical-quantum states, Eq.~\eqref{CLASSICAL_QUANTUM_STATES}, the corresponding relative entropy of discord under one-sided measurements $D^{\to}_R$ is just thermal discord and one-way deficit $D^{\to}_R =\tilde D_{\thm} = \tilde \Delta^{\to}$ \cite{arXiv:quant-ph/0410090}. For a given measurement $\{\Pi_a\}$, $D_R$ is related to discord as $D = D^{\to}_R - L_{\rho}$ \cite{arXiv:0911.5417}. However, the optimizations of the two are not the same. Some of the entropic costs form closed loops in Fig.~\ref{ALLSTATES} giving rise to the additivity relations
\begin{align}
T_{\rho} + L_\rho = & D_R + C_\rho.
\end{align}
The same relation holds for dissonance.

%**************************************************************
\subsection{Geometric measures}\label{SEC_GEOMETRIC_DISCORD}
%**************************************************************

%**************************************************************
\begin{figure}[t]
\resizebox{8 cm}{7.34 cm}{\includegraphics{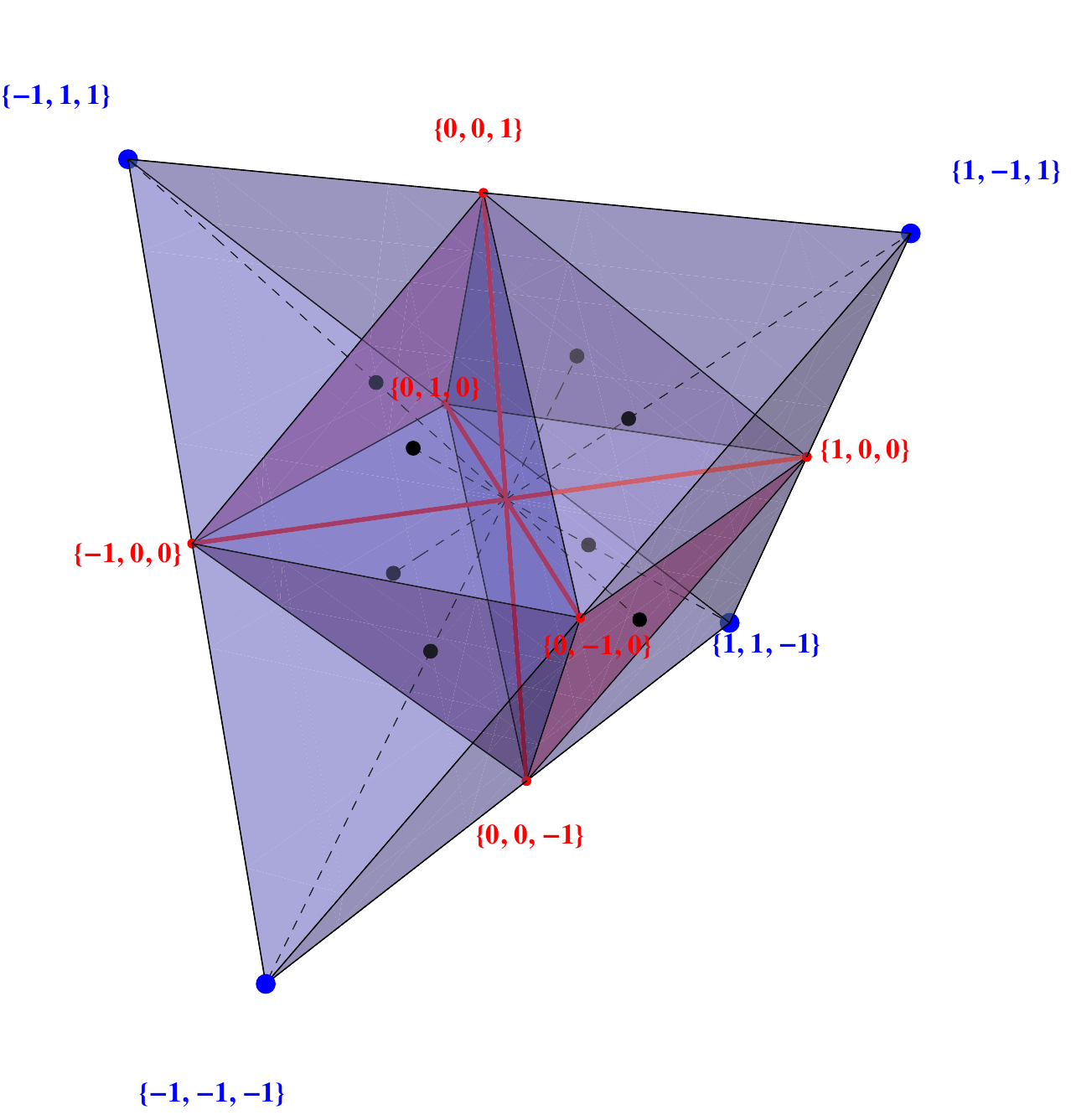}}
\caption{\label{FIG_DG} (Color online.) \emph{Geometric discord.} The set of two-qubit states with maximally-mixed marginals, the so-called Bell diagonal states: On the axes we plot $T_{11}$, $T_{22}$ and $T_{33}$ of decomposition in Eq.~\eqref{2QUBIT_BLOCH}. Physical states belong to the tetrahedron, among which the separable ones are confined to the octahedron \cite{PhysRevA.54.1838}. The states with vanishing geometric discord are labeled by the red lines. It is therefore clear that almost all states have finite discord \cite{arXiv:0908.3157}. The states with maximal $D_G$ are the four Bell states corresponding to vertices of the tetrahedron. Among the set of separable states, those which maximize geometric discord correspond to the centers of octahedron faces and are given by Eq.~\eqref{DG_MAXD_SEP}. This figure is reproduced from \cite{arXiv:1004.0190}.}
\end{figure}

Relative entropy is technically not a metric, e.g. it is not symmetric. We now discuss a measure based on a proper distance metric---the Hilbert-Schmidt distance \cite{arXiv:1004.0190}:
\begin{gather}
D_G \equiv \min_{\chi \in \mathcal{C}} \| \rho - \chi \|^2 = \min_{\chi \in \mathcal{C}} \tr \left[ ( \rho - \chi)^2 \right].
\end{gather}
If $\mathcal{C}$ is the set of classical-quantum states, Eq.~\eqref{CLASSICAL_QUANTUM_STATES}, this measure is known as \emph{geometric quantum discord}. Like the relative entropy of discord, the geometric measure gives the Hilbert-Schmidt distance to the state after the (optimal) measurement \cite{PhysRevA.82.034302}:
\begin{gather}\label{D_G_MIN_OVER_A}
D_G = \min_{\{\Pi_a\}} \|\rho - \rho'\|^2,
\end{gather}
where $\rho' = \sum_a \Pi_a \rho \Pi_a$. We shall prove it in a simple way. Assume $\chi_\rho$ is the closest classical state to $\rho$, i.e. $\|\rho - \chi\|^2 - \|\rho - \chi_\rho\|^2 \ge 0$ for any classical $\chi$. We show that the closest state is given by $\rho$ dephased in the eigenbasis of the closest state $\chi_\rho = \sum_k \lambda_k \ket{k} \bra{k}$. To this end, consider $\chi = \sum_k \ket{k} \bra{k} \rho \ket{k} \bra{k}$, and note that this form implies $\tr(\rho \chi) = \tr(\chi^2)$ and similarly $\tr(\rho \chi_\rho) = \tr(\chi \chi_\rho)$. This gives $\|\chi_\rho - \chi \|^2 \le 0$, which must vanish for $\chi_\rho = \sum_{k} \ket{k} \bra{k} \rho \ket{k} \bra{k}$. The same argument applies to measurements of the form $\Pi_a \otimes \openone$ and therefore to classical-quantum states.

Recently \cite{arXiv:1112.6370} study a unified version of geometric discord in a manner similar to the study of \cite{arXiv:0911.5417}. They found that the closest product state to a given quantum state is not the product of the marginal states, which makes computing the total correlations with a geometric measure nontrivial. On the other hand, the result above shows that the closest classical state is obtained by dephasing the quantum state. Putting it all together, they find that unlike for the relative entropy measures, geometric measures of correlations are not additive. They give an additivity expression for correlations as function of the original state for X-states, given in Eq.~\eqref{xstate}.

%**************************************************************
\subsubsection{Analytic formulas}\label{sec:GDanalytic}

The advantage of the geometric measure is that the minimization present in the definition can be performed explicitly. Consider first general two-qubit states. They admit the representation
\begin{gather}
\rho_{AB} = \frac{1}{4} \sum_{\mu = 0}^3 \sum_{\nu = 0}^3 T_{\mu \nu} \sigma_\mu \otimes \sigma_\nu,
\label{2QUBIT_BLOCH}
\end{gather}
where $\sigma_{\mu} = \{\openone,\sigma_x,\sigma_y,\sigma_z\}$ is the $\mu$th Pauli operator and the reals $T_{\mu \nu} \in [-1,1]$ are experimentally-accessible averages $T_{\mu \nu} = \tr(\rho \sigma_\mu \otimes \sigma_\nu)$. The geometric discord of a quantum state $\rho_{AB}$ equals:
\begin{gather}
D_G = \frac{1}{4}\sum_{k=1}^3 \sum_{\nu = 0}^3 T_{k \nu}^2 - \lambda_{\max},
\label{DG_LAMBDAMAX}
\end{gather}
where $\lambda_{\max}$ is the largest eigenvalue of the matrix $L = \vec a \vec a^T + \hat T \hat T^T$, built from the local Bloch vector $\vec a = (T_{10},T_{20},T_{30})$ and correlation matrix $\hat T$ having as entries $T_{kl}$ for $k,l=1,2,3$ \cite{arXiv:1004.0190}. For an explicit form of $\lambda_{\max}$ see \cite{arXiv:1110.5083}. This reveals, for example, that separable Bell-diagonal states with maximal discord have a simple symmetric form
\begin{gather}
\sigma_{j_1 j_2 j_3} \equiv \frac{1}{4}\left(\openone \otimes \openone + \frac{1}{3} \sum_{k=1}^3 (-1)^{j_k} \sigma_k \otimes \sigma_k\right),
\label{DG_MAXD_SEP}
\end{gather}
with $j_k = 0,1$, see Fig.~\ref{FIG_DG}. Intuitively this should be the case as they are evenly-weighted mixtures of ``maximally nonorthogonal" states.

\cite{arXiv:1010.1920, arXiv:1106.4488} claim similar results for more general bipartite states. \cite{arXiv:1107.2958} give an analytic formula for symmetric geometric discord for two-qubit systems. Geometric discord can be established directly from experimental data measured on up to six copies of a quantum state \cite{arXiv:1110.5693}. The idea is to rephrase the discord in terms of functions of powers of density operators and use known circuits for their implementation \cite{PhysRevLett.89.127902}. The methods utilized in the above articles are highly technical and we therefore forego the details here.

%**************************************************************
\subsubsection{Bounds on geometric discord}

\cite{arXiv:1110.5083} introduce a remarkably-tight lower bound on geometric discord $D_G$ of two qubits:
\begin{gather}
Q = \frac{1}{12}\left[ 2 \tr(L) - \sqrt{6 \tr(L^2) - 2 \tr(L)^2}\right],
\end{gather}
where $L$ is defined below Eq.~\eqref{DG_LAMBDAMAX}. A similar bound exists for systems in $2 \times d$ dimensions. The value of $Q$ (numerically) upper bounds the negativity of two-qubit states squared, i.e. $\mathcal{N}^2 \le Q \le D_G$, with equalities for pure states \cite{PhysRevA.84.052110}. In terms of quantum discord, the geometric discord of two qubits admits the bound $D_G \ge \frac{1}{2} D^2$ \cite{PhysRevA.82.034302}. \cite{PhysRevA.83.052108} give another lower bound on geometric discord, in terms of the correlation tensor of a general bipartite state.

%**************************************************************
\subsection{Continuous-variable discord}\label{SEC_GAUSSIAN_DISCORD}
%**************************************************************

A wide class of infinite-dimensional physical systems, of considerable experimental relevance, are describable using Gaussian states and Gaussian operations: A state is defined as Gaussian if its Wigner function (or equivalently characteristic function) is Gaussian. Gaussian operations are those operations which map Gaussian states to Gaussian states \cite{PhysRevA.36.3868, PhysRevA.49.1567}.

Gaussian quantum discord is defined as in Eq.~\eqref{DISCORD} with the restriction that the measurement of $A$ is a general single-mode \textit{Gaussian} POVM \cite{arXiv:1003.4979, arXiv:1003.3207}. These measurements are all executable using linear optics and homodyne detection \cite{PhysRevA.66.032316}.

A general form of Gaussian quantum discord is obtained for two-mode Gaussian states, i.e. both measurements and states are Gaussian \cite{arXiv:1003.4979, arXiv:1003.3207}. Such states $\rho_{AB}$ are fully specified, up to local displacements, by covariance matrix $\gamma$ with entries $\gamma_{kl} = \tr[\rho_{AB}(R_k R_l+R_l R_k) ]$, where $\vec R = (x_A,p_A,x_B,p_B)$ is the vector of phase-space operators \cite{JPhysA.40.7821}. Local unitary operations correspond to local symplectic operations of the covariance matrix and since quantum discord is not affected by them it is sufficient to study two-mode states in the standard form with diagonal sub-blocks
\begin{gather}
\gamma = \left(
\begin{array}{cc}
A & C \\
C & B
\end{array}
\right),
\end{gather}
where $A = a \openone$, $B = b \openone$ and $C = \mathrm{diag}(c,d)$.
In these terms the Gaussian quantum discord reads
\begin{align}
D^\to_{\rm CV}(\gamma) =& f(\sqrt{\det A}) - f(\lambda_-) - f(\lambda_+) \nonumber\\
&+ \min_{\gamma_0} f(\sqrt{\det \beta}),
\end{align}
where $\lambda_{\pm}$ are symplectic eigenvalues defined by $2 \lambda_{\pm}^2 = S \pm \sqrt{S^2 - 4 \det \gamma}$ with $S = A+B+2C$, and $\beta = B - C(A - \gamma_0)^{-1} C^T$ is the single-mode covariance matrix of $B$ after the measurement on $A$, and $f(x) = \frac{x+1}{2} \log(\frac{x+1}{2}) - \frac{x-1}{2} \log(\frac{x-1}{2})$. Minimization is over all covariance matrices $\gamma_0$ corresponding to pure single-mode Gaussian states. Given a two-mode covariance matrix $\gamma$ a closed formula is known for Gaussian quantum discord \cite{arXiv:1003.4979}.

%**************************************************************
\subsubsection{Properties of Gaussian discord}

A remarkable conclusion from these studies is that all nonproduct Gaussian states have nonclassical correlations according to Gaussian discord. This is somewhat ironic given the history of the Gaussian states. All Gaussian states were at first considered classical because of the nonnegativity of their Wigner function. On the other hand if classical states are defined as having nonnegative-regular Sudarshan-Glauber functions, then almost all two-mode Gaussian states are nonclassical, as found in numeric studies \cite{JOptB.2.L19}. Finally, according to Gaussian discord all nonproduct states are nonclassical.

Other properties of Gaussian discord include: (a) separable Gaussian states admit the bound ${D}^\to_{\rm CV} \le 1$; (b) Gaussian entanglement of formation $E_{\rm CV}$ tightly bounds Gaussian discord: $F_{\mathrm{\downarrow}}(E_{\rm CV}) \le D^\to_{\rm CV} \le F_{\mathrm{\uparrow}}(E_{\rm CV})$, for exact functions of the bounds see \cite{arXiv:1003.4979}; (c) let ${D}^{\max}_{\rm CV} \equiv \max \left\{ {D}^{\to}_{\rm CV}, {D}^{\leftarrow}_{\rm CV} \right\}$ and similarly $D^{\min}_{\rm CV} \equiv \min \left\{ D^{\to}_{\rm CV}, {D}^{\leftarrow}_{\rm CV} \right\}$, then the asymmetry-property ${D}^{\max}_{\rm CV} - D^{\min}_{\rm CV} \le {D}^{\min}_{\rm CV}/[{\rm exp}({D^{\min}_{\rm CV}})-1]$ is numerically given in \cite{arXiv:1003.4979}.

Other quantum-correlations measures are adapted to continuous-variable systems. \cite{arXiv:1110.2532} discuss geometric discord for Gaussian states  and \cite{arXiv:1012.4302, arXiv:1111.1101} study symmetric discord in detail. The latter reveals that non-Gaussian measurements such as photocounting may minimize quantum correlations both for Gaussian and non-Gaussian states. This raises the question of whether Gaussian discord overestimates quantum correlations.

%**************************************************************
\subsection{Generalized measurements}\label{POVM}
%**************************************************************

Many of our discussions up to now involved projective measurements when optimization is required for defining measures. Naturally, one would like to know if projective measurements are optimal. We now present results for different measures showing that extremal rank-one POVM measurements are optimal, and orthogonal projective measurements are sometimes not enough.

%**************************************************************
\subsubsection{Positive-operator-valued measure}

A positive-operator-valued measure (POVM), denoted as $\{E_a\}$, is a set of positive operators $E_a$ called \emph{POVM elements} that sum to identity, reflecting positivity and normalization condition for probabilities. As positive operators, each $E_a$ can be diagonalized and the number of its nonzero eigenvalues gives the rank of the POVM element. Rank-one POVMs are of special interest and they are defined to be POVMs with only rank-one elements. These elements are proportional to projectors, but these projectors need not be orthogonal. The set of POVMs is convex, i.e. if $E_{a}^{(1)}$ and $E_{a}^{(0)}$ are elements of a POVM, then the convex combination of elements $E_{a} \equiv p E_{a}^{(1)} + (1-p) E_{a}^{(0)}$ defines a valid POVM . This structure reflects an experimentalist's freedom to randomly choose one of many measuring apparatuses. A POVM is called \emph{extremal} if it cannot be represented as a convex combination of other POVMs. A rank-one POVM is extremal if and only if its elements $E_a$ are linearly independent \cite{arXiv:quant-ph/0408115}.

Every POVM element can be written as $E_{a} = M_a^\dagger M_a$ where $M_a$ is called measurement operator. This decomposition is not unique and therefore knowledge of POVM elements is not sufficient to describe post-measurement states. The full physical evolution is codified by the measurement operators. The post-measurement state, ignoring the measurement outcome, is given by the map $\rho' = \mathcal{E}(\rho) = \sum_a M_a \rho M_a^\dagger$. Due to the nonuniqueness it happens that, e.g. a nonextremal POVM can admit an extremal map \cite{JMathPhys.52.082202}.

%**************************************************************
\subsubsection{Symmetric discord}

Consider measures based on mutual information such as $D_S$ of Eq.~\eqref{AMID}. The goal is to maximize classical mutual information of the results of general local measurements. Here we show that coarse-grained measurements reveal less mutual information than fine-grained measurements \cite{arXiv:1105.4920}. Consider coarse-grained POVM elements of $A$: $E_a = \sum_k E_{ak}$, where $E_{ak}$ are the fine-grained elements.
The coarse-grained element can always be fine-grained to the rank-one level by writing it in terms of its spectral decomposition. The coarse-grained measurement has outcomes $a$ whereas the fine-grained outcomes are $a$ and $k$. Similarly for $B$, the fine-grained measurement gives outcomes $b$ and $l$. Since mutual information cannot increase when dropping local variables $I(a,k:b,l) \ge I(a:b)$, it is optimal to choose rank-one POVM.

The optimal POVM has to be extremal due to the joint convexity property of relative entropy.
Classical relative entropy $S(p_{ab}\|p_a p_b)$ is equal to mutual information, and joint convexity means that
\begin{align}
S & \left(p \, p_{a b}^{(1)} + (1-p) \, p_{ab}^{(0)}\|p \, p_{a}^{(1)} p_b + (1-p) \, p_{a}^{(0)} p_b \right) \nonumber \\
& \le p \, S \left( p_{a b}^{(1)} \| p_{a}^{(1)} p_b \right) + (1+p) \, S \left(p_{a b}^{(0)} \| p_{a}^{(0)} p_b \right),
\end{align}
where it is presented for $A$'s convex POVM $E_{a} \equiv p E_{a}^{(1)} + (1-p) E_{a}^{(0)}$ that yields probabilities of measurement results $p_a = p \, p_{a}^{(1)} + (1-p) \, p_{a}^{(0)}$. The same reasoning applies to $B$'s measurement and we conclude that mutual information of measurement results is maximized by an extremal POVM.

\cite{arXiv:0905.2123, arXiv:1105.4920} give an explicit example using a qubit-qutrit system for which the maximum of classical correlations is attained by a genuine rank-one POVM, and not a projective measurement onto orthogonal states. The example is related to studies of maximal accessible information and reads
\begin{gather}
\rho_{AB} = \frac{1}{3} \sum_{a=1}^3 \Pi_a \otimes \rho_{B|a},
\label{EX_POVM_NECCESSARY}
\end{gather}
where $\Pi_a$ are orthogonal projectors spanning the qutrit Hilbert space, and $\rho_{B|a}$ are three states of a qubit with corresponding Bloch vectors forming an equilateral triangle.
It turns out that the optimal measurement of $A$ is in the basis of $\Pi_a$'s, in which case $B$ has an even mixture of $\rho_{B|a}$ states. \cite{ProblPeredachiInf.9.31, ProblInfTransm.9.110} study this exact situation show that a certain three-outcome POVM extracts strictly more information than any two-outcome projective measurement on $B$'s qubit. Note that it is not known whether for two qubits orthogonal projective measurements maximize the classical correlations.

%**************************************************************
\subsubsection{Quantum discord}

The following argument that quantum discord is optimized by rank-one POVM is due to \cite{arXiv:0807.4490}. The quantity to be minimized is the classical-quantum version of conditional entropy $S(B|\{E_a\}) = \sum_a p_a S(\rho_{B|a})$. If instead of $E_a$ one considers its fine-graining $E_a = \sum_k E_{ak}$, the corresponding classical-quantum conditional entropy is $S(B|\{E_{ak}\}) = \sum_{a,k} p_{ak} S(\rho_{B|ak})$ with the state of $B$ after measurement of $A$ being $\rho_{B|ak} = \tr_A(E_{ak} \rho_{AB})/p_{ak}$. Since
\begin{align}
\rho_{B|a} =& \, \tr_A(E_a \rho_{AB})/p_a \nonumber\\
=& \sum_k \tr_A(E_{ak} \rho_{AB})/p_a = \sum_k p_{k|a} \rho_{B|ak},
\end{align}
where we used $p_{k|a} = p_{ak}/p_a$, and due to concavity of entropy we have $S(B|\{E_a\}) \ge S(B|\{E_{ak}\})$. It is therefore optimal to choose a rank-one POVM. Note that the POVM formalism is perfectly suited here, because to calculate the discord we do not need the post-measurement states of $A$: only the probabilities $p_a$ are important and the post-measurement states of $B$.

A similar line of reasoning shows that the optimal rank-one POVM has to be extremal \cite{PhysRevA.70.052325}. Let us denote by $\rho_{B|a}^{(j)} = \tr_A \left(E_a^{(j)} \rho_{AB} \right) / p_{a}^{(j)}$ where we also abbreviated $p_{a}^{(j)} = \tr_{AB}\left(E_a^{(j)} \rho_{AB} \right)$ for $j=0,1$. In this notation
\begin{align}
p_a = & p \, p_{a}^{(1)} + (1-p) \,p_{a}^{(0)}, \nonumber \\
\rho_{B|a} = & p \, \frac{p_{a}^{(1)}}{p_a} \rho_{B|a}^{(1)} + (1-p) \, \frac{p_{a}^{(0)}}{p_a} \rho_{B|a}^{(0)}.
\end{align}
Note that $\rho_{B|a}$ is now represented as a convex mixture. Plugging this into $S(B|\{E_a\})$ and using concavity of the entropy we find
\begin{gather}
S(B|\{E_a\}) \ge p S(B|\{E_a^{(1)}\}) + (1-p) S(B|\{E_a^{(2)}\}).
\label{POVM_DISCORD_CQ-COND-ENT}
\end{gather}
It is therefore optimal to choose an extremal POVM giving the smaller of the classical-quantum conditional entropies on the right-hand side. As an application, note that any POVM with more than four elements acting on a two-dimensional Hilbert space is not extremal \cite{arXiv:quant-ph/0408115} and therefore cannot optimize the discord.

The example of Eq.~\eqref{EX_POVM_NECCESSARY} can be adopted to show that discord is optimized by rank-one POVM which is not a set of projectors onto orthogonal states \cite{arXiv:1105.4920}. Furthermore, quantum discord is optimized by a genuine rank-one POVM already for some two-qubit examples \cite{arXiv:quant-ph/0403167, arXiv:1107.2005, PhysRevA.84.042313, arXiv:1110.6681}. However, orthogonal projective measurements give a pretty-tight upper bound on discord, and there is only a tiny set of states for which numerics shows the difference \cite{arXiv:1107.2005}. They also show that for rank-two states (with only two nonzero eigenvalues) orthogonal projective measurements are optimal.

%**************************************************************
\subsubsection{Demons}\label{SEC_POVM_DEMONS}

Maxwell's demons and goblins are discussed later in Sec.~\ref{maxwell}. Here we take up the matter of demons and goblins making generalized measurements. Demon discord is defined as the difference in work extractable by a demon having access to the whole system versus the work that can be extracted by local goblins, having access only to subsystems, under various communication scenarios \cite{arXiv:quant-ph/0112074}. Discussions about general measurements in this context are quite rare in the literature and therefore we first briefly explain the physical picture corresponding to work extraction via a general measurement. We then focus on thermal discord being a particular instance of demon discord, for which the total communication between the goblins is constrained to the measurement outcome of one of them. A similar reasoning holds for other demon discords \cite{arXiv:1105.4920}. It turns out that rank-one extremal measurements are optimal, but it is not known whether the goblins can extract more work using nonorthogonal projectors.

Consider a demon that has access to the whole state $\rho_{AB}$ of dimension $d_{AB}$. In order to allow for general measurements on $AB$ and keep track of the entropy flow, we allow the demon to introduce an ancillary system $M$, of arbitrary dimension $d_M$, initially in the state $\ket{0}$. A general measurement on $AB$ can now be implemented as a unitary evolution of the principal system and ancilla followed by a projective measurement on the ancilla \cite{Book.Nielsen.Chuang}. The post-measurement state is $\rho_{ABM}' = \sum_m p_m \rho_{AB|m} \otimes \Pi_m$, where $\rho_{AB|m} = M_m \rho_{AB} M_m^\dagger$ is described using general measurement operators $M_m$. The work extracted from the post-measurement state reads $W^+ = \log d_{AB} + \log d_M - \sum_{m} p_m S(\rho_{AB|m})$, whereas the work that has to be performed in order to erase the ancillary system and the demon's record of the measurement outcomes is $W^- = \log d_M +S(\{p_m\})$. Therefore, the net work gain from the total state, $W_t = W^+ - W^-$, is given by
\begin{gather}
W_t = \log d_{AB} - \sum_m p_m S(\rho_{AB|m}) - S(\{p_m\}).
\label{POVM_DD_W}
\end{gather}
Note that this expression is exactly the same if we forget about the need for an ancillary system and, regardless of whether $M_m$ are orthogonal projectors or not, say that the work extracted after a measurement is $\log d_{AB} - \sum_m p_m S(\rho_{AB|m})$, and erasure of the demon's knowledge about the measurement outcomes consumes $S(\{p_m\})$ bits of work. In conclusion, it is perfectly legitimate to allow general measurements for demons and goblins.

Another question is whether general measurements can do better than orthogonal projective measurements. For demons having access to the whole system $AB$, the latter are optimal.
To see this, consider a measurement scenario with an ancillary system, and note that the terms subtracted in Eq.~\eqref{POVM_DD_W} are given by the entropy of post-measurement state $S(\rho_{ABM}')$. Since (local) projective measurements do not decrease entropy $S(\rho_{ABM}') \ge S(\rho_{ABM}) = S(\rho_{AB})$, where $\rho_{ABM}$ is the state before the measurement, i.e. after unitary evolution of $\rho_{AB} \otimes \Pi_0$. Therefore, the maximal work, $W_t = \log d_{AB} - S(\rho_{AB})$, is attained by the measurement in the eigenbasis of $\rho_{AB}$.

This may be different for the work extractable by the local goblins. We shall focus now on the thermal discord, but the same argument applies to other discords as well \cite{arXiv:1105.4920}. If goblin $A$ performs a general measurement on its system, with measurement probabilities $p_a$, the work extracted locally is $W^+_A = \log d_A - \sum_a p_a S(\rho_{A|a})$. By communicating the measurement results $a$, the state of goblin $B$ becomes $\rho_{B|a}$ and the work extracted by him is $W^+_B = \log d_B - \sum_a p_a S(\rho_{B|a})$.
Since only goblin $A$ made a measurement, to erase its record one performs $W^- = S(\{p_a\})$ bits of work. The net effect under a general measurement, $W_l = W_A^+ + W_B^+ - W^-$, reads
\begin{align}
W_l =& \log d_{AB} - \sum_a p_a S(\rho_{A|a}) \nonumber\\
&- \sum_a p_a S(\rho_{B|a}) - S(\{p_a\}).
\label{POVM_THD_W}
\end{align}
Thermal discord is given by the difference $D_\thm = W_t - W_l$ and we now show that in order to maximize $W_l$ goblin $A$ should use rank-one extremal POVM.

Following \cite{arXiv:1105.4920}, we first prove that the work of Eq.~\eqref{POVM_THD_W} can be extracted using rank-one local POVM. After performing a general measurement, instead of extracting work goblin $A$ conducts further projective measurement with elements $\{ \Pi_{k|a}\}$, the eigenbasis of $\rho_{A|a}$. The overall measurement operators are given by the product $N_{ka} \equiv \Pi_{k|a} M_a$, where $M_a$ describe the initial general measurement,
and the corresponding POVM elements $N_{ka}^\dagger N_{ka}$ are of rank one. Since the post-measurement states of $A$ are pure, the work extracted by $A$ and $B$ is $W^+_A + W^+_B = \log d_{AB} - \sum_a p_a S(\rho_{B|a})$. However, the goblin has more detailed measurement record whose erasure consumes $W^- = S(\{p_{ka}\}) = S(\{p_{a}\}) + \sum_a p_a S(\{p_{k|a}\})$ bits of work. Since $\Pi_{k|a}$ form the eigenbasis of $\rho_{A|a}$ the overall rank-one POVM gives the same net work gain as Eq.~\eqref{POVM_THD_W}.

This work is optimized by an extremal POVM. To this end, consider a nonextremal POVM with elements $E_{a} = p E_{a}^{(1)} + (1-p) E_{a}^{(0)}$. We denote the maps corresponding to $E_{a}^{(1)}$ and $E_{a}^{(0)}$ as $\mathcal{E}_{a}^{(1)}$ and $\mathcal{E}_{a}^{(0)}$, respectively, and therefore the map of the nonextremal POVM can be written as $\mathcal{E}_{a} = p \mathcal{E}_{a}^{(1)} + (1-p) \mathcal{E}_{a}^{(0)}$. The same reasoning as that which leads to Eq.~\eqref{POVM_DISCORD_CQ-COND-ENT} shows now that $S(A|\{ E_a \}) \equiv \sum_a p_a S(\rho_{A|a})$ is concave in $\{ E_a \}$. This, together with concavity of classical-quantum conditional entropy $S(B|\{ E_a \})$ of Eq.~\eqref{POVM_DISCORD_CQ-COND-ENT} and concavity of $S(\{p_a\})$, gives
\begin{gather}
W_{l}(\{ E_a \}) \le p W_{l}(\{E_{a}^{(1)} \}) + (1-p) W_{l}(\{ E_{a}^{(2)} \}),
\end{gather}
where e.g. $W_{l}(\{ E_a \})$ is the work extracted using measurement $\{ E_a \}$. It is therefore optimal to choose extremal POVM.

%**************************************************************
\subsubsection{Quantum deficit}\label{SEC_POVM_DEFICIT}

The types of measurements that are permitted for optimization of quantum deficit are specified by the operations local parties are allowed to perform. Under CLOCC no particles can be added or removed. The local parties are restricted to local unitary transformations and sending particles via a dephasing channel. This is equivalent to allowing, in addition to local unitary transformations, only orthogonal local projective measurements and sending particles via perfect quantum channels.

A broader class of operations allows local pure ancillas to be borrowed under the constraint that they must be returned in pure states at the end of the protocol \cite{arXiv:quant-ph/0112074, PhysRevA.71.062303}. In this way an effective general local measurement on the principal system can be performed by an orthogonal projective measurement on an ancillary system after a suitable local unitary evolution. The orthogonal projective measurement on the ancilla can be seen as classical communication via a dephasing channel, giving one party information about the distant measurement result. Under this class of operations, one-way quantum deficit becomes equivalent to the thermal discord described above, and in this case Eq.~\eqref{POVM_THD_W} gives the localizable information of Sec.~\ref{SEC_DEFICIT}.

With this broader class of operations it is possible to define a measure similar to the deficit, based on the {distillable local purity} \cite{ PhysRevA.71.062303}. The amount of purity that can be distilled from a system of entropy $S(\rho)$ is, in the many copy limit, $K(\rho)=\log d-S(\rho)$. If one is to distill local purity from a bipartite system using one-way communication, the distillable (local) purity reads $K^\to(AB)=\log d_Ad_B-S(A)-S(B)-J(B|A)$. The difference is given by the quantum discord \cite{ arXiv:1002.4913}, $K(AB)-K^\to(AB)=D(B|A)$.

The zero-way quantum deficit also admits a corresponding demon discord. The localizable information is obtained by two independent goblins who communicate with each other only when erasing their classical records. The work goblins extract locally under general measurement is now given by $W^+_A + W^+_B = \log d_A - \sum_a p_a S(\rho_{A|a}) + \log d_B - \sum_b p_b S(\rho_{B|b})$, whereas the work required to erase their records using classical communication is $W^- = S(\{p_{ab}\})$. Therefore, the corresponding demon discord or equivalently zero-way quantum deficit under general measurements reads
\begin{align}
\Delta^{\emptyset} =&\sum_a p_a S(\rho_{A|a}) + S(\{p_{ab}\})\nonumber\\
&+ \sum_b p_b S(\rho_{B|b}) - S(\rho_{AB}),
\end{align}
where one should minimize the first three terms over independent general local measurements.
Clearly, under projective measurement we recover Eq.~\eqref{DEFICIT_ZERO-WAY} of Sec.~\ref{SEC_DEFICIT}.

The protocols allowed for calculation of two-way quantum deficit lead to the following minimization problem. Along with the initial state $\rho_{AB}$, consider some ancillary systems of total dimension $d_{M}$, all initialized in the $\ket{0}$ state. At the end of the protocol, $A$ $(B)$ has access to state $\rho_{A}'$ ($\rho_{B}'$), which may contain both the principal and ancillary systems. After both parties draw their work locally, the ancillas end up in a completely-mixed state, and therefore their erasure consumes $\log d_{M}$ bits of work. Taking this into account, the two-way quantum deficit is
\begin{gather}
\Delta_{\rm PC} = \min[S(\rho_A') + S(\rho_B')] - S(\rho_{AB}),
\end{gather}
and differs from the CLOCC two-way deficit of Eq.~\eqref{DEFICIT_DEL_MIN} in that now one minimizes over local unitary transformations and dephasing-channel entropies for the principal system and ancillas together. This class of operations allows for many intermediate effective POVM measurements on the principal system.

%**************************************************************
\subsubsection{Distance-based measures}\label{povmREDGD}

If the set of classical states in the diagrammatic approach, described in Sec.~\ref{SEC_DISSONANCE}, is chosen as a set of CC or CQ states, the proofs of \cite{arXiv:0911.5417} show that it is optimal to perform orthogonal projective measurements in order to minimize relative entropy of discord and dissonance. This is because the closest classical state, in terms of relative entropy, to an arbitrary state $\rho$ is shown to be $\rho$ itself dephased in the eigenbasis of a classical state, which is equivalent to orthogonal projective measurements.

In Sec.~\ref{SEC_GEOMETRIC_DISCORD} we prove the same result for geometric discord \cite{PhysRevA.82.034302}. Therefore, geometric discord is also optimized by orthogonal projective measurements.

One could consider states that after the measurement are classical in a larger space (e.g. the Neumark extended space). This then leads to the notion of generalized classical states, see Sec.~\ref{classicalstates}. In this case, it may happen that POVMs optimize the corresponding relative entropy or Hilbert-Schmidt distance.

%**************************************************************
\subsubsection{Gaussian discord}

There are only a few analytic results concerning optimization over general measurements in the definition of quantum correlations. Gaussian discord, described in Sec.~\ref{SEC_GAUSSIAN_DISCORD}, is a rare example having a closed-form expression for discord \cite{arXiv:1003.4979, arXiv:1003.3207}. It is optimized over generalized local Gaussian POVMs. It turns out that in many cases the optimum is achieved for projections onto coherent states, i.e. not an orthogonal projective measurement \cite{arXiv:1012.4302}.

%**************************************************************
\subsection{Evaluation of quantum discord for two qubits}\label{sec:evalD42qubits}
%**************************************************************

The optimization involved in computing discord poses a challenge to evaluating the quantity for general states, similar to that for entanglement of formation for example. Analytical results have only been obtained for a few specific cases, and many studies of discord rely on numerical optimization for determining the measurement basis that maximizes the classical correlations. In general, an attempt to derive a formula for discord in a specific case proceeds in three steps: simplification of the family of states to a normal form equivalent up to local unitary transformations; an efficient parametrization of the post-measurement states; and an optimization over the measurement variables including careful considerations of all constraints and symmetries.

For bipartite mixed states, the first analytic results are obtained for the three-parameter family of Bell-diagonal states, arbitrary mixtures of $ \ket{\Phi^{\pm }} = \tfrac{1}{\sqrt{2}}(\ket{00} \pm \ket{11})$, $\ket{\Psi^{\pm }} = \tfrac{1}{\sqrt{2}}(\ket{01} \pm \ket{10})$, which is also the family of two-qubit states having maximally-mixed marginals \cite{PhysRevA.77.042303}. These states have the general form $\rho_{BD} = \frac{1}{4} \left(\openone + \sum_{j=1}^{3}c_{j}\sigma _{j}\otimes \sigma _{j}\right) $ up to local-unitary transformations. \cite{PhysRevA.77.042303} parametrizes projective measurements on one party as $B_{0}=V \ketbra{0}{0} V^{\dag }$ and $B_{1}=V \ketbra{1}{1} V^{\dag }$, where the rotation $V=tI+i\left( y\cdot \sigma \right) $ is subject to the constraint $t^{2}+y_{1}^{2}+y_{2}^{2}+y_{3}^{2}=1$ from unitarity. Denoting a general post-measurement ensemble by $\left\{ p_{k},\rho _{k}\right\} $, it can be shown that $p_{0}=p_{1}=1/2$, and optimization of the conditional entropy quickly reduces to that over one parameter leading to the result, $J=\left( \frac{1-c}{2}\right) \log _{2}\left( 1-c\right) +\left( \frac{1+c}{2}\right) \log _{2}\left( 1+c\right) $, where $c=\max \left\{ \left\vert c_{1}\right\vert,\left\vert c_{2}\right\vert,\left\vert c_{3}\right\vert \right\} $. The mutual information and discord are computed from the eigenvalues of
$\rho _{BD}$: $\left\{ \frac{1}{4}\left( 1-c_{1}-c_{2}-c_{3}\right)\right.$, $\frac{1}{4}\left( 1-c_{1}+c_{2}+c_{3}\right) $, $\frac{1}{4}\left( 1+c_{1}-c_{2}+c_{3}\right)$, and $\left. \frac{1}{4}\left( 1+c_{1}+c_{2}-c_{3}\right) \right\}$. If $c=\left\vert c_{1}\right\vert $, the optimal projections are given by the $x$-basis, for example. The same result was found for the symmetric discord when optimizing over projective measurements \cite{JStatPhys.136.165}.

The next family of states tackled in the literature are the two-qubit X-states, which include the Bell-diagonal states as special cases. Labeling the basis elements as $\ket{1} = \ket{00} $, $\ket{2} = \ket{01} $, $\ket{3} = \ket{10} $, and $\ket{4} = \ket{11} $, an X-state is defined as having nonzero elements only on the diagonal and anti-diagonal:
\begin{gather}\label{xstate}
\rho_X=\begin{pmatrix}
\rho_{11} & 0 & 0 & \rho_{14} \cr
0 & \rho_{22} & \rho_{23} & 0 \cr
0 & \rho_{32} & \rho_{33} & 0 \cr
\rho_{41} & 0 & 0 & \rho_{44} \cr
\end{pmatrix}.
\end{gather}
The conditions $\sum_{i}\rho _{ii}=1$ and $\rho _{22}\rho _{33}\geq \left\vert \rho _{23}\right\vert ^{2}$, $\rho _{11}\rho _{44}\geq \left\vert \rho _{14}\right\vert ^{2}$ must be satisfied for $\rho_X$ to be a density matrix. The X-states are described by seven parameters before simplification by local-unitary transformations. Terms on the anti-diagonal can always be made real, and hence five parameters only suffice \cite{PhysRevA.84.042313}.

A first attempt to evaluate the discord for the two-qubit X-states in a closed form was reported in \cite{arXiv:1002.3429}, extending the method of \cite{PhysRevA.77.042303}. Simplifying the extremizing procedure, the authors argue that it is sufficient to check a few specific measurements for optimality. However, the specified algorithm turns out not to be reliable in every case. In fact \cite{arXiv:1009.1476} prove that no finite set of orthogonal projective measurements can be universal for the full family of X-states: Arbitrary rotations $\exp \left( i\varphi ^{A}\sigma _{z}^{A}/2\right) \otimes \exp \left( i\varphi ^{B}\sigma _{z}^{B}/2\right) $ maintain the X-state form. If a finite set of optimal measurements existed, the measurements would have to be of the form $\left( I\pm \sigma _{z}^{A}\right) /2$, but this is already contradicted by the Bell-diagonal states. Specific counterexamples to the algorithm have been given in \cite{arXiv:1009.1476, PhysRevA.84.042313}. On the other hand, \cite{PhysRevA.84.042313} confirm the algorithm for specific situations: for a real X-state rearranged such that $\left\vert \rho _{23}+\rho _{14}\right\vert \geq \left\vert \rho _{23}-\rho _{14}\right\vert $, the optimal measurement (for measurements on $A$) is: (a) $z$-basis if $\left( \left\vert \rho _{23}\right\vert +\left\vert \rho _{14}\right\vert \right) ^2\leq \left( \rho _{11}-\rho _{22}\right) \left( \rho _{44}-\rho _{33}\right) $; (b) $x$-basis if $\left\vert \sqrt{\rho _{11}\rho _{44}}-\sqrt{\rho _{22}\rho _{33}}\right\vert \leq \left\vert \rho _{23}\right\vert + \left\vert \rho _{14}\right\vert$. The reason for the discrepancy with the work of \cite{arXiv:1002.3429} is disputed, and arguments have been put forth concerning the treatment of all constraints, as well as the complete identification of extrema.

The family of X-states represents a small subfamily of the full set of two-qubit states, which in general can be parameterized by nine variables after simplification by local-unitary transformations. So far, the most compact formulation of the problem of evaluating discord for the general case has been given in \cite{PhysRevA.83.052108}, optimizing over orthogonal projective measurements. This formulation simplifies the problem by using a normal Bloch form, a Bloch sphere parametrization for a general measurement, and careful consideration of all symmetries and constraints. Extremization of the conditional entropy leads to a pair of transcendental equations, providing strong evidence that no general closed form is achievable, and demanding numerical treatment.

An alternative approach is statistical, and attempts to identify whether a fixed set of measurements can be optimal for computing discord with high probability. In \cite{arXiv:1009.1476}, the authors study statistically the usefulness of a measurement termed the maximal-correlations-direction measurement (MCDM). This is defined as the $x$-basis after the state $\rho$ in question has been cast in a form (by local unitary transformations) for which the matrix $T_{kl} \equiv \tr(\rho \sigma_k \otimes \sigma_l)$ is diagonal and ordered according to $\tr(\rho \sigma_x \otimes \sigma_x)\ge\tr(\rho \sigma_y \otimes \sigma_y)\ge\vert\tr(\rho \sigma_z \otimes \sigma_z)\vert$. Sampling two-qubit states randomly according to the Hilbert-Schmidt measure, suggests that the optimal measurements tend to be either the MCDM, or close to it, and the upper bound to discord obtained with MCDM tends to be very close to the real value. For the specific case of X-states, the states are generated from randomly-sampled two-qubits states, for which the off-diagonal components are set to zero. These numerical studies show the MCDM to be optimal 99.4\% of time, validating the procedure of \cite{arXiv:1002.3429} in a statistical sense. Conclusions along similar lines were reported by \cite{arXiv:1106.4488}. However, it should be emphasized that the X-states in these studies were not generated truly randomly, and direct sampling seems to find the MCDM to be optimal in a much lower proportion of cases \cite{PC.Vinjanampathy}.

Next we note a series of papers, \cite{arXiv:1102.4888, arXiv:1107.4457, arXiv:1110.6681}, which aim to interpret and augment results on two-qubit discord using a geometric interpretation based on the concept of a quantum steering ellipsoid \cite{PhD.Verstraete}. For a given state $\rho_{AB}$, a quantum steering ellipsoid is a visual representation of all possible post-measurement states of $B$ due to (POVM) measurements by $A$. It turns out, that for post-measurement ensembles minimizing the average entropy, the ensemble elements must all live on the surface of the corresponding ellipsoid. This provides a route to developing a geometric picture for the maximization step for evaluating classical correlations. To finish, we note that little progress has been made with evaluating discord beyond the case of two qubits. Some results are available for families of qubit-qudit states (with measurements on the qubit), where the optimization process can be simplified \cite{arXiv:1008.4013, arXiv:1106.4488}, as well as for the highly-symmetric Werner and isotropic states in $d \times d$ dimensional bipartite systems \cite{arXiv:1110.3057}.

In Sec.~\ref{SEC_KOASHI} a formula relating quantum discord to entanglement of formation is given by Eq.~\eqref{koashiwinter}. The immediate consequence of this formula is that computing discord of $D(A|B)$ is equivalent to computing entanglement of formation $E_F(A:C)$, where $C$ purifies the density operator of $AB$ having dimension ${\rm dim}(C) = {\rm dim}(AB)$. This implies that computing discord of two-qubit states is the same as computing the entanglement of formation of a qubit-quartit state, which is an open problem.

%**************************************************************
\subsubsection{Examples}\label{Sec:SimpleExamples}

To illustrate different optimization strategies we give discord, thermal discord and asymmetric MID for two simple bipartite states \cite{arXiv:quant-ph/0105072, arXiv:1002.4913}.

We begin with the Werner state, defined as $\rho_W = \frac{1-q}{4} \openone + q \ketbra{\Psi^-}{\Psi^-}$, where $\ket{\Psi^-} = \frac{1}{\sqrt{2}} (\ket{01} -\ket{10})$.
The classical correlations of this state are the same for all measurements of $A$ and therefore the maximization is straightforward. If $A$ measures along the standard basis, then conditional states for $B$ are: $\rho_{B|a}=\frac{1-q}{4} \openone + \frac{q}{2} \ketbra{1-a}{1-a}$ with $p_a = 1/2$ for $a=0,1$. The discord and classical correlations are \begin{align}
J(B|A)=& 1 - h(q_+)-h(q_-), \\
D(B|A)=& 1 + h(q_+)+ h (q_-) \nonumber \\
& -h ( q_+/2+ q/2 )- 3 h (q_-/2 ),
\end{align}
where $h(x) = -x \log(x)$ and $q_\pm = \frac{1 \pm q}{2}$.

\begin{figure}[t]
\resizebox{8 cm}{4.97 cm}{\includegraphics{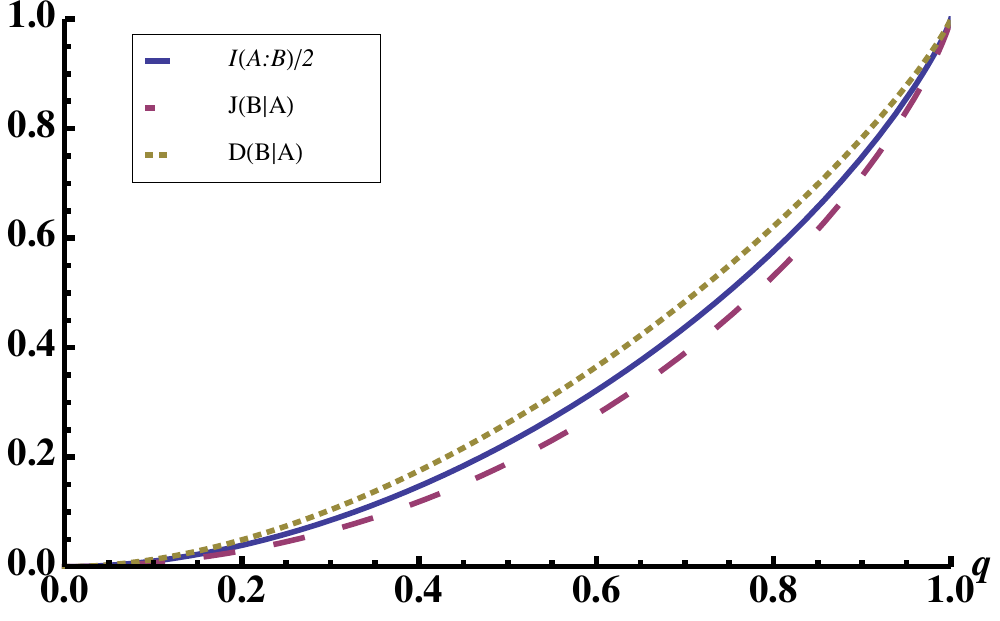}}
\caption{\label{fig:werner}(Color online.) \emph{Werner state correlations.} Mutual information (halved), discord, and classical correlations for the Werner state $\frac{1-q} {4} \openone +q \ket{\Psi^-} \bra{\Psi^-}$ with $\ket{\Psi^-} =\frac{1} {\sqrt{2}} (\ket{10} - \ket{01})$. Entanglement vanishes when $q \le \frac{1}{3}$ but all other correlations remain nonzero for $q>0$.}
\end{figure}

Since $\rho_A$ is completely mixed the measurement is always in its eigenbasis and therefore discord and thermal discord are the same. The discord vanishes only when the state is completely mixed, $q=0$. The MID for this state is not well defined since the local states do not have a well-defined basis, i.e. the local states are fully mixed. The discord, classical correlations, and half of the mutual information for the 2-qubit Werner state are plotted in Fig.~\ref{fig:werner}.

%**************************************************************

Let us now look at a state that is separable but not classical:
\begin{gather}
\rho_{AB}=\frac{1}{4}\left(\ketbra{00}{00}+\ketbra{11}{11} + 2\ketbra{+}{+} \otimes \frac{\openone}{2}\right)
\end{gather}
where $\ket{+} = \frac{1}{\sqrt{2}}(\ket{0} + \ket{1})$ and therefore $\rho_A=\frac{1}{2}(\ketbra{+}{+} + \frac{1}{2} \openone)$ and $\rho_B = \frac{1}{2} \openone$.

The discords for a measurement on A are
\begin{align}
& D(B|A)=0.05, \quad D_\thm (B|A)=0.20, \nonumber\\
& D(B|\{\Pi^{\rm Eig}_a\})=0.21,
\end{align}
and the discords for a measurement on B are zero.

As a last example we mention that for pure states the discord is symmetric and is equivalent to the unique measure of entanglement
\begin{align}
D(\ket{\psi}) & = D_\thm(\ket{\psi}) = M(\ket{\psi}) = E(\ket{\psi}) = \frac{1}{2} I(\ket{\psi}) \\
&=S[\tr_A(\ket{\psi}\bra{\psi})]=S[\tr_B(\ket{\psi}\bra{\psi})].\nonumber
\end{align}
Note that for pure states the conditional entropy ${\rm min}_{\{E_a\}}S(B\vert{\{E_a\}})$ vanishes \cite{PhysRevA.74.062308}.

%**************************************************************
%**************************************************************
\section{Unification of different measures}\label{SEC_UNIFICATION}
%**************************************************************
%**************************************************************

In the previous section we described the measures of quantum correlations other than entanglement most often discussed in the literature. This section presents known relations between them and discusses a framework for their unification.

%**************************************************************
\subsection{Entropic classification}\label{SEC_ENTROPIC_CLASSIFICATION}
%**************************************************************

We begin with the classification due to \cite{arXiv:1105.4920}. It deals with the bipartite scenario and measures that are of the form of a difference between a quantity $\mathcal{Q}$ for a quantum state, and its classical counterpart $\mathcal{C}$ which is maximized over various measurement strategies applied to the quantum state. In other words, using probabilities of measurement results allows $\mathcal{C}$ to be calculated, while $\mathcal{Q}$ is chosen among the following information-theoretic candidates: (1) mutual information; (2) conditional entropy; and (3) joint entropy. Each of these is then studied for three types of measurements: (a) in the eigenbases of reduced operators; (b) unconditional local measurements; (c) conditional local measurements, i.e. measurement of $B$ may depend on the outcome of $A$. Since these strategies satisfy the hierarchy (a) $\subset$ (b) $\subset$ (c), the value of $\mathcal{C}$ increases with every set, or equivalently there are less quantum correlations if more general measurements are allowed.

This classification enumerates different measures of quantum correlations by the choice of $\mathcal{Q}$ and measurement strategy. For example, a measure denoted as $\mathcal{M}_{(1a)}$ is a difference between quantum mutual information and mutual information given the probabilities of local eigenstates and therefore is just MID given in Eq.~\eqref{MID}. In this notation the measures discussed before are \cite{arXiv:1105.4920}:
\begin{align}
M = & \mathcal{M}_{(1a)} = \mathcal{M}_{(2a)} = \mathcal{M}_{(3a)}, \\
D_S = & \mathcal{M}_{(1b)}, \quad
D = \mathcal{M}_{(2c)}, \quad
\tilde D_{\thm} = \mathcal{M}_{(3c)}.
\end{align}
Discord appears here as a consequence of measurement strategy (c). If for every measurement outcome of $A$, $B$ measures in the eigenbasis of $\rho_{B|a}$, the measure depends only on the measurements of $A$: $\mathcal{M}_{(2c)} =\min_{\{E_a\}} S(B|\{E_a\}) - S(B|A)$, which is exactly quantum discord as discussed below Eq.~\eqref{DISCORD}.

The relations established by this classification and the works of \cite{arXiv:1008.4136} and \cite{arXiv:0707.0848} allow formulation of the following hierarchy:
\begin{gather}
I \ge M \ge D_R = \tilde \Delta^{\emptyset} \ge D_S \ge D.
\end{gather}
It is also demonstrated that $D_R \ge \mathcal{M}_{(3b)} \ge \mathcal{M}_{(2b)} \ge D_S$,
where the two measures $\mathcal{M}_{(2b)}$ or $\mathcal{M}_{(3b)}$ are not known yet to be reducible to any of those presented in Sec.~\ref{SEC_MEASURES} \cite{arXiv:1105.4920}. Finally, it is interesting whether relation
\begin{gather}
M \ge D^{\to}_R = \tilde{D}_{\thm} \ge D
\end{gather}
can be incorporated into the hierarchy \cite{arXiv:1002.4913}.

%**************************************************************
\subsection{Diagrammatic unification}\label{SEC_DIAGRAM_UNIFICATION}
%**************************************************************

%**************************************************************
\begin{figure}[th!]
\resizebox{8 cm}{14.09 cm}{\includegraphics{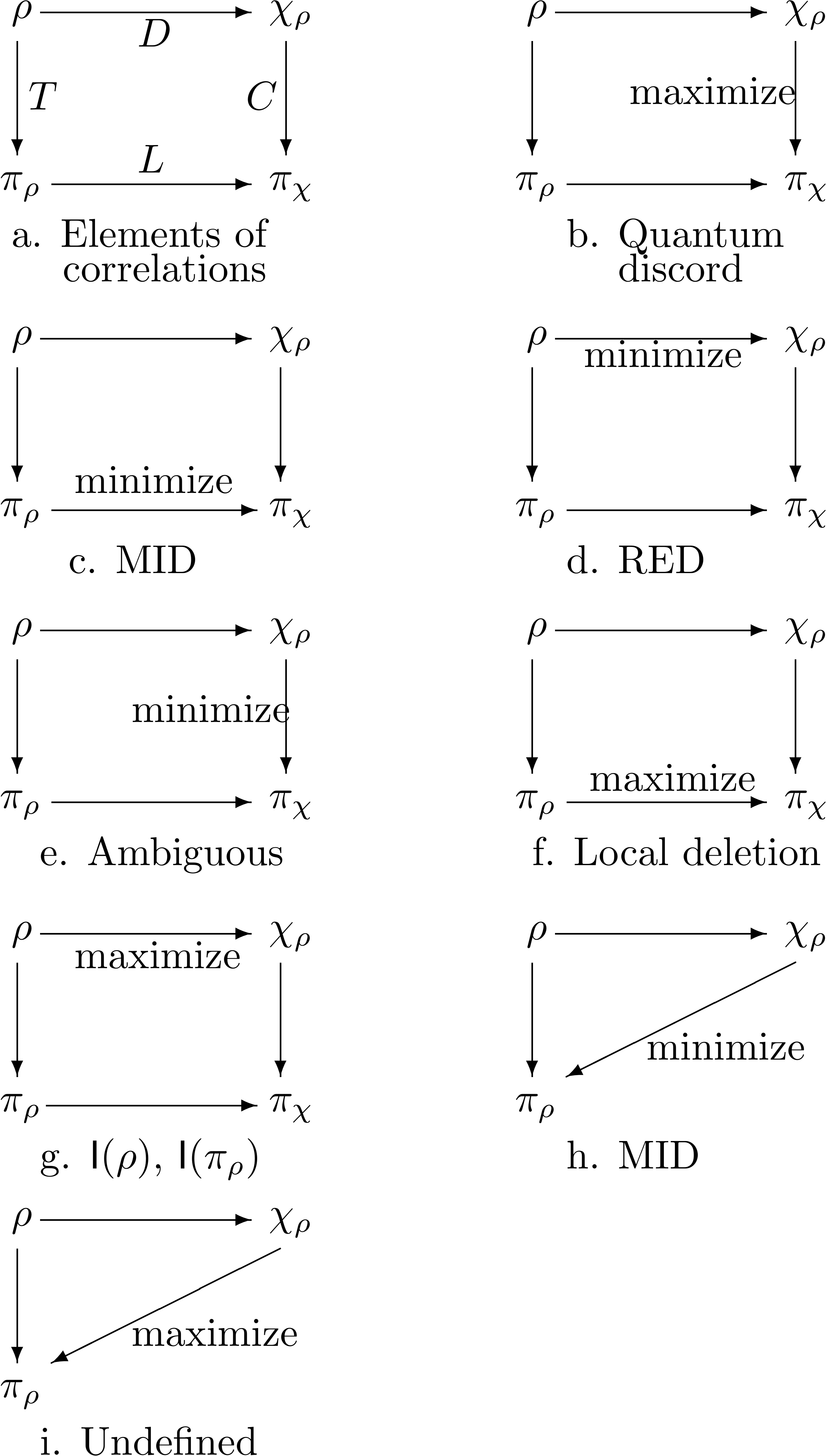}}
\caption{\label{ALLSTATES2} \emph{Diagrammatic unification.} Various possibilities to construct measures of quantum correlations. a. The fundamental elements needed in defining a measure of correlations, namely the discord $D$, classical correlations $C$, total correlations $T$ and a forth quantity $L$. In each case, the distance is measured using relative entropy, and the closest state is given by maximizing or minimizing one distance as we see in diagrams b - i. The relationship between $\rho$ and $\pi_\rho$ is fixed, so the total correlations are in all cases given by the mutual information. b. Quantum discord is obtained when classical correlations are maximized. c. Measurement-induced disturbance (MID) is obtained when the distance between the marginal states is minimized. d. Relative entropy of discord (RED) is obtained by minimizing the distance between $\rho$ and its classical state $\chi$. e. This measure is not well defined. f. Deleting local information by local measurements, however some correlations may survive. g. Information as defined in Sec.~\ref{SEC_DEFICIT}. h. Measurement-induced disturbance (MID). i. This can be well-defined, but has not been used previously. Some parts of this figure are reproduced from \cite{arXiv:1104.1520}.}
\end{figure}

In many cases it is desirable to compare various kinds of correlations present in a quantum state. Since different measures are often based on different concepts and use different mathematical entities, their direct comparison may be meaningless. For example, comparing concurrence \cite{PhysRevLett.78.5022, PhysRevLett.80.2245} with quantum discord has to be additionally motivated to make sense out of the resulting numbers. For this reason a unified approach is presented for which all the measures are defined by the same entity --- relative entropy \cite{arXiv:0911.5417}. The resulting measures and their mutual relations are described in Sec.~\ref{SEC_DISSONANCE} and in Fig.~\ref{ALLSTATES}. Note that this unification also incorporates entanglement.

This approach admits other unifying features. All the relations of Fig.~\ref{ALLSTATES} stay unchanged independently of whether, for classical, we assume the CC states of Eq.~\eqref{CLASSICAL_CLASSICAL_STATES} or the CQ states of Eq.~\eqref{CLASSICAL_QUANTUM_STATES}. In the theorems of \cite{arXiv:0911.5417}, one just replaces one-sided measurements with two-sided measurements and all the steps are unchanged. One can think about the diagram of Fig.~\ref{ALLSTATES} as a template: Once the meaning of classicality is chosen, it gives the relations for suitable measures accordingly. Furthermore, it is independent of the number of particles in the correlated state.

Extension of the diagrammatic approach to other measures is presented in \cite{arXiv:1104.1520}. As can be seen already from Fig.~\ref{ALLSTATES}, there are four fundamental states involved in the studies of quantum correlations. These are: the initial state $\rho$, its marginals $\pi_{\rho}$, the classical state $\chi_{\rho}$ obtained by dephasing $\rho$ in some basis, and the marginals of $\chi_{\rho}$ denoted as $\pi_{\chi}$. They are presented in Fig.~\ref{ALLSTATES2}a. It turns out that different measures of quantum correlations put different constraints on the relations between these states. The relative entropy distance is minimized between state $\rho$ and the classical states. In this way relative entropy of discord $D_R$, (or $D_R^{\to}$ depending on the notion of classicality), is obtained as presented in Fig.~\ref{ALLSTATES2}d. By replacing $\rho$ with its closest separable state $\sigma$ we obtain relative entropy of dissonance.

For $D$ one maximizes the classical correlations $S(B) - \sum_a p_a S(\rho_{B|a})$, see Eq.~\eqref{HV_CL_CORR}. This can also be regarded as relative entropy distance. Indeed, since $S(\rho_{B|a}) = S(\Pi_a \otimes \rho_{B|a})$ and using $\sum_a p_a S(\Pi_a \otimes \rho_{B|a}) = S(\sum_a p_a \Pi_a \otimes \rho_{B|a}) - S(\sum_a p_a \Pi_a) = S(\chi) - S(\tr_B(\pi_{\chi}))$, we find that classical correlations are given by maximizing the mutual information $I(\chi)=S(\pi_{\chi}) - S(\chi) = S(\chi\|\pi_\chi)$. This is illustrated in Fig.~\ref{ALLSTATES2}b. Finally, for MID we dephase the state in the eigenbasis of the reduced operators, i.e. $\pi_\chi = \pi_\rho$ and effectively $M$ minimizes the distance of $\chi$ from $\pi_\rho$, as shown in Fig.~\ref{ALLSTATES2}c and \ref{ALLSTATES2}h.

One advantage of this unification scheme is that it allows us to define all of these quantities for multipartite states, see Sec.~\ref{multipartite} for details. Other diagrams may also be interesting and are given in Fig.~\ref{ALLSTATES2}e--i. Some of these quantities are not well-defined and others have not been explored. For instance, the quantity in Fig.~\ref{ALLSTATES2}f is interesting as it attempts to delete all local coherence while maintaining some correlations, which means that all information in the multipartite state is stored in the correlations only. Fig.~\ref{ALLSTATES2}e is ambiguous as one can always completely decohere $\rho$ to get $\chi=\chi_A \otimes \chi_B \otimes \dots$, achieving the said minimization with vanishing minimized quantity. Finally, the quantities given in Figs.~\ref{ALLSTATES2}g and~\ref{ALLSTATES2}h are known quantities: the information content from Sec.~\ref{SEC_DEFICIT} and MID respectively. In Fig.~\ref{ALLSTATES2}g maximization is achieved always by dephasing in a way that gives $\chi=\openone/d$ so $S(\chi)=\log(d)$.

%**************************************************************
\subsubsection{Interpreting relative entropy}\label{SEC_REL_ENT_INTERPRETATION}

The operational meaning of relative entropy known from statistical inference extends to the quantum domain and gives meaning to measures based on this quantity \cite{RevModPhys.74.197, PhysRevA.57.1619, arXiv:quant-ph/9703025, arXiv:1104.1520}.
In the classical scenario, say, of two different coins with a probability of heads being $p$ and $q$ respectively, relative entropy quantifies how easy it is to confuse these coins. It answers the question: what is the probability that when coin $p$ is tossed $n$ times, the experimenter says it is coin $q$? For large $n$ this probability equals $P_n(p \to q) = 2^{-n S(q\|p)}$ where $S(q\|p)$ is the classical relative entropy. It is now clear why relative entropy is asymmetric: it makes a difference which coin is being tossed. This is illustrated by the example of $p=1$ and $q=1/2$, i.e. $S(q\|p) = \infty$ and $S(p\|q) = 1$: The experimenter estimates that coin $p$ is tossed if all $n$ outcomes are heads, otherwise she estimates it is coin $q$. If coin $p$ is tossed, she always sees heads and therefore correctly estimates coin $p$ from the first toss on, and indeed the probability of confusion $P_n(p \to q) = 0$. If coin $q$ is tossed there is a nonzero probability that a string of $n$ heads is observed which would be the case of wrongly estimating that $p$ is tossed. This happens with probability $1/2^n$ which is exactly $P_n(q \to p)$.

These ideas generalize to the quantum domain where quantum relative entropy $S(\rho\|\sigma)$ determines the probability of confusing state $\sigma$ for state $\rho$. Accordingly, relative entropy of entanglement tells us how easy it is to confuse a separable $\sigma$ with an entangled $\rho$. This naturally generalizes to the other measures. For example, relative entropy of dissonance $Q_R = \min S(\sigma\|\chi)$ quantifies how easy it is to pretend we have a nonclassical $\sigma$ when in fact we possess classical $\chi$.

%**************************************************************
\subsubsection{Tsallis entropy}

Clearly, relative entropy distance is not the only possible choice for the distance measure. In Sec.~\ref{SEC_GEOMETRIC_DISCORD} we discussed the geometric measure of discord given by the Hilbert-Schmidt distance. One may wonder whether this measure relates to the diagrams and information approach. This task is partly accomplished by \cite{PhysRevA.82.052342, arXiv:1112.1587} who essentially show that the geometric discord is given by the deficit related to the information measure introduced in \cite{PhysRevLett.83.3354, TheoMathPhys.151.693}. Let us confirm this in a different way. According to \cite{PhysRevA.82.034302} geometric discord is given by $D_G = \tr(\rho - \chi_\rho)^2$ where $\chi_\rho = \sum_{k} \ket{k} \bra{k} \rho \ket{k} \bra{k}$ is the closest classical state. (Note that this form is exactly the same as that obtained for relative entropy in the diagrammatic approach.) This implies
\begin{gather}
D_G = \tr(\rho^2 - \chi_\rho^2).
\label{GEN_ENT_DG}
\end{gather}
 One can also introduce the quantum Tsallis entropy \cite{JStatPhys.52.479}:
\begin{gather}
S_\alpha(\rho) \equiv \log (d) \frac{1 - \tr (\rho^\alpha)}{1 - d^{1-\alpha}}, \quad \alpha > 1,
\end{gather}
normalized such that the completely-mixed state of a $d$-level system admits $\log d$ bits of entropy for every $\alpha$, and this is the maximum of the entropy. The information content of a quantum state $\rho$ is then defined as $\mathsf{I}_\alpha \equiv \log d - S_\alpha(\rho)$. It reduces to the von Neumann information for $\alpha \to 1$ and for $\alpha = 2$ it gives the information measure discussed in \cite{PhysRevLett.83.3354}. In analogy to information deficit of Sec.~\ref{SEC_DEFICIT}, one can now study information deficit according to this measure. For $\alpha=2$, the one-way and zero-way deficit read:
\begin{gather}
\tilde \Delta_{2} = \frac{d \log (d)}{d-1} \min \tr(\rho^2 - \rho'^2),
\end{gather}
where the minimum is over one-sided or two-sided projective measurements, respectively, and $\rho'$ is the post-measurement state. The minimum is, up to a constant, given by the geometric discord Eq.~\eqref{GEN_ENT_DG}. The problem with generalizing the diagrammatic approach to different entropic measures, and making it an even broader template, is the lack of a satisfying definition generalizing quantum relative entropy, which would reduce to suitable differences of entropies for a wide range of parameter $\alpha$.

%**************************************************************
\subsection{Multipartite generalizations}\label{multipartite}
%**************************************************************

%**************************************************************
\subsubsection{Distance-based measures}

The measures discussed in Secs.~\ref{SEC_DISSONANCE},~\ref{SEC_GEOMETRIC_DISCORD},~\ref{SEC_DIAGRAM_UNIFICATION} have obvious multipartite generalization. These measures of correlations are based on (pseudo) distance measures, and as such do not discriminate between bipartite and multipartite states. The only caveat is to define the multipartite state in an unambiguous manner \cite{arXiv:0911.5417}. Once the classical state is defined, there are several proposals defining the multipartite quantum correlations \cite{arXiv:1101.6057, arXiv:1104.1520, arXiv:1105.2548}.

However, measures discussed in Sec.~\ref{SEC_ENTROPIC_CLASSIFICATION}, including discord, MID, and other measures defined using mutual information, do not have unique generalizations to the multipartite case. Essential properties of mutual information, i.e. its nonnegativity and the fact that it operationally captures all of the correlations in the bipartite state, are naturally generalized by the total information \cite{PhysRevA.72.032317}:
\begin{gather}
T(A_1 \dots A_N) \equiv \sum_{j=1}^N S(A_j) - S(A_1 \dots A_N),
\label{TOTAL_INFO}
\end{gather}
being just relative entropy $S(\rho\|\pi_\rho)$ between the initial $N$-partite state and its marginals.

To see this consider first a bipartite system. By operations of $A$ alone, it is possible to bring any initial state to a product state. The erasure of correlations consumes an amount of randomness, in the form of choices $A$ makes between different decorrelating operations, that is equal to mutual information. Similarly for an $N$-partite system, first $A_1$ decorrelates herself from the rest of parties, this consumes $I(A_1:A_2\dots A_N)$ bits of randomness, next decorrelating $A_2$ from the rest consumes $I(A_2: A_3 \dots A_N)$ bits of randomness and so on. The sum of all these mutual informations gives the right-hand side of the total information. Since this is just relative entropy, the diagram of Fig.~\ref{ALLSTATES} gives relations also between multipartite measures of quantum correlations.

%**************************************************************
\subsubsection{Quantum dissension}

\cite{arXiv:1006.5784} give another route to multipartite discord. They argue that multipartite quantum correlations are too complex to be captured by a single number, there should rather be a set of numbers --- a vector-like quantity. Their example is the generalization of quantum discord to the tripartite setting which is called \emph{quantum dissension}. The starting point is the classical three-variable mutual information:
\begin{gather}
I(A:B:C) \equiv I(A:B) - I(A:B|C).
\label{IABC}
\end{gather}
This quantity can be negative because knowledge of $C$ may enhance correlations between $A$ and $B$. However, since we understand the meaning of this negativity, it should not be regarded as a drawback of this definition. There are different ways to generalize this quantity to a quantum domain, which give rise to a vector-like quantity for quantum correlations. For details see \cite{arXiv:1006.5784}.

%**************************************************************
\subsection{Entanglement and discord}\label{disent1}
%**************************************************************

Entanglement is one of the most fascinating phenomena in nature. For pure states it has a well-defined measure given by the entropy of a subsystem. For mixed states however there are several measures for entanglement, each relating to a different task. The various measures of quantum correlations discussed in Sec.~\ref{SEC_MEASURES} are sometimes considered the natural extension of entanglement into domain of mixed states. As we see in the rest of the review, this notion is justified for some tasks. On the other hand, for some tasks quantum discord has been related to various measures of entanglement. This relation is often derived from the Koashi-Winter relation and the purification process (any mixed state comes from a partial trace of a pure state). In this review we assume that the reader is at least familiar with the different measures of entanglement for mixed states, such as the entanglement of formation $E_F$, distillable entanglement $E_D$, etc. We refer the interested reader to the thorough review \cite{RevModPhys.81.865}. 

%**************************************************************
\subsubsection{Purification}

Any system in a mixed state can be seen as part of a larger pure state, and constructing a pure state from a given mixed state is called purification. This is an important feature of quantum mechanics, which can be used to distinguish quantum mechanics from other theories \cite{arXiv:1011.6451}. Consider the spectral decomposition of a mixed state $\rho_A = \sum_{a} p_a \ketbra{a}{a}$. A pure state can be constructed as $\ket{\psi_{AB}} =\sum_a \sqrt{p_a} \ket{a} \otimes \ket{b_a}$, where $\{\ket{b_a}\}$ are orthonormal: This is called the Schmidt decomposition.

%**************************************************************
\subsubsection{Koashi-Winter relation}\label{SEC_KOASHI}

Quantum conditional entropy is defined following the classical definition as $S(B|A) =S(AB)-S(A)$ or $S(B|\{E_a\})=\sum_a p_a S(\rho_{B|a})$. As noted in Sec.~\ref{SEC_DISCORD}, quantum discord is the difference in these two definitions of conditional entropies: $D(B|A) = \min_{\{E_a\}} S(B|\{E_a\}) - S(B|A)$.

While the classical conditional entropy is always a positive quantity, its quantum version $S(B|A)$ can become negative. A typical example is when the total system is pure and entangled, in which case $S(AB)=0$ and $S(A)>0$. Nevertheless, this quantity has proven to be very useful, for instance, the negativity is an entanglement witness \cite{PhysLettA.194.147, PhysRevA.54.2629}, and yet for a long time it lacked an operational interpretation. The key breakthrough came in the form of a task known as quantum state merging \cite{Nature.436.673}.

The second definition of quantum conditional entropy suffers from classicalization, that is there must be a measurement on the state in order to determine its outcome \cite{JPhysA.34.6899}. This quantity is always positive and it is related to entanglement of formation due to the monogamy relation \cite{PhysRevA.69.022309}:
\begin{gather}\label{koashiwinter}
E_F(B:C) + J(B|A) = S(B)
\end{gather}
for any tripartite pure state $\ket{\psi_{ABC}}$. We can see this as follows: Let $A$ make a complete measurement on her state. For the $a$th measurement outcome the $BC$ state collapses to a pure state $\ket{\phi_{BC|a}}$ with probability $p_a$. The entropy of $B$ of the collapsed state is the entanglement of formation of that state \cite{PhysRevA.54.3824}: $S(\tr_C [\ket{\phi_{BC|a}} \bra{\phi_{BC|a}}]) =E_F(\ket{\phi_{BC|a}})$. Since $A$ is making a complete measurement, the minimum average entanglement in all such ensembles of $BC$ is the entanglement of formation of $BC$:
\begin{gather}
E_F(B:C)\equiv \min_{ \{p_a, \ket{\phi_{BC|a}} \}} \sum_a p_a E_F(\ket{\phi_{BC|a}}).
\end{gather}
This means that $\min_{\{E_a\}}S(B|\{E_a\}) = E_F(B:C)$. These two facts can be used to give discord an operational meaning (see Sec.~\ref{disent}).

%**************************************************************
\subsubsection{Conservation law}

The Koashi-Winter monogamy is related to the asymmetry of quantum discord. \cite{arXiv:1006.2460} make use of this relation to give
\begin{gather}\label{fanchineq}
E_F(A:B) + E_F(A:C) = D(A|B) + D(A|C).
\end{gather}
They call it a \emph{quantum conservation law}:
\begin{quotation}
``Given an arbitrary tripartite pure system, the sum of all possible bipartite entanglement shared with a particular subsystem, as given by the $E_F$, cannot be increased without increasing, by the same amount, the sum of all discord shared with this same subsystem."
\end{quotation}
Similarly, the difference in discord as measured by a single party can be understood as the difference in entropies of the unmeasured parties:
\begin{gather}
D(B|A) - D(C|A) = S(B) - S(C).
\end{gather}
Finally, \cite{arXiv:1106.0289} give the \emph{discord chain rule}, which expresses entanglement of formation in terms of different discords:
\begin{gather}
E_F(A:B) = D(A|B) + D(B|C) - D(C|B).
\end{gather}

%**************************************************************
\subsubsection{General bounds for discord}

\cite{arXiv:0807.4490, JPhysA.44.375301} prove a very general bound relating discord to the von Neumann entropy of the measured subsystem: $D(B|A) \leq S(A)$. Determining which states saturate this bound is more demanding, and was done in \cite{arXiv:1111.3837}. The inequality is saturated if and only if there is a decomposition of the Hilbert space for B, $H_B=H_{B^L} \otimes H_{B^R}$ for which $\rho_{AB} = \ketbra{\psi_{AB^L}}{\psi_{AB^L}} \otimes \rho_{B^R}$. In this case, $D(A \vert B) = D(B \vert A) = E_F(\rho_{AB})$, where $E_F$ denotes the entanglement of formation (generalizing the result for pure states). Furthermore, for a purification $\psi_{ABC}$, it must also hold that $\rho_{AC} = \rho_{A} \otimes \rho_{C}$: the maximal quantum correlations of the measured system precludes any further correlations with $C$. For a two-qubit system, the equality case is immediately excluded other than when $\rho_{AB}$ is pure. Proof of all these results rests largely on the strong subadditivity inequality for the von Neumann entropy, and the form of the states which saturate the inequality \cite{arXiv:quant-ph/0304007}, as well as on the Koashi-Winter relation and the quantum conservation law discussed above.

Next we note two papers that present much-stricter bounds on discord. Firstly, \cite{arXiv:1102.1301} present computable bounds for discord $D(B \vert A)$ for $2 \times d$ dimensional states $\rho_{AB}$. A key observation of the paper is that for a purification $\ket{\psi_{ABC}}$ of $\rho_{AB}$ in a $2 \times d \times2d$ dimensional system, $\rho_{BC}$ is rank two (as can be seen from the Schmidt decomposition), and there are closed expressions for the corresponding concurrence and tangle. Furthermore, the entanglement of formation, concurrence and tangle are all defined by optimizing a scalar quantity over all ensemble decompositions of $\rho_{BC}$. A lower bound for the discord is therefore achieved using the Koashi-Winter relation by bounding the entanglement of formation for $\rho_{BC}$ by a function of the concurrence. An upper bound is obtained using the measurement on $A$ which induces the optimal decomposition of $\rho_{BC}$ with respect to the tangle. For the full mathematical expressions we refer to \cite{arXiv:1102.1301}. The bounds hold for both POVM and orthogonal-projective measurements on $A$, and coincide for special cases; they can be considered to be tight, but are slightly weaker for $d>2$. A further work \cite{arXiv:1107.5458} presents bounds for discord that apply to arbitrary finite dimensional $\rho_{AB}$. These bounds are much weaker than the previous type, but are experimentally accessible and can be measured by joint measurements on two-fold copies of an unknown state. Lower and upper bounds are derived using the Koashi-Winter relation and bounds on the entanglement of formation which follow from bounding the concurrence. Again these results hold for both optimization of the discord with respect to POVM or orthogonal-projective measurements.

%**************************************************************
\subsubsection{Rank-two states of qubit-qudit system}

The Koashi-Winter relation and the relation between concurrence and $E_F$ \cite{PhysRevLett.80.2245} give an explicit algorithm for calculating the quantum discord of rank-two states of $2 \times d_B$ dimensional systems \cite{arXiv:1006.4727, arXiv:1105.6321, arXiv:1106.0289, arXiv:1107.2005}. Due to Koashi-Winter, the discord of a state $\rho_{AB}$ reads
\begin{gather}\label{KW}
D(B|A) = E_F(B:C) - S(B|A),
\end{gather}
where system $C$ purifies $AB$. For rank-two states $\rho_{AB} = \sum_{ab=1}^2 \lambda_{ab} \ket{\psi_{ab}} \bra{\psi_{ab}}$ the purification reads $\ket{\psi_{ABC}} = \sum_{ab=1}^2 \sqrt{\lambda_{ab}} \ket{\psi_{ab}} \ket{\psi_c}$ where $\{ \ket{\psi_c} \}$ is any orthonormal basis of $C$ and accordingly $C$ is a qubit. Therefore, $BC$ is a state of two qubits and Wootters' formula can be applied for calculation of discord.

The final algorithm looks as follows: Find the eigenvectors and eigenvalues of the state $\rho_{AB}$ and construct $\rho_{BC} = \tr_A(\ket{\psi_{ABC}} \bra{\psi_{ABC}})$. Entanglement of formation, and therefore discord, is given by \cite{PhysRevLett.80.2245}:
\begin{gather}
E_F(B:C) = h \left[ \tfrac{1}{2} \left( 1+\sqrt{1-\mathcal{C}^2} \right) \right],
\label{2D_EF}
\end{gather}
where $h(x) = -x \log x - (1-x) \log(1-x)$ is the binary entropy and $\mathcal{C}$ is the concurrence of state $\rho_{BC}$: $\mathcal{C}$ is given by $\max(0,\eta_1 - \eta_2 - \eta_3 - \eta_4)$,
where $\eta_i$'s are the eigenvalues in decreasing order of the Hermitian matrix $\sqrt{\sqrt{\rho_{BC}} \tilde \rho_{BC} \sqrt{\rho_{BC}}}$, and $\tilde \rho_{BC} = (\sigma_y \otimes \sigma_y) \rho_{BC}^* (\sigma_y \otimes \sigma_y)$, with $\sigma_y$ the Pauli matrix and $\rho_{BC}^*$ denoting complex-conjugated $\rho_{BC}$ (when the latter is written in the standard basis).

%**************************************************************
\subsubsection{Monogamy of discord}

One of the most important properties of entanglement is its monogamy. The monogamy of entanglement reads: ``If $A$ and $B$ are maximally entangled then neither $A$ nor $B$ is entangled with any other party $C$''. A quantitative formulation of entanglement monogamy
\begin{gather}
\sum_{i=1}^n E(A:B_i) \le E(A:B_1,B_2,\dots,B_n),
\end{gather}
does not hold for all entanglement monotones $E$. For $n=2$ the square of concurrence satisfies the monogamy relation above \cite{PhysRevA.61.052306}, and some entanglement monotones satisfy the monogamy for $n>2$ \cite{PhysRevLett.96.220503}. However, entanglement of formation does not. \cite{arXiv:1108.5168} investigate the same relationship for quantum discord. They show that for any tripartite state $\rho_{ABC}$, the inequality
\begin{gather}
D(A|B)+D(A|C) \le D(A|BC)
\end{gather}
holds if and only if $I(A:B:C) \ge J(A|BC)=J(A|B)-J(A:B|C)$. The formula for tripartite mutual information, $I(A:B:C)$, is given in Eq.~\eqref{IABC} and $J(A|BC)$ is given by Eq.~\eqref{JBA}, where now $BC$ make their measurements together. This formula is related to quantum dissension \cite{arXiv:1006.5784}. Interestingly, they find evidence that Greenberger–Horne–Zeilinger (GHZ)-type states are monogamous while W-type states are polygamous. \cite{arXiv:1110.3026} give a similar result for monogamy of MID. \cite{PhysRevA.84.054301} shows that entanglement monogamy and discord monogamy are the same for pure states. For tripartite pure states, \cite{arXiv:1109.4318} show light-cone-like behavior for monogamy of quantum deficit. \cite{arXiv:1111.5163} show that for tripartite pure states monogamy for the same measuring party, $D(B|A)+D(C|A) \le D(BC|A)$, is equivalent to $E_F(B:C) \le I(B:C)/2$ using the Koashi-Winter relation.

A very general theorem due to \cite{arXiv:1112.3967} shows that monogamy relations cannot hold for quantum measures that do not vanish for separable states. They work with quantum-correlations measures that satisfy the following criteria: (a) positivity; (b) invariance under local unitary transformations; and (c) nonincreasing when an ancilla is introduced. All correlations measures discussed in this review satisfy these conditions. Let us now sketch their theorem.

Consider a generic separable state $\rho_{AB}=\sum_c p_c \ketbra{\alpha_c}{\alpha_c} \otimes \ketbra{\beta_c}{\beta_c}$. This state could be seen as coming from $\rho_{ABC}=\sum_c p_c \ketbra{\alpha_c}{\alpha_c} \otimes \ketbra{\beta_c}{\beta_c} \otimes \ketbra{c}{c}$, where $\{\ket{c}\}$ form an orthonormal basis. With local unitary operations on $BC$, $\rho_{ABC}$ can be turned into $\sigma_{ABC}=\sum_c p_c \ketbra{\alpha_c}{\alpha_c} \otimes \ketbra{0}{0} \otimes \ketbra{c}{c}$. By condition (c) we have the inequality $Q(\sigma_{A:C}) \ge Q(\sigma_{A:BC}) = Q(\rho_{A:BC})$, the last equality is due to condition (b). Now if we additionally assume monogamy: $Q(\rho_{A:BC}) \ge Q(\rho_{A:B}) + Q(\rho_{A:C})$, this implies $Q(\sigma_{A:C}) \ge Q(\rho_{A:B}) + Q(\rho_{A:C})$. However $Q(\sigma_{A:C}) = Q(\rho_{A:C})$ since $\rho_{AC}=\sigma_{AC}$, and therefore due to condition (a) we must have $Q(\rho_{A:B})=0$. Note that \cite{arXiv:1112.3967} do not make any assumption about measurements.

This theorem proves that under some minimal assumptions quantum correlations in separable states are not monogamous. A way to deal with this might involve a hybrid approach to monogamy of correlations, when both quantum and classical correlations are present, as \cite{TeoretMatFiz.160.534} suggests. Finally, \cite{arXiv:1110.1054} argue that the relationship between quantum and classical correlations is at the heart of why entanglement of formation is not monogamous.

%**************************************************************
\subsubsection{State ordering under different discords}

Another property of various quantum-correlations measures that is related to entanglement measures is the lack of the same ordering \cite{PhysLettA.268.31} of states under different measures of quantum correlations. \cite{PhysLettA.268.31} prove that for all measures of entanglement that are equivalent for pure states they must have different ordering for mixed states. Their proof is easily extended to the case of discord. 

First note that all bipartite entropic measures of discord are equivalent for pure states \cite{arXiv:1105.4920}. Let us consider two measures of discord $\mathcal{D}_1$  and $\mathcal{D}_2$ such that $D_1(\ket{\psi}\bra{\psi})= D_2(\ket{\psi}\bra{\psi})$ for all pure states. Next take the following ordering: $D_1(\ket{\psi} \bra{\psi}) - \epsilon = D_1(\rho) = D_1(\ket{\phi}\bra{\phi}) + \epsilon$. Now suppose we have 
\begin{align}
D_2(\ket{\psi}\bra{\psi}) &\ge D_2(\rho) \ge D_2(\ket{\phi}\bra{\phi}) \nonumber \\
D_1(\ket{\psi}\bra{\psi}) &\ge D_2(\rho) \ge D_1(\ket{\phi}\bra{\phi}) 
\nonumber \\
D_1(\rho)+\epsilon &\ge D_2(\rho) \ge D_1(\rho)-\epsilon.
\end{align}
Letting $\epsilon \to 0$ tells us that $\mathcal{D}_1$ is the same measure of discord as $\mathcal{D}_2$. In other words, in general we will have  $\mathcal{D}_1(\rho_1) < \mathcal{D}_1(\rho_2)$ and $\mathcal{D}_2(\rho_1) > \mathcal{D}_2(\rho_2)$. This statement is implicitly present in many works on discords and is explicitly addressed in \cite{arXiv:quant-ph/0203007, PhysRevA.82.032340, arXiv:1109.4132}. One has to be careful in using this result, as it will not apply when comparing an entropic measure of discord to the geometric discord.

%**************************************************************
\subsubsection{Separable states vs classical states}

\cite{arXiv:0707.2195, PhysRevA.78.024303, arXiv:1105.4115} describe a unifying feature of certain classical and separable states. Separable states can be seen as \emph{shadows} of classical states. A state $\rho_{AB}$ is separable, i.e. can be represented as
\begin{gather}
\rho_{AB} = \sum_{i} p_i \; \rho^i_A \otimes \rho^i_B,
\end{gather}
if and only if there exists a state on a larger Hilbert space $\chi_{AA'BB'}$, that is classical-classical in the cut $AA'$ versus $BB'$, and has $\rho_{AB}$ as a subsystem. For a proof note that classical states have separable subsystems: tracing out $A'$ and $B'$ from the state $\chi_{AA'BB'} = \sum_{i} p_{i} \Pi^i_{AA'} \otimes \Pi^i_{BB'}$ gives a separable state in general. Conversely, starting with a separable state $\rho_{AB}$ two ancillary systems $A'$ and $B'$ can be added each with Hilbert space dimension equal to the highest value of index $i$ in the decomposition of $\rho_{AB}$. Now the state $\chi_{AA'BB'} = \sum_{i} p_{i} \rho^i_A \otimes \Pi^i_{A'} \otimes \rho^i_B \otimes \Pi^i_{B'}$ has $\rho_{AB}$ as a subsystem and by decomposing each $\rho^i_A$ and $\rho^i_B$ in their eigenbases it becomes clear the state is CC between the $AA'$ and $BB'$ division.

%**************************************************************
\subsection{Criteria for correlations}
%**************************************************************

The correlations measures described above are based on a number of fundamental concepts: the information gain from a measurement; the effects of measurements on a system; the notion of classical states; and the lack of correlations in product states. These concepts are used in various ways, and are sometimes defined in different ways to give measures of correlations. However there have been a few attempts at defining a stricter set of criteria for measures of correlations, similar to those for entanglement described in \cite{arXiv:quant-ph/0504163} and references therein.

%**************************************************************
\subsubsection{Criteria for classical correlations}

\cite{JPhysA.34.6899} specified a set of four criteria they expect a measure of classical correlations to satisfy. Based on the criteria for entanglement measures, these criteria are: (a) \emph{product states are uncorrelated}, (b) \emph{classical correlations are invariant under local unitary operations}, (c) \emph{classical correlations are nonincreasing under local operations} and (d) \emph{for pure states the classical correlations, just like the quantum, are given by the entropy of the reduced states}. They show that their measure $J(A|B)$ satisfies these axioms. A fifth property, \emph{symmetry under the interchange of the subsystems}, is conjectured but later found to be inconsistent in general, $J(A|B)\ne J(B|A)$.

While the criteria are motivated by those for entanglement, the general method is based on classical information-theoretic ideas, mainly that of conditional entropy, and the information gain from a measurement. While they give a very strong foundation for measures of correlations they are, as we have already seen, not the only way to construct correlations measures.

%**************************************************************
\subsubsection{Criteria for generalized discord}

In \cite{arXiv:1108.3649} a more general method for constructing correlations measures (for bipartite and multipartite systems) is presented based on the diagrammatic approach. This method leads to a set of criteria for measures of correlations which can be divided into three categories: (1) necessary conditions; (2) reasonable properties; and (3) debatable criteria. The correlations are measured using a generalized discord function $\mathcal{K}[\rho,\tau]$, and a set of measurements $\{\mathcal{M}\}$. For each state $\rho$ one can associate a classical state $\chi_\rho=\mathcal{M_\rho}(\rho)$, where the measurement is chosen according to some strategy. The quantum correlations are then given by $\mathcal{Q}(\rho) = \mathcal{K} [\rho,\mathcal{M}_\rho (\rho)]$, the classical correlations $\mathcal{C}(\rho) =\mathcal{K} [\mathcal{M}_\rho (\rho), \mathcal{M}_\rho (\pi_\rho)]$, and the total correlations $\mathcal{T}(\rho) = \mathcal{K} [\rho, \pi_\rho]$ where $\pi_\rho$ is the product of the marginals of $\rho$, $\pi_\rho = \rho_A \otimes \rho_B \otimes \dots$.

The five necessary conditions for any measure of correlations are then: (1a) product states have no correlations; (1b) all correlations are invariant under local unitary operations; (1c) all correlations are nonnegative; (1d) total correlations are nonincreasing under local operations; and (1e) classical states have no quantum correlations. The correlations measures presented in this review are all consistent with these requirements.

The next three criteria involving continuity, described as reasonable, are: (2a) continuity; (2b) strong continuity of the measurement basis; and (2c) weak continuity of the measurement basis. Strong continuity here means that the measurement which minimizes discord is changed continuously for small changes in the state, and weak continuity means that the measurement is not necessarily continuous but using the basis which optimizes a nearby state results in a small error for calculating the correlations. The measures based on measurements which do not affect the marginals (MID) are found to fail all of the above continuity requirements. No measure of correlations is found to obey the strong-continuity property. This indicates that the classical states associated with two nearby quantum states may be very different. However, all forms of discord with an optimization process, either maximizing classical correlations or minimizing quantum correlations are proved to be continuous. This includes all measures reviewed here except MID.

Finally a set of criteria based on entanglement measures and information-theoretic ideas are presented as debatable: (3a) for pure bipartite states the correlations can be defined by the marginals; (3b) correlations are additive $\mathcal{T=C+Q}$ or super-additive $\mathcal{T<C+Q}$; (3c) classical and/or quantum correlations are nonincreasing under local operations; and (3d) symmetry under the interchange of subsystems.

%**************************************************************
\subsubsection{Genuine multipartite correlations}

Generalizations of bipartite correlations to multipartite systems are not only computationally more complex but also involve conceptual difficulties when one attempts to characterize the correlations as \emph{genuine multipartite}. One would sometimes like to distinguish between the amount of correlations in an $n$-partite system and the part of the correlations that is genuinely $n$-partite.

In \cite{PhysRevA.74.052110} a set of postulates is formulated which every measure of genuine $n$-party correlations, $\mathcal{G}$, should satisfy: (a) $\mathcal{G} \ge 0$ for all quantum states; (b) $\mathcal{G} = 0$ for all biproduct states $\rho_1 \otimes \rho_2$; (c) invariant under local unitary operations, i.e. $\mathcal{G}(\rho) = \mathcal{G}(U_a \otimes \dots \otimes U_z \rho U_a^\dagger \otimes \dots \otimes U_z^\dagger)$; (d) $\mathcal{G}(\rho) = \mathcal{G}(\rho \otimes \pi)$ where $\pi = \pi_a \otimes \dots \pi_z$ is a state of fully-uncorrelated auxiliary systems; (e) nonincreasing under general local operations $\mathcal{G}(\mathcal{E}(\rho)) \le \mathcal{G}(\rho)$ with local trace-preserving quantum operations $\mathcal{E} = \mathcal{E}_a \otimes \dots \otimes \mathcal{E}_z$. An example of a measure satisfying these postulates is
\begin{gather}
\mathcal{G}_{\mathrm{cum}}^2(\rho) = \frac{1}{4} \tr(C^\dagger C),
\label{CUMULANT}
\end{gather}
where $C$ is the cumulant of state $\rho$, i.e. a particular linear combination of state $\rho$ and its reduced operators \cite{PhysRevA.74.052110}. The authors calculate it for various states which illustrates postulate (b) particularly nicely. It turns out that for pure three-qubit states $\mathcal{G}_\mathrm{cum}(\rho) = 0$ implies that $\rho$ is biproduct, but this is not the case for general mixed states. For example, $\mathcal{G}_{\mathrm{cum}}(\rho)$ vanishes for $\frac{1}{2} \ket{000} \bra{000} + \frac{1}{2} \ket{111} \bra{111}$, implying that this state does not have any genuine three-party correlations.

An example provided in \cite{PhysRevLett.101.070502} (also see \cite{PhysRevLett.104.068901, PhysRevLett.104.068902}) stimulated further study of the axioms that genuine multipartite-correlations measures or indicators should satisfy, and give rise to additional postulates presented in \cite{PhysRevA.83.012312}. The example consists of a mixed state
\begin{gather}
\rho = \frac{1}{2} \ket{W} \bra{W} + \frac{1}{2} \ket{\overline{W}} \bra{\overline{W}},
\label{W-WBAR}
\end{gather}
where $\ket{W} = \frac{1}{\sqrt{n}}(\ket{1 0 \dots 0} + \ket{0 1 \dots 0} + \dots + \ket{0 0 \dots 1} )$ and $\ket{\overline{W}}$ has exchanged roles of zeros and ones. For an odd number of particles this state is genuinely $n$-party entangled, i.e. it cannot be represented as a mixture of biproduct states, and simultaneously its covariances $\langle (M_1 - \langle M_1 \rangle) \dots (M_n - \langle M_n \rangle) \rangle$ vanish for all local measurements $M_1, \dots, M_n$. This is considered an indicator of the lack of genuine $n$-party classical correlations. One of the arguments supporting this view is that if all the covariances vanish for three-qubit states with completely-mixed marginals, then the cumulant also vanishes, i.e. at least in this case the covariance criterion implies the cumulant criterion. The authors conclude that, contrary to the bipartite case, in multipartite states genuine quantum correlations can exist without genuine classical correlations.

This conclusion is questioned in \cite{PhysRevA.83.012312}, where postulates (a) and (b) are replaced with the following postulates: (f) $\mathcal{G}(\rho) = 0 \to \mathcal{G}(\Lambda(\rho)) = 0$, with local trace-nonincreasing operations $\Lambda = \Lambda_a \otimes \dots \otimes \Lambda_z$ containing general local quantum operations and unanimous postselection; (g) $\mathcal{G}(\rho) = 0 \to \mathcal{G} (\rho_{\mathrm{sp}}) = 0$, where $\rho_{\mathrm{sp}}$ is the same density operator as $\rho$ but with the systems of some parties split into more parties. In essence it should not be possible to create genuine multipartite correlations by splitting subsystems. It turns out that covariance does not satisfy these new postulates, and therefore cannot be regarded as an indicator of any genuine correlations. For example, if additionally to the state in Eq.~\eqref{W-WBAR}, each party has a local ancilla in state $\ket{0}$ and performs a c-\textsc{not} gate on it and the initial system as the control qubit, the resulting $2n$-particle state has nonvanishing covariance along local $z$ measurements. Another argument shows that starting with the state in Eq.~\eqref{W-WBAR}, it is possible by operations allowed in postulate (f) to bring it to the state with arbitrary weights. In particular, in the limit of infinitely-many particles, the operations effectively project the initial state onto the $\ket{W}$ state with finite probability. Therefore, any indicator of genuine multiparty correlations which reveals that the state in Eq.~\eqref{W-WBAR} is not correlated should also reveal that $\ket{W}$ is not correlated to be in agreement with postulate (f). This is not the case both for the covariance indicator and for the cumulant measure due to the theorem mentioned below Eq.~\eqref{CUMULANT}.

\cite{PhysRevA.83.012312} propose a new candidate for a measure of genuine multipartite classical correlations. The idea uses concepts from the section on quantum deficit (see Sec.~\ref{SEC_DEFICIT}). If parties can extract more work with CLOCC operations and classical communication across any bipartite cut than with CLOCC operations, and without sending classical information across at least one cut, then the state has genuine multipartite classical correlations. Initial calculations with this measure suggest the existence of genuine tripartite classical correlations in the state defined by Eq.~\eqref{W-WBAR} for $n=3$.

A route towards quantification of genuine multipartite classical and quantum correlations based on relative entropy is taken in \cite{PhysRevLett.107.190501}. The authors define genuine tripartite correlations as $T^{(3)} \equiv T - T^{(2)}$, where $T$ is the total mutual information (see Sec.~\ref{multipartite}) and $T^{(2)}$ is the maximum among bipartite correlations, i.e. $T^{(2)} = \max[I(A:B), I(A:C),I(B:C)]$. Defined in this way, $T^{(3)}$ is equal to the lowest bipartite mutual information in a state, e.g. $I(AB:C)$, or equivalently it is the shortest relative-entropy distance to a state with no tripartite correlations. Since the genuine correlations are of the form of mutual information, measures of genuine classical $J^{(3)}$ and quantum $D^{(3)}$ correlations follow from the standard way of defining them using the difference between two versions of mutual information. On the other hand, we would expect that $J^{(3)} = J - J^{(2)}$ and $D^{(3)} = D - D^{(2)}$, where $J$ gives the classical correlations of the total state, i.e. the smallest distance between the closest classical state and its reduced density operators, and $J^{(2)}$ is the biggest bipartite classical correlation. The authors prove that these two definitions indeed coincide, at least for pure states of three qubits. Generally, any genuine $n$-partite correlations measure $T^{(n)}$ can similarly be phrased as a suitable mutual information while satisfying $J^{(n)} = D^{(n)} = T^{(n)}/2$ for pure multipartite states, which nicely generalizes the bipartite case. Note however, that so-defined genuine quantum correlations may increase under general local operations. In fact, any discord-like measure does not follow the postulates of \cite{PhysRevA.83.012312, PhysRevA.84.042124}, since classical states can be transformed to nonclassical states using general local operations.

%**************************************************************
\subsection{Quantum correlations without classical correlations}
%**************************************************************

The intriguing question of whether quantum correlations can exist without underlying classical correlations depends on how the correlations are measured. Here we briefly review the status of this phenomenon for various measures and bipartite systems.

%**************************************************************
\subsubsection{Symmetric discord}\label{III.F.1}

There can be no quantum correlations without classical correlations if the latter are characterized by the mutual information of measurement results maximized over local measurements on $A$ and $B$. The reason is that for every correlated state there is a local measurement of $A$ and $B$ with correlated outcomes, and therefore all correlated states possess some classical correlations.

To see this consider first a tomographically-complete POVM measurement of $A$. It can be chosen as a set of projectors on nonorthogonal states, i.e. with POVM elements $E_a = \eta_a \Pi_a$. By assumption, the probability distribution $p_a = \tr(\rho_A E_a)$ uniquely identifies a quantum state $\rho_A$. Similarly, the probability distribution $p_b = \tr(\rho_B E_b)$ uniquely identifies a quantum state $\rho_B$. Therefore, a joint probability distribution $p_{ab} = tr(\rho_{AB} E_a \otimes E_b)$ uniquely identifies the joint state $\rho_{AB}$, and we conclude that $p_{ab} = p_a p_b$ if and only if $\rho_{AB} = \rho_A \otimes \rho_B$. Furthermore, if $\rho_{AB}$ is not a product state, then there exist particular outcomes $\alpha$ and $\beta$ for which $p_{\alpha \beta} \ne p_\alpha p_\beta$. Plugging in the formulas for probabilities shows that $q_{\alpha \beta} \ne q_\alpha q_\beta$ where e.g. $q_\alpha = \tr(\rho_A \Pi_a)$, and therefore there also exist local projective measurements with correlated outcomes.

We end on a historical note: Lindblad conjectured that the correlations in every state are at least half classical, i.e. the mutual information of the post-measurement state is at least half the mutual information of the state before the measurement \cite{CommMathPhys.33.305, LectNotePhys.378.71}. Recently, \cite{JStatPhys.136.165, arXiv:0911.2848} disprove this conjecture.

%**************************************************************
\subsubsection{Quantum discord}

For the reason stated in the first paragraph of the previous section, there can be no quantum correlations without classical correlations if the latter are characterized by $J(B|A)$ of Eq.~\eqref{HV_CL_CORR}. Note that $J(B|A)$ is given by the mutual information of the post-measurement state. Since the measurement is performed on $A$ only, the data-processing inequality gives $J(B|A) \ge I(A':B')$, where $I(A':B')$ is the mutual information for a state after measurements by both $A$ and $B$. For correlated initial states, we proved in the previous section that $I(A':B') > 0$, and therefore also $J(B|A) > 0$. The Lindblad conjecture also does not hold for the asymmetric discord, that is there are states with more quantum correlations than classical correlations \cite{PhysRevA.84.042124}.

%**************************************************************
\subsubsection{Quantum deficit}

Surprisingly, for zero-way quantum deficit there exist quantum states, both separable and entangled, for which this quantity is equal to mutual information \cite{arXiv:0705.1370}. Therefore, they solely contain quantum correlations. For such states it is found that the optimal local dephasing is in the eigenbases of local density operators of the initial state (see next section).

For two-way quantum deficit, the question of whether quantum correlations can be greater than classical correlations is posed in \cite{arXiv:quant-ph/0410090}. It is an open problem whether states exist for which two-way deficit is greater than half of the mutual information: $\Delta > \frac{1}{2} I$.

%**************************************************************
\subsubsection{Diagrammatic approach}

If a set of CC states is chosen to be the relevant set of classical states, the relative entropy of discord is known to be equal to zero-way deficit. As already mentioned, for zero-way quantum deficit there are states with only quantum correlations \cite{arXiv:0705.1370}. The diagrammatic approach gives an intuitive understanding why the optimal dephasing for such states is in the eigenbases of the reduced operators. Namely, the lack of classical correlations means that the closest classical state $\chi$ is a product state $\pi_\chi$. Since the closest product state to $\rho$ is just a tensor product of reduced operators, we expect dephasing in their bases to be optimal. Otherwise the relative entropy of discord would be bigger than mutual information of the state.

%**************************************************************
\subsection{Maximally-discordant mixed states}
%**************************************************************

The maximally-discordant mixed states (MDMS) are defined in analogy to maximally-entangled mixed states \cite{PhysRevA.64.030302, PhysRevA.67.022110}. They have the highest-possible discord for a given value of a mixedness parameter, usually von Neumann \cite{arXiv:1008.4136} or linear \cite{arXiv:1007.1814} entropy of the state. The boundary of physically-allowed states is pimpled and multibranched in both cases. These features persist even if mixedness is replaced with classical correlations \cite{arXiv:1007.2174}. It turns out that the set of MDMS, at a fixed value of von Neumann entropy, coincides for symmetric discord $D_S$ and quantum discord $D$. These states are subclasses of X-states, given in Eq.~\eqref{xstate}, and are experimentally realized in \cite{PhysRevA.84.020304} and to some extent in \cite{NewJPhys.13.053038}.

%**************************************************************
\subsection{Other measures}
%**************************************************************

The measures discussed above do not exhaust all possibilities to address classicality. For example, there are various ways to define conditional density operators \cite{PhysRevLett.79.5194, PhysRevA.76.032327, JPhysA.40.11361, EPL.75.1, arXiv:1110.1085}, which naturally lead to various new definitions of quantum discords. These measures are not easily merged with discord-like measures. The relations between alternative definitions of conditional states and quantum correlations have not been explored much so far. An all encompassing theory of classical correlations as a singular concept within the quantum framework is missing or not possible.

%**************************************************************
%**************************************************************
\section{Classical states}\label{classicalstates}
%**************************************************************
%**************************************************************

The set of multipartite-classical states, having zero discord with respect to one or more parties, is important for several reasons: Vanishing discord corresponds to a key notion of classicality, for which maximal information about a subsystem can be obtained by some specific local measurement without altering correlations with the rest of the system. Zero-discord states have application to the theory of decoherence where they describe the classical correlations, between the pointer states of some measurement apparatus and the internal quantum states, which results from interaction with the environment, see Sec.~\ref{Einselection}. The dynamics of an open system is completely positive when the discord of the initial system-environment correlations is vanishing, see Sec.~\ref{dynamics}. The set of classical states can be used to define discord measures using a notion of minimum distance, as is the case for the relative entropy of discord and the geometric quantum discord. In practical terms, it is often necessary only to ascertain whether or not nonclassical correlations are present, and the precise values given by the various discord measures are less important. Therefore several nullity conditions have been proposed which avoid optimization. In what follows, we first summarize the key features of classical states before describing several nullity tests and experimental discord witnesses.

%**************************************************************
\subsection{Features of the set of classical states}\label{sec:FeaturesOfC}
%**************************************************************

We begin by stating a theorem which characterizes the zero-discord states: A state $\rho_{AB}$ satisfies $D(B|A)=0$ if and only if there exists a complete set of rank-one orthogonal projectors ${\Pi_a}$ on $A$, satisfying $\sum_{a} \Pi_{a} = \openone$ and $\Pi_{a} \Pi _{a'}=\delta_{aa'} \Pi_{a}$, such that
\begin{gather}\label{eq:BlockDiagDecomp}
\rho_{AB}=\sum_{a} p_{a} \Pi_{a}\otimes \rho_{B|a}.
\end{gather}
The set of states classical with respect to $A$ is denoted as $\mathcal{C}_{A}$. Restating the theorem in the language of dephasing channels, $D \left( B| A\right)=0$ if and only if there exists a quantum channel such that $\rho=\sum_{a}\Pi_a \rho \Pi_{a}$. Eq.~\eqref{eq:BlockDiagDecomp} gives rise to an obvious physical interpretation for zero-discord states: for any state in $\mathcal{C}_A$ there exists a basis for $A$ for which the locally-accessible information is maximal and, from the perspective of an external observer, this information can be obtained without disturbance to the combined system.

Proof of the backwards implication of the theorem is an immediate consequence of the definition of $D (B|A)$: Given that $\rho_{AB}$ has the block-diagonal decomposition Eq.~\eqref{eq:BlockDiagDecomp}, it follows for the joint entropy that $S(AB)=S(A) + \sum_{a} p_{a} S(\rho_{B|a})$, and hence that $\left\{ \Pi_{a}\right\} $ defines a measurement on $A$ for which $ I(A:B) = J(B:\left\{ \Pi _{a}\right\} )$.

However, proof of the forward statement is rather more involved. The approach given in \cite{arXiv:1003.5256} is to relate the problem to the strong-subadditivity property of the von Neumann entropy. First note that the post-measurement state $\rho_{AB}'=\sum_{a} M_{a} \rho_{AB} M_{a}^\dagger$, for a given measurement with $E_a = M_a^\dagger M_a$ being rank-one-POVM elements, is related by a partial trace to a tripartite state:
\begin{align}
\rho'_{\tilde{A}BC}=&\sum_{\tilde{a} \tilde{a}'} \braket{\tilde{a}| \rho _{AB}|\tilde{a}'} \ket{\tilde{a}}_{\tilde{A}}\bra{\tilde{a}'} \otimes \ket{\tilde{a}}_{C}\bra{\tilde{a}'},
\end{align}
where orthogonal projectors $\{\ket{\tilde{a}}\}$ form Neumark extensions of POVM elements $\{E_a\}$.
Discord in $\rho_{AB}$ is
\begin{align}
D(B|\{E_a\}) =& I(A:B)-J(B|\left\{ E_a \right\} ) \\
=& S \left( \rho'_{\tilde A C}\right) +S\left( \rho'_{ \tilde AB}\right) -S\left( \rho'_{ \tilde ABC}\right) -S(\rho'_{\tilde A}). \nonumber
\end{align}

The right-hand side of the last equation is the conditional mutual information. If the measurement sets the discord to zero, then $\rho'_{ABC}$ saturates the strong subadditivity inequality, and Eq.~\eqref{eq:BlockDiagDecomp} follows from the structure of $\rho'_{ABC}$ in this case \cite{arXiv:quant-ph/0304007}. For details see \cite{arXiv:1003.5256} and references therein. One implication of the above proof is that when discord is zero the measurement that minimizes it is a complete set of projections in the space of the system.

%**************************************************************
\subsubsection{Classical states and classical theories}

An important aspect of classical states is given by their role in the classical limit. \cite{arXiv:1101.5937} studies such a limit with $\hbar \rightarrow 0$. Using a model of interacting (colliding) spinning balls, he examines entanglement at the classical limit using an effective Planck constant $\hbar_{{\rm eff}}$. At the classical limit the states describing the balls are well approximated by a corresponding classical state.

Classical states are also the only states which are allowed in a classical probability theory. \cite{arXiv:1112.5066} examines the role of discord in generalized probabilistic theories \cite{arXiv:1011.6451}. He defines zero-discord states as those that are: (a) not entangled and (b) can be objectively measured with complete information on one subsystem. An objective measurement is defined as one that is repeatable and which does not induce a loss of information about the probability distribution; it is also complete if the resulting states are pure. This definition is a generalization of Eq.~\eqref{eq:BlockDiagDecomp} or equivalently $\rho=\sum_i\Pi_a\rho\Pi_a$ to other possible theories.

Null discord states are the only states allowed by a classical probability theory, i.e., one where all pure states are perfectly distinguishable. More importantly, if in a given theory all separable states have null discord, then all pure states are perfectly distinguishable. One may conclude that discord is not only a signature of ``quantumness'', rather it is a signature of nonclassicality.

%**************************************************************
\subsubsection{Generality of classical states}

We now consider whether alternative definitions of discord yield the same null set, as defined by the block-diagonal decomposition Eq.~\eqref{eq:BlockDiagDecomp}. Recalling the definition of $\tilde D_{\thm}$ in Sec.~\ref{SEC_THM_DISCORD}, which differs from $D$ by including the entropic cost for $A$ of measurement, we have $D \left( B| A\right)=0$ if and only if $\tilde D_{\thm}\left( B| A \right) = 0$. To prove this note that if $\rho_{AB}$ is of the form Eq.~\eqref{eq:BlockDiagDecomp}, the reduced state $\rho _{A}= \sum_a p_a \Pi_a$ has the same form before and after measurement in a basis $\left\{ \Pi_a \right\}$, and hence $\tilde D_{\thm} \left( B|A \right) = 0$. The converse follows because $\tilde D_{\thm} \left( B| A \right) \geq D \left( B | A\right)$ (the entropic cost of measurement for $A$ is always nonnegative). \cite{arXiv:1002.4913, arXiv:1012.0632} give alternative definitions that yield the same zero-discord states.

However not all formulations of discord define the same null set as $D$. In \cite{arXiv:1101.1717}, the author considers how the properties of $D$ change when the von Neumann entropy is replaced by one of several alternative (concave) entropy functions. In fact none of the alternatives discussed yields the null set $\mathcal{C}_{A}$ characterized by Eq.~\eqref{eq:BlockDiagDecomp}. As an example, the quadratic entropy function, $S_{Q}(\rho)=1-{\tr}\left(\rho^2\right) $, is proven to define a nonnegative discord measure; however, if we consider the completely-mixed state $\rho _{AB} = \frac{1}{d_{AB}}\openone$, which is clearly of the form Eq.~\eqref{eq:BlockDiagDecomp}, the modified discord is nonzero (except in the trivial cases that $d_{A}=1$ or $d_{B}=1$).

%**************************************************************
\subsubsection{Zero-measure of classical states}

\cite{arXiv:0908.3157} prove some important facts about the set of classical states. They are consequences of the sufficient condition for vanishing discord presented in Sec.~\ref{SUFF_FOR_ZERO_D}. First, by parameterizing the set of all density matrices $\rho_{AB}$ for parties $A$ and $B$ using the Bloch representation (for arbitrary finite dimensions), it is shown that states which satisfy the sufficient condition are parameterized by strictly-less independent parameters than the full Hilbert space. This subset, which includes the set of classical states, must therefore have volume zero (as defined by the Lebesgue measure). In particular, this implies that the probability for picking a zero-discord state at random from all possible states is zero. Secondly, it is proven in topological terminology that the set of states satisfying the sufficient condition for vanishing discord is closed and nowhere dense --- and therefore has no interior points. As a result, within any arbitrarily-small ``distance'' of a state $\rho_{AB}$ in $\mathcal{C}_{A}$, there is a state $\tilde{\rho}_{AB}$ for which discord does not vanish. Taken together, these two results for the classical states present a fundamental difficulty with defining the notion of vanishing discord in an operational manner: Any measurement procedure is subject to errors and cannot by itself prove a complete absence of nonclassical correlations. Any attempt to implement a discord witness must (implicitly) make additional assumptions about the form of the state being investigated. Several such witnesses are described in Sec.~\ref{sec:DiscordWitnesses}. As an aside, we also point out here that $\mathcal{C}_{A}$ is topologically path-connected, i.e., one can move in a continuous fashion between any two states in $\mathcal{C}_{A}$ without going outside this set. This follows from the fact that any density operator in $\mathcal{C}_{A}$ can be mixed with the maximally-mixed state (normalized-identity) with arbitrary weights to yield another density operator in $\mathcal{C}_{A}$.

%**************************************************************
\subsection{Generalized classical states}\label{sec:GenC}
%**************************************************************

Next, we summarize a generalized notion of classicality introduced in \cite{arXiv:1005.4348}, motivated by the task of unambiguous state discrimination. The aim of unambiguous state discrimination for pure states is to identify one candidate out of a set of (possibly nonorthongonal) states $\ket{\phi _{1}}, \cdots,\ket{\phi _{d}}$, using a POVM which makes no misidentifications but which can fail. This is possible if and only if $\left\{ \ket{\phi _{j}} \right\} $ are linearly independent, and in general a strategy using only projective von Neumann measurements is suboptimal. The following definition is given in \cite{arXiv:1005.4348} to extend the notion of classicality for multipartite states to allow for unambiguous state discrimination for the zero-discord parties: $\rho _{AB}$ is generalized classical (or is said to allow for nondisruptive local state identification) with respect to $A$ if there exist a decomposition $\rho _{AB}=\sum_{a} p_{a} \ketbra{\phi_a}{\phi_a} \otimes \rho_{B|a}$ where the set $\left\{ \ket{\phi_a} \right\} $ is linearly independent, and a local measurement $M_a$ with $\sum_{a} M_{a}^\dag M_{a} \leq \openone_{A}$ such that
\begin{gather}
M_{a} \ketbra{\phi_{a'}}{\phi_{a'}}M_{a}^\dag =\lambda \delta_{aa'} \ketbra{\phi_a}{\phi_a},
\end{gather}
where $0<\lambda \leq 1$. The case $\lambda =1$ reduces to the case defined by Eq.~\eqref{eq:BlockDiagDecomp}, and so classical states are also generalized classical. As an example, the state $\rho _{AB}=\mu \ketbra{0+}{0+} + (1-\mu )\ketbra{11}{11}$ is generalized classical but not classical with respect to party $B$. A local measurement optimal here with respect to nondisruptive local state identification would be given by
\begin{gather*}
M_{+}= \frac{1}{\sqrt{1 + \frac{1}{\sqrt{2}}}} \ketbra{+}{0},\;\;{\rm and}\;\;
M_{1}= \frac{1}{\sqrt{1 + \frac{1}{\sqrt{2}}}} \ketbra{1}{-},
\end{gather*}
and there is also an outcome for failure. Note however that the discord computed with respect to this POVM is nonzero. It is pointed out that all states $\rho _{AB}$ which are generalized classical with respect to either party are minimum-length states --- that is to say $\rho _{AB}$ can be written as a convex combination of a number of pure-product states equal to the rank of the density operator. The set of minimum-length states has measure zero and hence the probability of picking a generalized-classical state at random is zero, leading to the problem of detection mentioned for the set $\mathcal{C}_{A}$ defined by Eq.~\eqref{eq:BlockDiagDecomp}. A nullity condition for generalized-classicality is given in the next section.
	
%**************************************************************
\subsection{Nullity conditions}\label{sec:NullityConditions}
%**************************************************************

%**************************************************************
\subsubsection{Sufficient condition}
\label{SUFF_FOR_ZERO_D}

The form of Eq.~\eqref{eq:BlockDiagDecomp} immediately suggests some tests of the condition $D \left( B| A\right)=0$ for $\rho _{AB}$. If $D \left( B| A\right)=0$ and the spectrum of the reduced state $\rho_A$ is nondegenerate, then the eigenbasis of $\rho_{A}$ defines a measurement which is minimizing and it is sufficient to check this case. If the spectrum of $\rho_{AB}$ is nondegenerate, then in principle one could check whether the eigenvectors of $\rho_{AB}$ have a tree product form consistent with Eq.~\eqref{eq:BlockDiagDecomp}, that is to say the eigenvectors must be of the form $\Pi _{a}\otimes \Pi_{b|a}$, where the set of projectors $\left\{ \Pi_{b|a} \right\} $ diagonalizes $\rho_{B|a}$. However these methods fail in the case of degeneracy and are inefficient. For example, mixtures of orthogonal product states can have vanishing or finite discord depending on the relative weighting of the components \cite{arXiv:1002.4913}. A simpler test for nonzero discord, which follows directly from Eq.~\eqref{eq:BlockDiagDecomp} is \cite{arXiv:0908.3157}:
\begin{gather}	
\label{eq:CommuteMarginalTest}
D \left(B| A\right)=0 \Longrightarrow \left[ \rho_A \otimes \openone_B, \rho_{AB} \right]=0.
\end{gather}
For a simple example consider the density operator $\rho _{AB} = \mu \ketbra{0+}{0+} +\left( 1-\mu \right) \ketbra{11}{11} $ (where $\ket{+} =( \ket{0} +\ket{1}) / \sqrt{2}$ and $0<\mu <1$). By comparison with Eq.~\eqref{eq:BlockDiagDecomp} we see that $D \left( B| A\right) =0$. In addition, $D \left( A| B\right) \neq 0$ since $[\openone_{A}\otimes \rho_B,\rho_{AB}]\neq 0$. However, if we take any Bell state, we note that it commutes with its marginals which are maximally mixed, but it does have nonclassical correlations in the form of entanglement. The test of Eq.~\eqref{eq:CommuteMarginalTest} therefore constitutes a necessary but not sufficient condition for vanishing discord. The states satisfying Eq.~\eqref{eq:CommuteMarginalTest} are dubbed lazy states in \cite{arXiv:1004.5405} and a local discord witness based on this equation is discussed in Sec.~\ref{lazystates}. Alternative nullity conditions are presented below which overcome this problem.

%**************************************************************
\subsubsection{Commutator based}

We have already introduced a necessary nullity condition in Eq.~\eqref{eq:CommuteMarginalTest}. We now present a simple necessary and sufficient nullity condition for a state to have discord zero with respect to one party, first presented by \cite{arXiv:1005.4348}. The same condition is also presented in \cite{arXiv:1102.5249}, and applied to two-qudit circulant states in \cite{arXiv:1104.1804}. The condition can be applied for any finite number of parties and dimensionality, but for simplicity here we assume a bipartite state $\rho_{AB}$. Then, taking an arbitrary orthonormal basis $\left\{ \ket{b} \right\}$ for party $B$: $D \left( B| A\right) =0$ if and only if there exists a complete-orthonormal basis $\left\{ \ket{a} \right\} $ which simultaneously diagonalizes all the operators $\rho_{A|bb'} \equiv\braket{b | \rho _{AB} | b'}$, that is if and only if the operators $\rho _{A|bb'}$ commute. To check for classicality therefore, it is necessary to verify a number $O(d_{B}^{4})$ of commutation relations.

The proof of the first equivalence follows immediately from comparing Eq.~\eqref{eq:BlockDiagDecomp} with the expansion
\begin{align}
\rho_{AB} =&\sum_{bb'} \braket {b | \rho _{AB} | b'} \otimes \ketbra{b}{b'} \nonumber\\
=&\sum_{abb'}c_{abb'} \ketbra{a}{a} \otimes \ketbra{b}{b'}.
\end{align}
The semi-positivity of the operators $\rho _{B|a}=\sum_{bb'} c_{abb'}\ketbra{ b}{b'}$ follows from that of $\rho _{AB}$. To prove the second equivalence, we note that the conditions $\left[ \rho _{A|bb'},\rho _{A|b'b}\right] =0$
establish that the operators $\rho_{A|bb'}$ are normal, and can therefore be diagonalized individually using a unitary transform. The full set of operators $\rho_{A|bb'}$ are then simultaneously diagonalizable (using the same unitary matrix) if and only if each operator commutes with every other one.

The nullity theorem can be modified to test for generalized classicality: $\rho _{AB}$ is generalized classical with respect to party $A$ if and only if there exists a linearly-independent (but not necessarily orthogonal) basis $\left\{ \ket{a}\right\} $ which simultaneously diagonalizes all the operators $\rho _{A|bb'} = \braket{b | \rho _{AB} | b'} $. \cite{arXiv:1005.4348} provide an efficient semi-definite programming algorithm to implement this test.

%**************************************************************
\subsubsection{Singular-value-decomposition based}\label{svd}

An alternative nullity condition is proposed by \cite{arXiv:1004.0190}, and makes use of the singular-value decomposition. The idea is as follows: Given a state $\rho_{AB}$, of arbitrary finite dimensions, one first obtains the (real-valued) correlation matrix $R= \left( R_{nm}\right) $ by making the expansion $\rho =\sum_{nm}R_{nm}A_{n}\otimes B_{m}$, where $\left\{ A_{n}\right\} $ $\left( \left\{ B_{n}\right\} \right) $ defines a basis of Hermitian operators for party $A$ $(B)$. By the singular-value decomposition, $R$ can be diagonalized as $R=U^{T}rW$, where matrices $U$ and $W$ are orthogonal, and the diagonal entries of $r$ are the nonnegative singular values of $R$. Then $\rho _{AB}=\sum_{p}r_{pp}S_{p}\otimes T_{p}$ where $S_{p} = \sum_{n} U_{pn} A_{n}$ and $T_{p}=\sum_{m}W_{pm}B_{m}$. The existence of the block diagonalization of Eq.~\eqref{eq:BlockDiagDecomp} is equivalent to the simultaneous diagonalizability of the operators $\left\{ S_{n}\right\} $. This gives the nullity condition $D (B| A)=0$ if and only if the operators $S_{n}$ commute. The number of commutation relations to check is given by $(1/2) \times {\rm rank}(R) \times ({\rm rank}(R)-1)$, a number which has been substantially reduced by the singular-value decomposition. If rank$(R)$ is greater than the dimension of $A$, then $\rho _{AB}$ cannot be classical with respect to $A$.

%**************************************************************
\subsubsection{Other conditions}

An entirely different approach to the issue of nullity conditions is proposed by \cite{arXiv:1004.0434}. This nullity condition is based on the Cholesky decomposition and provides a necessary condition for vanishing discord for qubit-qudit states. (The test fails for arbitrary bipartite dimensions.) Since all states with vanishing discord must be separable, and have a positive partial transpose (PPT), the authors term their criterion as a requirement for strong PPT, in analogy to the Peres-Horodecki criterion for separability. We refer the interested reader to \cite{arXiv:1004.0434}.

%**************************************************************
\subsection{Discord witnesses}\label{sec:DiscordWitnesses}
%**************************************************************

In this subsection we discuss proposals for experimentally-practical witnesses for nonclassical correlations. The essential motivation comes from the concept of an entanglement witness, defined mathematically as a Hermitian operator $W$, satisfying ${\tr}\left( W\sigma \right) \geq 0$ for all separable states $\sigma$, and for which there exists an entangled state $\rho$ such that $\tr \left( W\rho \right) <0$. Repeated (or ensemble) measurements then yield an average value for $\rho $ which might distinguish it from the class of separable states. There are no such witnesses which are universal, that is to say able to detect all entangled states. Witnesses provide an operational characterization of entanglement, and an alternative to a full state tomography. To modify the definition of an entanglement witness for the purposes of detecting nonclassical correlations, one might try simply replacing the set of separable states with a set of zero-discord states, and suppose there exists a state of nonzero discord $\rho $ for which ${\tr}\left( W\rho \right) <0$. However, as first pointed out in \cite{arXiv:0911.3460, arXiv:1012.5718}, the mathematical properties of $W$ must change. In particular, since any separable state is a convex combination of product states having no correlations, any linear $W$ gives a nonnegative value for it, and cannot detect any nonclassical correlations.

%**************************************************************
\subsubsection{Sufficient discord witness}

Examples of nonlinear witnesses for nonclassical correlations have been given by several authors. In \cite{arXiv:0911.3460}, the following general form is proposed for a witness $W$ on bipartite states $\rho _{AB}$ for identifying any nonclassical correlations (i.e. $D \left( B| A\right) \neq 0$ or $D \left( A| B\right) \neq 0)$
\begin{gather}
W:\rho \mapsto c-{\tr}\left( \rho w_{1}\right) {\tr}\left( \rho w_{2}\right) \cdots {\tr}\left( \rho w_{m}\right)
\end{gather}
for $m\geq 2$, where the $w_{i}$ are positive Hermitian operators and $c$ is a proper constant. A witness of this form is suitable for implementation using NMR, which can implement global unitary operations and magnetization measurements of the nuclear spins. This allows values for the ${\tr}\left( \rho w_{i}\right) $ to be estimated in a experimental single run.

For example for the state $\rho _{AB}=\frac{1}{2}(\ketbra{0+}{0+} + \ketbra {11}{11})$
the authors suggest the witness
\begin{gather}
W: \; \rho_{AB} \mapsto 0.18-{\tr}\left( \rho_{AB} \ketbra{0+}{0+}\right) {\tr}\left( \rho_{AB} \ketbra{11}{11} \right),
\end{gather}
which assigns to it a value of $-0.07$. Recalling the discussion in Sec.~\ref{sec:FeaturesOfC}, $\rho_{AB}$ is generalized classical but not classical with respect to $B$.

%**************************************************************
\subsubsection{Sufficient classicality witness}

A different type of nonlinear witness is presented in \cite{arXiv:1012.3075}, for which the input is restricted to two-qubit states with Bloch representation
\begin{align}
\rho_{AB} =& \frac{1}{4}\sum_{i=1}^{3} \left(\openone \otimes \openone + x_i \sigma_i \otimes \openone + \openone \otimes y_i \sigma_i \right. \nonumber\\
&+ \left. T_{ii} \sigma _{i}\otimes \sigma _{i} \right),
\end{align}
where contributions from the off-diagonal components of the correlation tensor $T_{ij}$ are assumed to be $0$.  (Any two-qubit state admits this form for a suitable choice of local $x,y,z$ directions.) The witness is taken to be
\begin{gather}
W:\rho \mapsto \sum_{i=1}^{3}\sum_{j=i+1}^{4}\left| {\tr}\left( \hat{O}_{i}\rho \right) {\tr}\left( \hat{O}_{j}\rho \right) \right|,
\end{gather}
where $\hat{O}_{i}=\sigma _{i} \otimes \sigma_{i}$ and $\hat{O}_{4}= {\bf r} \cdot \sigma \otimes \openone + \openone \otimes {\bf s} \cdot \sigma $, where ${\bf r}$ and ${\bf s}$ are arbitrary unit vectors. An outcome $0$ for $W$ implies either that $T_{ii}=0$ for all $i$, or that exactly one component $T_{ii}$ is nonzero and the local Bloch vectors ${\bf x}={\bf y}={\bf 0}$. In both the cases $D \left( B|A\right) =D \left( A|B\right) =0$, and $W$ serves as a witness for classical-only correlations. This witness has been demonstrated experimentally using NMR \cite{arXiv:1104.1596} in agreement with earlier NMR studies \cite{arXiv:1004.0022}.

%**************************************************************
\subsubsection{Necessary-and-sufficient discord witness}

%**************************************************************
\begin{figure}[t]
\resizebox{8 cm}{6.01 cm}{\includegraphics{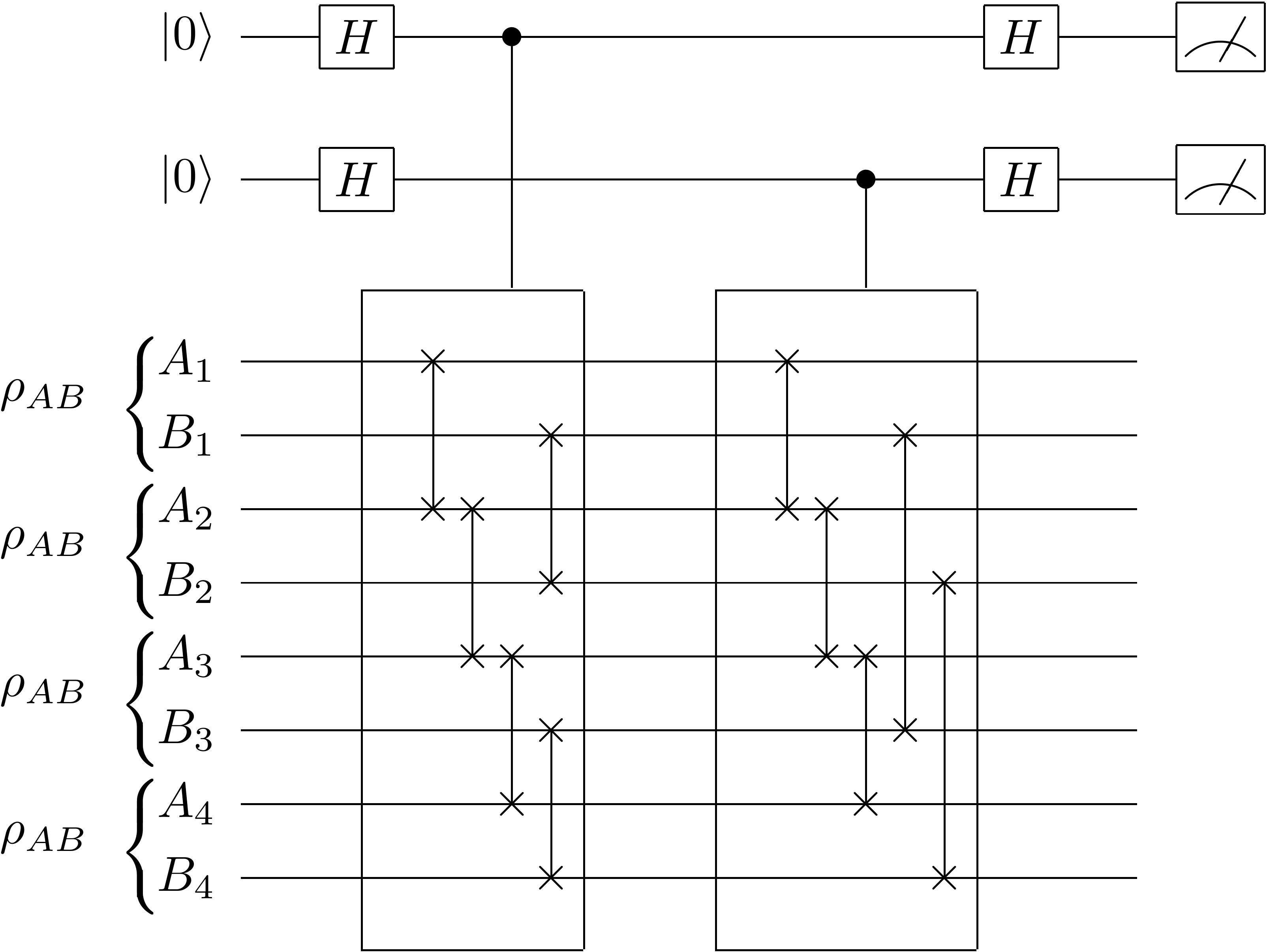}}
\caption{\label{fig:SecIVCFig} \emph{Discord witness.} A quantum circuit implementing the universal discord witness for bipartite states proposed by \cite{arXiv:1102.4710}, using two ancillas, a series of controlled-swap gates and four copies of the input state. The witness can be evaluated using ${\tr}\left( W\rho _{AB}^{\otimes k}\right) =\braket{\sigma _{z}^{2}} - \braket{\sigma _{z}^{1}} $. A zero value corresponds to $D(B|A)\left( \rho _{AB}\right) =0$. The figure is a slightly modified version of a figure from \cite{arXiv:1102.4710}.}
\end{figure}

A different approach to achieving the nonlinearity required for a discord witness has been proposed by \cite{arXiv:1102.4710, arXiv:1101.5075}, who use a Hermitian observable acting on multiple copies of the input state $\rho_{AB}^{\otimes k}$. This approach is similar in spirit to entanglement estimation using a few copies of a quantum state \cite{Nature.440.1022, PhysRevLett.98.140505}. To construct witnesses that are invariant under local unitaries, the authors use the invariants ${\tr}\left( U_A \otimes U_B \rho _{AB}^{\otimes k}\right)$, where the $U_{A(B)}$ are permutation operators for party $A$ ($B$). The witness introduced in \cite{arXiv:1102.4710} is {\it universal} and works for bipartite states of arbitrary finite dimension: it requires $k=4$ and is defined by
\begin{gather}
W=\frac{1}{2}\left( X_{A}+X_A^{\dag }\right) \otimes \left( V_{B}^{13}V_{B}^{24}-V_{B}^{12}V_{B}^{34}\right),
\end{gather}
where $X_{A}$ denotes the cyclic permutation operator with cycle $(1234)$ and $V_{B}^{ij}$ is the swap operator for input states $i$ and $j$. A quantum circuit implementing $W$ is illustrated in Fig.~\ref{fig:SecIVCFig}. To prove that $D(B|A)=0$ if and only if
\begin{gather}
{\tr}\left( W\rho _{AB}^{\otimes k}\right) =0,
\end{gather}
\cite{arXiv:1102.4710} provide the following argument: Define a complete set of $d_{B}^{2}$ observables $\left\{ G_{\mu }\right\} $ for party $B$ by
\begin{align}
& G_{m}=\ketbra{m}{m}, \quad G_{mn}^{+}= \frac{1}{\sqrt{2}} \left( \ketbra{m}{n} + \ketbra{n}{m} \right), \nonumber\\
& {\rm and} \quad G_{mn}^{-} = \frac{1}{i\sqrt{2}} \left( \ketbra{m}{n} - \ketbra{n}{m} \right),
\end{align}
 and make the expansion $\rho _{AB}=\sum_{\mu }\rho _{A|\mu }\otimes G_{\mu }$ using Hermitian operators $\rho _{A|\mu } =\tr_{B}\left( \rho _{AB}G_{\mu }\right) $. Then
\begin{gather}
{\tr}\left( W\rho _{AB}^{\otimes k}\right) =\frac{1}{2}\sum_{\mu,\nu }{\tr}\left( \left[ \rho _{A|\mu },\rho _{A|\nu }\right] ^{2}\right),
\end{gather}
and ${\tr}\left( W\rho _{AB}^{\otimes k}\right) =0$ if and only if the
$\rho_{A|\mu }$ all commute, or equivalently $\rho _{AB}$ has a block diagonal representation with respect to party $A$. This looks very much like the nullity condition presented in Sec.~\ref{svd}.

%**************************************************************
%**************************************************************
\section{Quantum correlations in quantum information}
%**************************************************************
%**************************************************************
\label{SEC_QI}

Since quantum discord has its roots in quantum information theory, it is desirable to see what role it plays in information-theoretic tasks. There are several major examples of how quantum correlations play a role in quantum communications tasks. Here its role comes in different guises as: a condition for a no-go theorem, a resource for locking of classical correlations, determining entanglement consumption and creation, and differences in coding capacities. What this shows is that the role of quantum correlations is not singular but rather varied. This hints at the fundamental nature of quantum correlations of mixed states, very much like entanglement for pure states. We begin with an important no-go theorem which generalizes the celebrated no-cloning theorem.

%**************************************************************
\subsection{No local-broadcasting}\label{NLBC}
%**************************************************************

The task of quantum cloning \cite{Nature.299.802, PhysLettA.92.271} is achieved by a unitary operation that makes a copy of an unknown state from a given set $\{\ket{\psi_i}\}$: $U \ket{\psi_i} \otimes \ket{0} = \ket{\psi_i} \otimes \ket{\psi_i}$. This is an impossible task in the quantum formalism unless the set $\{\ket{\psi_i}\}$ is a set of orthonormal states. A generalization of quantum cloning is quantum broadcasting \cite{PhysRevLett.76.2818} where, instead of unitary operations, linear operations are allowed to copy a set of density operators $\{\rho_i\}$: $\Gamma (\rho_i \otimes \omega_0) = \sigma_i$, where $\omega_0$ is the initial `blank' state. The goal is to achieve a final state such that $\tr_1(\sigma_i) = \tr_2(\sigma_i) = \rho_i$. \cite{PhysRevLett.76.2818} show that broadcasting is possible if and only if the operators $\{\rho_i\}$ commute with each other.

\cite{arXiv:0707.0848} consider an even more general version of this problem that they call local broadcasting. They consider $A$, $B$, $C$, etc. sharing a multipartite-correlated state $\rho$. Their task is to broadcast $\rho$ using local operations (but no communication is allowed). That is, letting $\omega= \omega_A \otimes \omega_B\otimes...$ be the multipartite blank state, their task is to act with local operations $\Gamma=\Gamma_A \otimes \Gamma_B \otimes...$ onto the state $\rho \otimes \omega$, and produce $\sigma$ such that $\tr_1(\sigma) = \tr_2(\sigma) = \rho$. This is a more complicated task, as each party not only has to broadcast his or her local state, but also has to act collectively to broadcast the correlations. \cite{arXiv:0707.0848} demonstrate that local broadcasting is possible if and only if $\rho$ is a fully-classical state:
\begin{align}\label{FULLY_CLASSICAL_QUANTUM_STATES}
\chi=\sum_{abc\dots} p_{abc\dots} \Pi_a \otimes \Pi_b \otimes \Pi_c \dots
\equiv \sum_{\mathbf{z}} p_{\mathbf{z}} \mathbf{\Pi}_{\mathbf{z}}
\end{align}
where $\{\Pi_a\}$ forms a rank-one orthonormal basis on space of $A$, and similarly for the other parties, $\mathbf{z}=(a,b,c,\dots)$, and $\mathbf{\Pi}_\mathbf{z} = \Pi_a \otimes \Pi_b \otimes \Pi_c \otimes \cdots$.

They begin by making the observation that, under generalized local operations, the quantum mutual information is a decreasing function: $I(\rho) \geq I(\sigma)$. Since $\sigma$ comprises two copies of $\rho$, $I(\sigma) \geq I(\rho)$. Therefore we must have $I(\sigma) = I(\rho)$. Next they make use of Petz's theorem \cite{RevMathPhys.15.79}, which says that $I(\rho)=I(\Gamma(\rho))$ if and only if the action of $\Gamma$ can be inverted, i.e. there exists a $\Lambda$ such that $\Lambda(\Gamma(\rho))=\rho$. Putting it all together, they show that a state $\rho$ can be locally broadcasted if and only if it is fully classically correlated or $I(\rho)=I(\tr_1(\sigma))=I(\tr_2(\sigma))$. Furthermore, this result encapsulates the standard no-broadcasting theorem as consequence. Consider a bipartite CQ state Eq.~\eqref{CLASSICAL_QUANTUM_STATES}, $\rho_{AB}=\sum_a p_a \Pi_a \otimes \rho_{B|a}$, and assume that the set $\{\rho_{B|a}\}$ can be broadcast. Then by the virtue of no local-broadcasting theorem we have $\rho$ must be a fully classical state, i.e. the elements of the set $\{\rho_{B|a}\}$ must commute.

%**************************************************************
\subsubsection{Uni-local and probabilistic broadcasting}

\cite{LettMathPhys.92.143} considers the scenario that lies between broadcasting and local broadcasting, dubbing it uni-local broadcasting. In uni-local broadcasting only one party acts, aiming to broadcast the correlations in a bipartite state. A bipartite state is shown to be uni-local-broadcastable if and only if it is a CQ state, classical with respect to the broadcasting party. Going further, \cite{PhysRevA.79.054305, PhysRevA.82.012338} show that no-broadcasting, no-uni-local broadcasting, and the no-local-broadcasting theorems are all equivalent to each other, i.e., one implies the others under appropriate settings.

%**************************************************************
\subsubsection{Discrimination}

Note that the no-local-broadcasting theorem is strictly about deterministic local broadcasting. In \cite{arXiv:1005.4348} generalized-classical states (see Sec.~\ref{sec:GenC}) are defined as those states that can be identified under nondisruptive local state identification (defined in terms of the well-known task of unambiguous state discrimination.) They point out that these generalized-classical states can be probabilistically broadcast. However, whether the converse statement holds, that is to say that only-generalized classical states can be probabilistically broadcast, remains an open question, as does the question of the efficiency of probabilistic broadcasting. More recently, \cite{arXiv:1011.2785} study quantum discord in the context of channel discrimination and state discrimination \cite{PhysRevLett.107.080401, arXiv:1111.2645}. These studies find that entanglement is unnecessary for discrimination, while discord (or dissonance) is conjectured to be the necessary ingredient.

The no-cloning theorem is considered to be one of the most fundamental statements of quantum information theory. It is one of the earliest examples of a task that differentiates the quantum world from the classical world. It is striking that when the no-cloning theorem is considered in its most general form, the notion of quantum correlations different from entanglement arises naturally (one can locally broadcast only fully-classical states). We should note that the no-local-broadcasting theorem does not strictly make use of quantum discord, rather it straddles the same quantum-classical boundary as quantum discord. On the other hand, we would like to have an operational meaning of quantum discord, rather than just the quantum-classical divide. Below we show that quantum discord measures the entanglement consumption in a process called extended state merging and other protocols which follow.

%**************************************************************
\subsection{Discord and entanglement}\label{disent}
%**************************************************************
Discord and entanglement are closely related as they both measure quantum correlations. Here we show explicit links between several different discords and entanglement. In this subsection we use the results of Sec.~\ref{disent1} extensively.

%**************************************************************
\subsubsection{Entanglement consumption in state merging}\label{entsm}

%**************************************************************
\begin{figure}[t]
\resizebox{8. cm}{4.38 cm}{\includegraphics{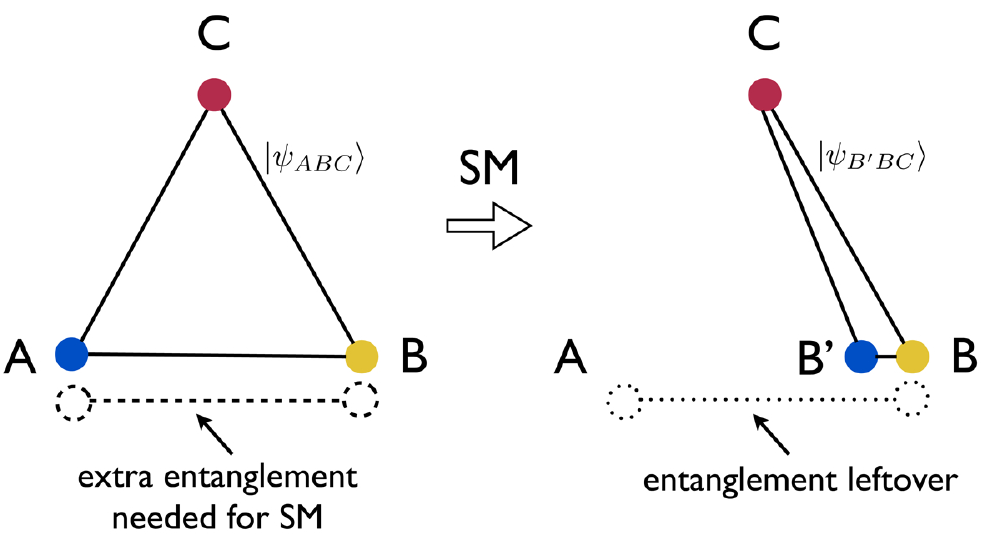}}
\caption{\label{triangle} (Color online.) \emph{State merging.} Quantum state merging. A state $\ket{\psi_{ABC}}$ is shared by $A$, $B$, and $C$. The task is to transfer the state of $A$ to $B$ using LOCC and possibly extra entanglement. Extended state merging takes into account the resources required to build the initial state between $A$ and $B$. Quantum discord between $A$ and $C$, as measured by $C$, is equal to the entanglement consumed in extended state merging with the state of $A$ going to $B'$, Eq.~\eqref{opmean1}. The dotted line represents possible left-over entanglement between $A$ and $B$ after state merging. This figure is reproduced from \cite{arXiv:1008.3205}.}
\end{figure}

Quantum state merging is defined in the following manner: Consider a known pure state of parties $ABC$. The task is for $A$ to transfer her state to $B$ using LOCC and shared entanglement without disturbing the coherence with $B$ or $C$: $\ket{\psi_{ABC}} \to \ket{\Psi_{B'BC}}$, i.e. the density operator for $ABC$ should be the same as $B'BC$ (see Fig.~\ref{triangle} for an illustration). One easy way to do this is by teleportation, but that turns out to be an overkill if shared entanglement is a precious resource. The protocol laid out in \cite{Nature.436.673} proves that, (in the many-copy limit), $S(A|B)$ is the number of ebits required for $A$ and $B$ to complete quantum state merging. If $S(A|B)$ is a positive number, $A$ and $B$ must consume that many ebits, and if $S(A|B)$ is negative, they can perform state merging with LOCC with $-S(A|B)$ ebits leftover. Before we interpret discord in terms of state merging we need to look at the other definition of conditional entropy.

\cite{arXiv:1008.3205} derive the following expression for discord using Eq.~\eqref{KW}
\begin{gather}\label{opmean1}
D(A|C)=E_F(A:B)+S(A|B) \equiv \Gamma(A|B),
\end{gather}
where $S(A|B)=-S(A|C)$ due to the fact that for a tripartite pure state $S(AC)=S(B)$ and $S(C)=S(AB)$. The same equation is first noted in \cite{arXiv:1006.2460, arXiv:1007.0228} in different contexts (see Secs.~\ref{SEC_REGULARIZATION} and~\ref{purdqc1}). The first term quantifies the amount of entanglement needed to construct the state of $AB$, while the second term quantifies the amount of entanglement needed to perform state merging from $A$ to $B$. Together, $\Gamma(A|B)$ quantifies the total amount of entanglement consumed in a protocol called extended state merging (state formation plus state merging). This relationship between discord and entanglement shows that discord records the resources in state merging from a distance, due to monogamy of correlations. That is to say, state merging from $A$ to $B$ is related to the discord in the state of $AC$ as measured by $C$. This is one of the first task-oriented operational interpretations of quantum discord.

One of the seemingly-troubling features of quantum discord is its asymmetry under party exchange, i.e. $D(A|B) \neq D(B|A)$. This is because it implies that $A$ is more correlated to $B$ than $B$ is to $A$. However this is really not the case, since the total correlations are determined by mutual information, a symmetric quantity under party exchange, (the correlations between $A$ and $B$ are the same). However, the proportion of quantum and classical correlations in each differ, i.e. $D(B|A) \ne D(A|B)$ and $J(B|A) \ne J(A|B)$, while $D(B|A)+J(B|A)=D(A|B)+J(A|B)$.

\cite{arXiv:1008.3205} also give an interpretation for the asymmetry of quantum discord. $D(A|C)$ quantifies the total entanglement consumption in extended state merging from $A$ to $B$. The difference in discords as measured by $A$ and as measured by $C$ is the difference in consumption of entanglement in extended state merging from $A$ to $B$ and from $C$ to $B$ (see Eq.~\eqref{opmean1}):
\begin{gather}
D(A|C) - D(C|A) = \Gamma(A|B) - \Gamma(C|B).
\end{gather}
Similarly, one can work out the difference in discords as onto a single party, say $C$, as measured by $A$ and $B$ in terms of extended state merging
\begin{gather}
D(C|A) - D(C|B) = \Gamma(C|B) - \Gamma(C|A).
\end{gather}
Above, the consumption of entanglement when $C$ merges with $B$ versus when $C$ merges with $A$ is given by the difference in discord between $AC$ and $BC$ as measured by $A$ and $B$ respectively.

%**************************************************************
\subsubsection{Entanglement generation in measurements}

Now we show that quantum discord and one-way deficit are related to entanglement generation between a bipartite system and a measuring apparatus.

\cite{arXiv:1012.4903} consider rank-one POVMs on one part of a bipartite system. Any bipartite state can be written as $\rho_{AB}=\sum_{bb'} \rho_{A|bb'} \otimes \ketbra{b}{b'}$. A measurement on system $B$, in the Neumark-extended basis $\{\ket{b}\}$, can be written as $\rho_{ABE} = \rho_{AB}\otimes\ketbra{0}{0} \to \tilde \rho_{ABE}=U_{BE} \rho_{AB} \otimes \ketbra{0}{0} U^\dag_{BE} =\sum_{bb'} \rho_{A|bb'} \otimes \ketbra{b\;e_b}{b'\;e_{b'}}$, where $\{\ket{e_b}\}$ forms an orthonormal basis on $E$. System $E$ is descibing a measuring device and the unitary operation potentially entangling $B$ and $E$ is sometimes called a \emph{premeasurement}. Taking the trace of the last equation with respect to system $E$ gives $\tilde \rho_{AB}=\sum_{b} \rho_{A|b} \otimes \Pi_b$. Note that this state is the same as the post-measurement state, $\rho_{AB}' = \tilde \rho_{AB}$. Furthermore, $S(\rho_{AB})= S(\rho_{ABE})= S(\tilde \rho_{ABE})$.

Consider the state corresponding to a complete measurement in basis $\{\ket{e_b}\}$ on system $E$: $\sigma_{ABE} = \sum_b \ketbra{e_b}{e_b} \tilde \rho_{ABE} \ketbra{e_b}{e_b} = \sum_b \rho_{A|b} \otimes \ketbra{b\;e_b}{b\;e_b}$, this is a separable state for any split. The relative entropy of entanglement of $\tilde \rho_{ABE}$ can be bounded from above as
\begin{align}
E_R(\tilde \rho_{AB:E}) &\leq S(\tilde \rho_{ABE} \| \sigma_{ABE}) \nonumber\\
&= S(\rho_{AB}') -S(\tilde \rho_{ABE}) \nonumber\\
&\leq E_D(\tilde \rho_{AB:E}),
\end{align}
where the last inequality comes from \cite{ProcRSocLond.461.207}. Here $E_D$ and $E_R$ are the entanglement of distillation and relative entropy of entanglement respectively. We also used $-\tr[\tilde \rho_{ABE} \log(\sigma_{ABE})]=S(\sigma_{ABE})=S(\rho'_{AB})$. But in general we have $E_R \geq E_D$, therefore we must have
\begin{align}
E_R(\tilde \rho_{AB:E}) = E_D(\tilde \rho_{AB:E})
= S(\rho'_{AB}) -S(\tilde \rho_{ABE}).\nonumber\\
\end{align}

This shows that the premeasurement process creates entanglement between the measuring apparatus $E$ and the whole system $AB$. If we minimize the entropy of the $AB$ system after the measurement, then we have
\begin{align}
\min_{\{\ket{b}\}} E_{D,R}(\tilde \rho_{AB:E}) &= \min_{\{\ket{b}\}} S(\rho'_{AB}) -S(\rho_{AB}) \nonumber\\
&= \Delta^\leftarrow(\rho_{AB}).
\end{align}
The measurement is a function of the Neumark-extended basis $\{ \ket{b} \}$. This means that distillable entanglement between the system and the measurement apparatus is only created when quantum deficit is nonzero. On the other hand, quantum deficit quantifies the minimum distillable entanglement generated between the whole system and the measurement apparatus in measuring one subpart of the total system.

In fact, different types of quantum discords quantify different types of entanglement that are generated in the division $AB:E$. The same argument can be carried out with partial entanglement defined as
\begin{gather}
P_E(\tilde \rho_{A|B:E}) = E_D(\tilde \rho_{AB:E}) - E_D(\tilde \rho_{B:E}).
\end{gather}
It is the amount of entanglement left between $B$ and $E$ when $A$ is traced out. In this way quantum discord quantifies the minimum partial entanglement generated between the system and the measurement apparatus in measuring subsystem $B$:
\begin{gather}\label{Dis_PE}
D(A|B)=\min_{\{\ket{b}\}} P_E(\tilde \rho_{A|B:E}).
\end{gather}

%**************************************************************
\subsubsection{Entanglement activation and discord}

Using very similar tools as in the last subsection an interpretation for relative entropy of discord can be attained \cite{PhysRevLett.106.220403}. Consider an $n$-partite mixed state $\rho_{A}$ and an ancillary state $\rho_{A'}$ of $n$ pure qubits in state $\ket{\bf{0}}=\ket{000\dots}$. An adversary is allowed to perform local unitary transformations $\{U_i\}$ on each of the qubits of $A$. After the adversary has performed these local unitary transformations, c-NOT gates are performed, each gate acting on an ith qubit of A and the corresponding ith qubit of A', see Fig. 7. The challenge to the adversary is to minimize the entanglement (any monotone) between $A$ and $A'$ at the end of the game. He knows the state of $A$ and therefore he can plan an optimal strategy. A measure of quantum correlations is defined as $Q_x(\rho_{A}) =\min_{\{U_i\}} E_x(A:A')$, where $x$ denotes the type of entanglement monotone considered.

%**************************************************************
\begin{figure}[t]
\resizebox{8. cm}{4.84 cm}{\includegraphics{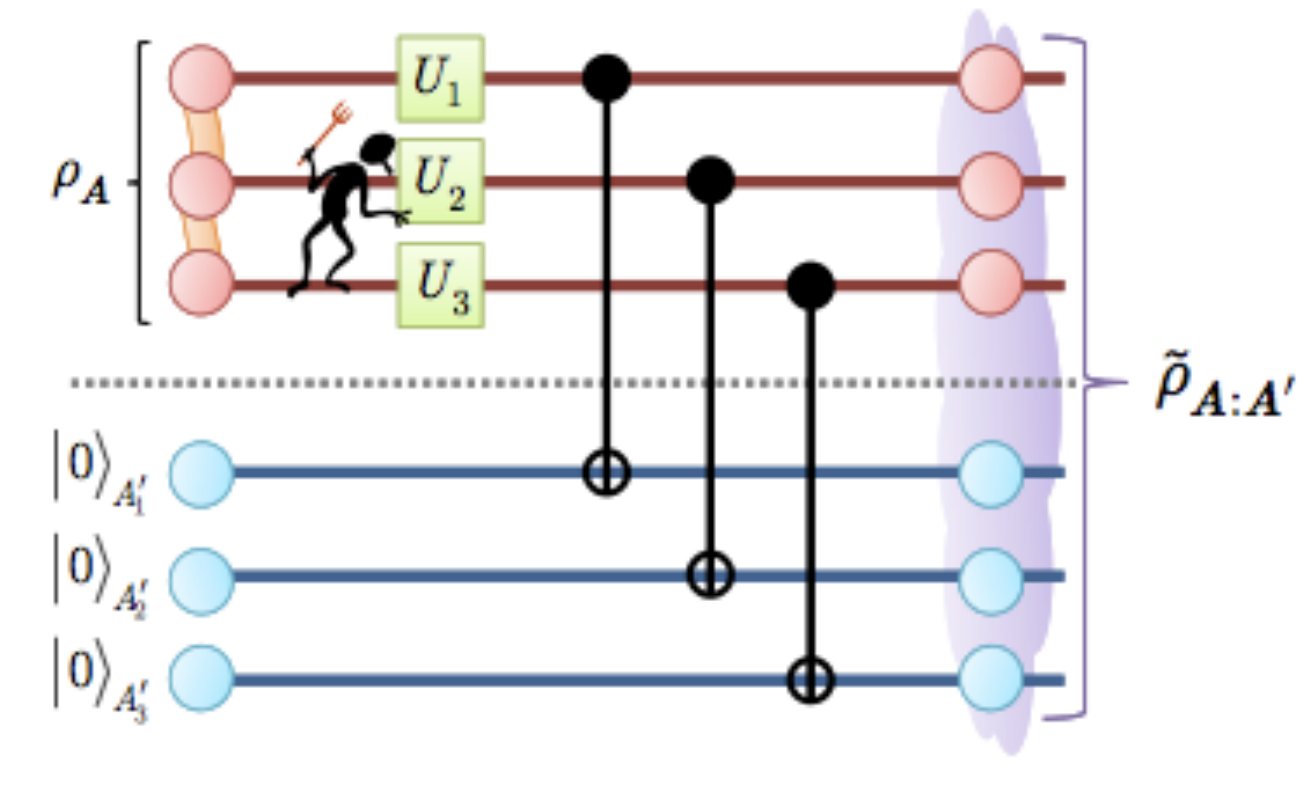}}
\caption{\label{pianired} (Color online.) \emph{Entanglement activation.} The entanglement-activation game. $\rho_A$ is the initial state and $A'$ is in pure state $\ket{0}$. The adversary performs local unitary transformations on subparts of $A$. Systems $A$ and $A'$ are then correlated via c-\textsc{not} gates to yield $\tilde \rho_{A:A'}$. The adversary tries to minimize the entanglement between $A$ and $A'$ at the end of the process by choosing the right unitary operations $\{U_1, U_2,U_3\,\dots\}$. Success of this task is related to the relative entropy of discord. This figure is reproduced from \cite{PhysRevLett.106.220403}.}
\end{figure}

It is now shown that the entanglement is vanishing if and only if the initial state of A is fully classical: For ``only if", express the state of $AA'$ after the adversary has implemented the unitary operations in the basis of the c-\textsc{not} gates: $\rho_{A} \otimes \rho_{A'}=\sum_{\mathbf{zz'}} q_{\mathbf{z}\mathbf{z'}} \ketbra{\mathbf{z}}{\mathbf{z'}} \otimes \ketbra{\bf{0}}{\bf{0}}$. Applying the c-\textsc{not} gates yields $\tilde \rho_{A:A'} =\sum_{\mathbf{zz'}} q_{\mathbf{zz'}} \ketbra{\mathbf{z}}{\mathbf{z'}} \otimes \ketbra{\mathbf{z}}{\mathbf{z'}}$, and a state of this form is called a maximally-correlated state. Next, using the result of \cite{PhysRevA.70.030302}, the entanglement of distillation of the maximally-correlated state is $E_D(\tilde \rho_{A:A'}) =\min_{\ket{\mathbf{a}}} S\left(\sum_{\mathbf{a}} \ketbra{\mathbf{a}} {\mathbf{a}} \sigma_{A} \ketbra{\mathbf{a}}{\mathbf{a}}\right) - S(\tilde \rho_{A:A'})$, which is the same as the relative entropy of entanglement \cite{PhysRevA.57.1619}, and the relative entropy of discord of $A$ \cite{arXiv:0911.5417}.

Now we show that $\tilde \rho_{A:A'}$ is classical in the $A$ and $A'$ division if and only if $\rho_{A}$ is fully classical. Note that the reduced states of $A$ and $A'$ are the same $\tilde \rho_{A'}= \tilde \rho_{A}=\sum_{\mathbf{z}} q_{\mathbf{zz}} \ketbra{\mathbf{z}} {\mathbf{z}}$. Therefore $\tilde \rho_{A:A'}$ under measurement in basis $\ket{\mathbf{zz'}}$ is invariant if and only if $\tilde \rho_{A:A'}$ is classical \cite{arXiv:1002.4913}. This implies $q_{\mathbf{zz'}}=q_{\mathbf{zz'}}\delta_{\mathbf{zz'}}$ and $\sigma_{AB}$ is classical if and only if $\rho_{A}$ is fully classical. The proof presented here is different from the one in \cite{PhysRevLett.106.220403}.

The measure of quantum correlations defined by entanglement of distillation:
\begin{align}
Q_D(\rho_{A}) \equiv& \min_{\{U_i\}}E_D(\tilde \rho_{A:A'})= \min_{\{U_i\}}E_R(\tilde \rho_{A:A'})\nonumber\\
=& D_R(\rho_A)
\end{align}
gives the relative entropy of discord an operational interpretation in terms of entanglement activation.

Other measures of quantum correlations can be defined using the game above by choosing different entanglement monotones. An example given in \cite{PhysRevLett.106.220403} using negativity is
\begin{align}\label{mepot}
Q_N(\rho_{A}) &\equiv \min_{\{U_i\}}E_N(A:A') \nonumber\\
&= \min_{\{\ket{\mathbf{a}}\}} \sum_{\mathbf{a} \neq \mathbf{a'}} \frac{ \left| q_{\mathbf{aa'}} \right|} {2}.
\end{align}
Another example using entanglement of formation and MID may be worked out using the result of \cite{arXiv:1008.3205} in Eq.~\eqref{opmean1}:
\begin{align}
Q_F(\rho_{A}) \equiv& \min_{\{U_i\}}E_F(A:A') \nonumber\\
=& D(A|P) - M(\tilde \rho_{A:A'}),
\end{align}
where the first term is the discord in $\tilde \rho_{A:P}$, where $P$ is the purification of $\rho_{A}$. We also used the fact that $M(\tilde \rho_{A:A'})=S(\tilde \rho_A)-S(\tilde \rho_{A:A'})$, which is MID from Sec.~\ref{SEC_MID}. Once again, the relationships between a variety of entanglement measures and discord measures, and knowledge of entanglement theory, give nice tools to compute various properties for quantum discord \cite{arXiv:1105.3419} and vice versa.

\cite{arXiv:1112.4672} make use of the protocol of \cite{PhysRevLett.106.220403} in a realistic system (considering losses due to the environment) of cavity quantum optomechanics \cite{Physics.2.40} in order to convert the available mechanical quantum correlations into optomechanical entanglement. They point out that mechanical quantum correlations may be very difficult to measure, but this prescription allows for an indirect detection. Lastly, we should point out that the scheme outlined in \cite{arXiv:1112.4672} falls in the continuous-variable regime.

A hybrid approach of the last two subsections is taken in \cite{arXiv:1110.2530}. They show a hierarchy of quantum correlations in a measurement chain, linking the microscopic object being measured to the macroscopic observer.

%**************************************************************
\subsection{Discord as communication cost}
%**************************************************************

%**************************************************************
\subsubsection{State merging}\label{SMCC}

A different approach to understanding discord operationally in terms of state merging is taken in \cite{arXiv:1008.4135}. The authors notice that the general quantum operation (including a measurement) on party $A$ can be implemented using a unitary transformation between $A$ and a party $E$ in state $\ket{0}$:
\begin{align}
\rho_{AB} \to \rho_{AB}' =& \sum_a M_a \rho_{AB} M_a^\dagger \nonumber\\
=& \tr_E [U_{AE}\rho_{AB} \otimes \ket{0}\bra{0}U^\dag_{AE}].
\end{align}
Therefore, $S(AB) = S(A'BE')$ since $E$ is in a pure state and similarly the conditional entropy $S (B|A) = S (B|A'E')$. Discarding the ancillary system we have $S(B|A'E')\leq S (B|A')$. To minimize $S(B|A')$ without any ambiguity, the quantum measurement must be restricted to be rank-one POVM. This restriction is important otherwise it would be best to use the identity operation (a trivial full-rank POVM) for $U_{AE}$ so $S(B|A') =S(B|A)$, i.e. nothing changes, see \cite{arXiv:1105.4920} for discussion.

The minimization over rank-one POVM gives a nice relation $\min_{\{U_{AE}\}} S(B|A') = \min_{\{E_a\}}S(B|\{E_a\})$. The conditional entropy $S(B|A)$ quantifies the entanglement required for state merging (see Sec.~\ref{entsm} and Fig.~\ref{triangle}) from $B$ to $A$ in the initial state $\rho_{AB}$, while the conditional entropy $\min_{\{E_a\}}S(B|\{E_a\})$ quantifies the entanglement required for state merging from $B$ to $A$ in the post-measurement state $\rho'_{AB}$. The difference in the conditional entropy is the increase in the cost of state merging due to the measurement on $A$, and is equal to quantum discord $D(B|A)$.

%**************************************************************
\subsubsection{Dense-coding capacity}

Conditional entropy $S(A|B)$ also describes the usefulness of a quantum state $\rho_{AB}$ as a resource for dense coding \cite{PhysRevLett.69.2881}. Dense coding is a procedure where $A$ sends her subsystem to $B$, and by doing so she is able to transmit more classical information than she could if the system is classical. In the most general dense-coding scenario \cite{QuantInfComp.1.70, JMathPhys.43.4341, arXiv:quant-ph/0701134}, $A$ encodes her message by means of general quantum operations $(\Lambda_A \otimes \mathcal{I}_B) [\rho_{AB}]= \rho_{A'B}$, and the quantum operation changes the dimension of $A$ from $d_A$ to $d_{A'}\leq d_A^2$. If the encoding is applied to single copies of $\rho_{AB}$, then the single-copy dense-coding capacity is
\begin{gather}\label{DCCAP}
\Xi(A|B)=\log_2 d_{A'}-\min_{\{\Lambda_A\}}S(A'|B),
\end{gather}
where the minimization is over all quantum operations with output dimension $d'_A$ and $S(A'|B)$ is the conditional entropy of $\rho_{A'B}$.

If we let $B$ make an optimized rank-one POVM then the dense-coding capacity becomes $\Xi(A|B')=\log_2 d_{A'}-\min_{\{\Lambda_A,E_b\}}S(A'|\{E_b\})$. The difference in the two capacities is the quantum discord of the state $\rho_{A'B}$
\begin{align}
\Xi(A|B)-\Xi(A|B')=& \min_{\{\Lambda_A\}} \left[ \min_{\{E_b\}} S(A'|\{E_b\}) - S(A'|B)\right] \nonumber\\
=&D(A'|B).
\end{align}
Again note that $B$ is restricted to rank-one POVM. In fact, consider any protocol that has a cost of the type $S(A|B) + C_{\{E_b\}}$, where $C_{\{E_b\}}$ is a term that is invariant under measurements on $B$. After an optimized rank-one POVM by $B$ the cost becomes $S(A|B') + C_{\{E_b\}}$. The increase in the cost is then given by $D(A|B)$. An analysis of such a protocol, called a \emph{mother protocol}, is given in \cite{arXiv:1107.0994}.

%**************************************************************
\subsubsection{Dense-coding capacity and asymmetry}

Let us go back to the purification scenario considered in Sec.~\ref{entsm} with $\ket{\psi_{ABC}}$. Consider $C$ sending a message, via dense coding, to either $A$ or $B$. The differences in quantum part of the dense-coding capacity, from $C$ to $A$ versus from $C$ to $B$, is related to the difference of two
discords. The dense-coding capacity in Eq.~\eqref{DCCAP} depends on the output dimension $d_{A′}$, but $\log(d_{A′})$ can be considered as the classical contribution, while the quantum advantage of dense coding is $\Xi_Q(A|B) = - \min_{\Lambda_A} S(A′|B)$. \cite{arXiv:1008.3205} show that if $C$ sends the message to $A$ versus $B$ the difference in capacity is captured by discord between $AC$ and $BC$ both
measured by $C$:
\begin{gather}
D(A|C) - D(B|C) = \Xi_Q(C|A) - \Xi_Q(C|B).
\end{gather}

All possible asymmetries in quantum discord are captured and expressed operationally here and in Sec.~\ref{entsm}. Since state merging and dense coding are not symmetric tasks, it is reasonable to have asymmetric quantifiers for the resource. The asymmetry of quantum discord captures this notion and quantifies the differences in the resource needed due to the asymmetry.

%**************************************************************
\subsection{Quantum locking of classical correlations}
%**************************************************************

Quantum locking of classical correlations \cite{arXiv:quant-ph/0303088} is a remarkable effect of quantum mechanics. Here $A$ and $B$ share a large amount of classical correlations: $\rho_{AB} = \frac{1}{2^m} \sum_{a=0}^{2^{m-1}} \Pi_a \otimes \Pi_a$. The classical correlations in this state are given by the mutual information $I(A:B)=m$ bits. Next, a random key $K = \{0,\dots,d-1\}$ is generated and a control unitary on $B$ is applied based on the value of the key:
\begin{gather}
\rho_{ABK} = \frac{1}{d \; 2^m} \sum_{a=0}^{2^{m}-1} \sum_{k=0}^{d-1} \Pi_a \otimes U_k \Pi_a U^\dag_k \otimes \ketbra{k}{k}.
\end{gather}
The value of $K$ is known to $A$ but not to $B$.

The set of unitary transformations $\{U_k\}$ can contain as few as two elements: $U_0$ being identity and $U_1$ satisfying $\braket{a|U_1|a'} = 1/2^m$. In this case the mutual information $I(AK:B) \approx m/2$, which is the accessible information of $B$ (the information $B$ can gain by simply measuring his system), while the total mutual information $I(AK:BK)=m+1$. This means that the classical correlations between $A$ and $B$ are locked away due to the presence of $K$. If $A$ reveals the value of $K$ to $B$ --- one bit of information --- the classical correlations become $m$ once again. Hence, by using one bit of communication $A$ can unlock $m/2$ bits of classical correlations, with $m$ being arbitrarily large. More generally, let us say that there are $K$ unitary transformations. Then the amount of information available to $B$ with the knowledge of the key minus the amount of information available to $B$ plus the size of the key is the amount of correlations locked:
\begin{gather}
D_L = I_{\rm acc}(AK:BK) - (I_{\rm acc}(AK:B) + |K|).
\end{gather}
In classical locking schemes, e.g. a one-time pad, the size of the key is equivalent to the size of the message $|K_{\rm cl}| \approx m$. However for quantum locking $|K| \ll m$, a remarkable feature of quantum locking of classical correlations.

This problem, just by construction, looks related to quantum discord and the classical correlations therein. It is analyzed in terms of MID in \cite{arXiv:0811.4003} and symmetric discord in \cite{arXiv:0905.2123}. More recently, a complete proof of the equality of the locked correlations with quantum discord is given in \cite{arXiv:1105.2768}. There, the CQ states $\rho_{AB}=\sum_a p_a \Pi_a \otimes \rho_{B|a}$ are considered, and the asymptotic regime is assumed as the quantities are only achievable in a many-copy limit.

After the key is revealed to $B$, the information accessible to $B$ is equal to the total mutual information: $I_{\rm acc}(AK:BK)=I(AK:BK)=m+|K|$. Again, the accessible information is the information $B$ can gain by measuring his system, in this case with the knowledge of the value of $K$. On the other hand $I(AK:BK) \leq I(AK:B)+|K| \leq m + |K|$. The first inequality comes from the no-signaling condition which here means that the mutual information should not increase more than $|K|$ bits when $|K|$ bits are sent from $A$ to $B$ \cite{Nature.461.1101}. Therefore, we have $I_{\rm acc} (AK:BK) =I (AK:BK) =I (AK:B) +|K|$. The amount of correlations locked is then given by
\begin{gather}
D_L = I(AK:B) - I_{\rm acc}(AK:B),
\end{gather}
which is quantum discord $D(AK|B)$.

%**************************************************************
\subsection{Regularization and entanglement irreversibility}
\label{SEC_REGULARIZATION}
%**************************************************************

Most of the tasks discussed above are defined asymptotically and therefore the quantities involved, e.g. discords, entanglements and entropies, should be considered in their regularized form, i.e.
\begin{gather}
\Omega^{\rm reg} (\rho) \equiv \lim_{n \to \infty} \frac{1}{n} \Omega(\rho^{\otimes n}).
\end{gather}
For instance, an operational measure of quantum correlations based on the notion of regularized broadcasting along with mutual information is proposed in \cite{arXiv:0901.1280}. In \cite{arXiv:1008.3205} gave a regularized form of entanglement consumption and quantum discord for extended state merging as well as dense-coding capacities. In \cite{arXiv:1008.4135}, a regularized cost of quantum communication and quantum discord are given and the discords considered in \cite{arXiv:1012.4903, PhysRevLett.106.220403} are related to entanglement and are therefore easy to regularize. In our opinion, the regularized forms of discords and related quantities do not lead to any significant clarification of the subject at hand, with one exception.

Using the Koashi-Winter relationship in Eq.~\eqref{koashiwinter}, and the definition of quantum discord in Eq.~\eqref{DISCORD}, an equation relating quantum discord, conditional entropy, and entanglement of formation is given by \cite{arXiv:1007.0228}: $D(A|C)=E_F(A:B)+S(A|B)$. Here once again the three parties $ABC$ together share a pure state. Note that this equation is the same as the one used for the operational interpretation in terms of extended state merging in Eq.~\eqref{opmean1}. When considering the regularized version of this equation, the entanglement of formation is replaced by the entanglement cost, while the conditional entropy gives the lower bound of the negative of entanglement of distillation in a protocol called hashing \cite{ProcRSocLond.461.207}.

%**************************************************************
\begin{figure}[t]
\resizebox{8 cm}{3.38 cm}{\includegraphics{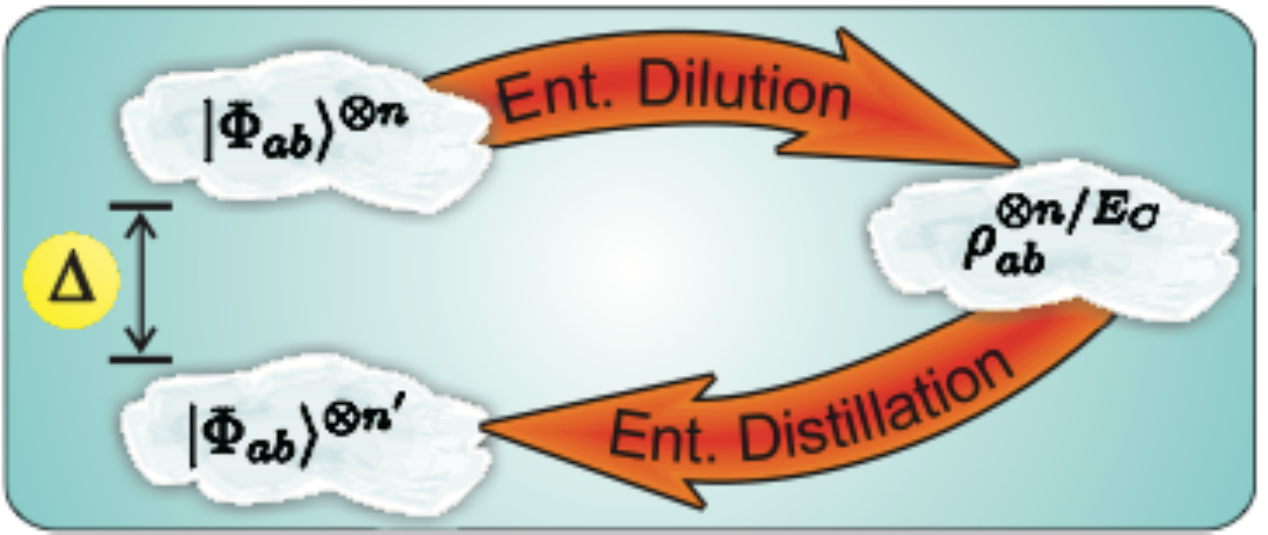}}
\caption{\label{disreg} (Color online.) \emph{Entanglement irreversibility.} The entanglement cost of a bipartite state $\rho$ is the optimal rate for converting $n$ ebits into $m$ copies of $\rho$ under LOCC, while entanglement of distillation is the optimal rate for converting $m$ copies of $\rho$ into $n'$ ebits via LOCC. Entanglement irreversibility is the difference in these two quantities given by $\Delta$ in this figure, which is equal to regularized discord when a hashing protocol is used for distillation. This figure is reproduced from \cite{arXiv:1007.0228}.}
\end{figure}

The entanglement cost of a bipartite state $\rho$ is the optimal rate for converting $n$ ebits into $m$ copies of $\rho$ under LOCC. Entanglement of distillation is the reverse process: it is the optimal rate for converting $m$ copies of $\rho$ into $n'$ ebits via LOCC. Here $m$, $n$, and $n'$ are large numbers \cite{RevModPhys.81.865}. The difference in entanglement cost and distillation is called the irreversibility of entanglement, see Fig.~\ref{disreg}. It is apparent that the regularized version of quantum discord with a purification $C$, i.e. $D^{\rm reg}(A|C) = \lim_{n \to \infty} \frac{1}{n} D(A^{\otimes n}|C^{\otimes n}) $ quantifies the entanglement irreversibility in hashing \cite{arXiv:1007.0228}:
\begin{gather}\label{entir}
D^{\rm reg}(A|C)=E_C(A:B)-E^H_D(A|B),
\end{gather}
where the superscript $H$ denotes that this is the distillable entanglement in the hashing protocol, which may not be optimal.

This is a powerful result which has some immediate applications. In \cite{arXiv:1007.0228} the authors prove that if a mixed-entangled state has additive entanglement of formation for some finite number of copies, and if it is possible to attain the best distillation rate, $E_D$, operating only on a finite number of copies before performing hashing, then the entanglement is irreversible. They further show that, for an entangled state $\rho_{AB}$ with purification party $C$ such that $\rho_{AC}$ is separable, entanglement between $AB$ is irreversible.

Another application of Eq.~\eqref{entir}, given in \cite{arXiv:1007.0228}, considers the tripartite pure state
\begin{gather}
\ket{\Psi}=\sum_b \alpha_b \ket{a_b, b, c_b},
\end{gather}
where $\{\ket{b}\}$ forms an orthonormal basis. The resulting reduced state of $AB$ is called a one-way maximally correlated state. It is shown that entanglement for such states is irreversible and the following holds:
\begin{align}
E_C(A:B) \ge& E_D(A:B) = D(A|B)\nonumber\\
=& D^{\rm reg}(A|B)=-S(A|B).
\end{align}
This implies that $D(A|B)=0$ for a one-way maximally-correlated state if and only if it is separable. A similar more recent study is \cite{arXiv:1110.6873}.

%**************************************************************
%**************************************************************
\section{Correlations in quantum algorithms}\label{Sec:discinQC}
%**************************************************************
%**************************************************************

The advantage associated with quantum algorithms is often believed to be related to the ability to create and manipulate quantum correlations. In general, classically-correlated states are zero measure in the set of all states and should require less resources to be simulated on a classical computer. In the case of pure states, the total correlations are proportional to entanglement, and the set of unentangled pure states is zero measure in the set of all states. \cite{arXiv:quant-ph/0201143} show, using a rigorous analysis, that a classical algorithm can in fact efficiently simulate all quantum computations using pure states for which the entanglement remains bounded throughout. The sense of simulation here refers to the ability to reproduce the measurement statistics efficiently to an arbitrary precision using a classical algorithm.

More specifically, the authors use the notion of $p$-blocked form, which can be applied for both pure and mixed states: a state $\rho$ is said to be $p$-blocked if it can be written as $\varrho=\bigotimes^k_{i=1}\rho_i$, where each $\rho_i$ is a state of at most $p$ qubits. They prove that if a computational register retains a $p$-blocked structure throughout the computation, (and $p$ does not scale with the problem), then the computation can be efficiently simulated. For pure states, the $p$-blocked structure puts a limit on entanglement, since at most $p$ qubits can be
entangled within each block. Therefore, it can be concluded that the amount of entanglement at some point in the computation has to scale with the size of the problem in order to achieve an exponential speedup. A similar conclusion is drawn in \cite{arXiv:quant-ph/0301063} where it is shown that any algorithm where the bipartite entanglement, (over all possible bipartitions), remains low throughout the computation can be efficiently simulated classically. However unbounded entanglement is not sufficient for a speedup as is demonstrated by the Gottesman-Knill theorem \cite{Book.Nielsen.Chuang}.

For mixed states, the $p$-blocked form forbids any correlations (quantum or classical) between the blocks. Thus the result for mixed states is not as clear cut, but some kind of correlations must scale with the size of the problem to yield a quantum advantage. \cite{arXiv:quant-ph/0201143, arXiv:quant-ph/0301063} suggest that the large volume of separable mixed states might allow efficient quantum computation without entanglement.

%**************************************************************
\subsection{Mixed-state quantum computation}
%**************************************************************

The ability to simulate pure-state quantum computation with no entanglement, and the idea that entanglement may not be necessary for an advantage in mixed-state quantum computation, leads to two natural questions. First: what is the natural extension of entanglement for mixed states? Second: what are the restrictions on (mixed) states if one wants to classically simulate the computation? As we see in this section, a possible candidate for the first question is discord. The first step in exploring the role of discord in simulating quantum computation is to explore the computational resources required to simulate computations without discord.

%**************************************************************
\subsubsection{Simulating concordant computation}

A computation, (in the circuit model), is termed concordant if the state is fully classical, as in Eq.~\eqref{FULLY_CLASSICAL_QUANTUM_STATES}, before and after the operation of each quantum gate. \cite{arXiv:1006.4402} shows that concordant quantum computation can be efficiently simulated classically assuming: a product input state diagonal in the computational basis, only one- and two-qubit gates, and terminal measurements on each qubit. Concordant computation is more general than classical computation since the local product basis is not restricted to the computational basis and can change at any point during the computation. The essential idea of Eastin's simulation method is that each step of the computation is equivalent to a permutation of the eigenvalues and a change of local basis: thus it is enough to keep track of the permutation and the change of eigenbasis at each step. Eastin outlines an algorithm for calculating these permutations and rotations, showing that operations which manipulate discord are necessary to achieve a quantum advantage in this specific instance of mixed-state quantum computation.

The method above is only the first step in simulating states with little or no discord. Its extension to gates operating on more than two qubits (e.g. Toffoli gates) is subject to fundamental difficulties. This restriction confines the types of concordant computation which can be simulated, at least using the ideas presented so far. Moreover, there is evidence that a computation involving many-qubit gates can achieve a nontrivial speedup over classical computation even when the computation is concordant. An example is the case where the \emph{deterministic quantum computation with one qubit} (DQC1) algorithm generates no discord \cite{arXiv:1004.0190}, discussed in detail in next section.

\cite{arXiv:1109.5549} outline a geometric approach to understanding the role of discord in quantum computation. They use this approach to illustrate the idea that since the set of concordant states is zero measure and nowhere dense in the space of all states, it is probably not useful for gaining a significant advantage over classical computation. The set of classical states is path connected, since any classical state $\chi$ has an adjacent classical state $(1-\epsilon)\chi+\epsilon/d\openone$ which is closer to the center of the Bloch hypersphere. Thus one can go from any classical state to any other via the center of the hypersphere. However if we allow finite-discord states in an intermediate step, we can take a ``shortcut''. This gives an idea of why discordant states may help speed up the computation.

%**************************************************************
\subsubsection{Distributed algorithms and restricted gates}

One feature of quantum gates is the ability to create entanglement. Entangling gates are an essential resource in the circuit model of quantum computation, however their ability to entangle is not always evident in mixed-state algorithms. \cite{arXiv:1009.2571} suggest a way of identifying the entanglement resources in quantum computation with separable states using a paradigm of distributed quantum gates, where the input (and output) states are distributed between two separate parties $A$ and $B$. Some entanglement resources are required in order to implement a unitary gate operation $U$ even if the input (and output) states are separable.

They require $A$ and $B$ to implement a map, $G(\rho_{AB})=U\rho_{AB} U^\dagger$ for a limited predetermined set of separable states $\rho\in\mathcal{L}$, using LOCC: $G(\varrho) = \sum_j K_j \varrho K^\dagger_j$ for any state $\varrho$ with $\sum_j K_j\rho K^\dagger_j=U\rho U^\dagger$, when $\rho\in\mathcal{L}$. It is not possible to implement the gate without entanglement for certain sets of separable inputs (and outputs).

Implementing the c-\textsc{not} gate on the set of four separable nonorthogonal states in Table~\ref{t:restricted}, which remain separable at the end, requires some entanglement. An implementation without entanglement would allow $A$ and $B$ to distinguish between these nonorthogonal states in a deterministic way and is therefore impossible.

\begin{table}[t]
\begin{tabular}[v]{cl|cl} \hline
\# & State & \# & {State} \\ \hline
$a$&$\ket{z_-}\ket{y_+} \to i\ket{z_-} \ket{y_-}$ & $c$ &$\ket{y_+}\ket{x_-} \to\ket{y_-}\ket{x_-}$\\
$b$&$\ket{z_+}\ket{y_+} \to\ket{z_+} \ket{y_+}$ & $d$ & $\ket{y_+}\ket{x_+}
\to\ket{y_+}\ket{x_+}$ \\ \hline
\end{tabular}
\caption{\label{t:restricted} A distributed implementation of the above four $input\to output$ states is impossible without shared entanglement. $A$ and $B$ each hold one unknown qubit randomly selected from the set $\{a,b,c,d\}$. They can only implement the desired operation, i.e. achieve the output state above, if they share entanglement. $\ket{x_\pm}$, $\ket{y_\pm}$, and $\ket{z_\pm }$ are the $\pm 1$ eigenstates of the Pauli matrix $\sigma_x$, $\sigma_y$, and $\sigma_z$ respectively.}
\end{table}

More generally, for qubits, a set including at least one separable pure state, the completely-mixed state, and a quantum-quantum state requires some shared entanglement for the implementation of a generic unitary operation which can change the discord. For any set of separable pure states, entanglement is required if the operation changes the discord of any mixture associated with these states \cite{arXiv:1009.2571}.

%**************************************************************
\subsubsection{Almost-completely-mixed states}

The role of discord is especially interesting when analyzing algorithms involving states which are very close to the completely-mixed state like the {\it pseudo-pure states} used in \emph{liquid-state NMR}. In NMR quantum computation, a spin system at temperature $T$ is described by the density matrix $\rho=\frac{1}{Z}e^{-H/(k_BT)}$. At room temperature these states can be very close to the completely-mixed state
\begin{gather}
\rho = \frac{1-\alpha}{2^n} \openone + \alpha \; \eta
\end{gather}
with $\alpha \ll 1$ and $\eta$ an arbitrary state. Entanglement in these states is vanishingly small throughout any computation and vanishes for $\alpha<2/4^n$ for a large ensemble size $n$ \cite{PhysRevLett.83.1054}. However, one can see that the number of parameters required to specify such a state is the same as the number of parameters that specify $\eta$.

Some algorithms exploit the smallest amount of purity in these states to provide an advantage over known classical algorithms. \cite{arXiv:quant-ph/0110029} were the first to consider the role of correlations other than entanglement and they identify quantum discord as a possible indicator of quantum advantage. In a more recent paper, \cite{arXiv:0906.3656} presents a similar line of thought, where he discusses results concerning different algorithms (the Deutch-Jozsa algorithm, Grover's search algorithm, and DQC1), discussed in detail below. More generally, quantum correlations other than entanglement and indistinguishable quantum states are found to be closely related to the efficiency of various algorithms using highly-mixed states.

%**************************************************************
\subsection{Deterministic quantum computation with one qubit}
%**************************************************************

The discovery by \cite{arXiv:0709.0548} that the DQC1 algorithm, which estimates the normalized trace of a unitary operator, produces bounded amounts of entanglement on the one hand and discordant states on the other, is the first real evidence that mixed-state quantum computation may have an advantage over classical computation even when entanglement is absent (or at most bounded), as predicted by \cite{arXiv:quant-ph/0201143, arXiv:quant-ph/0110029}. The algorithm provides an exponential speedup over any known classical algorithm. \cite{arXiv:0807.0668, arXiv:1105.2262} have experimentally demonstrated the algorithm in optics and liquid-state NMR respectively.

%**************************************************************
\subsubsection{The DQC1 model}

\cite{PhysRevLett.81.5672} introduce DQC1 in the following fashion: A $n+1$ qubit system is prepared in the state $\Omega$, with $n$ qubits called the target ($T$) in the completely-mixed state $\frac{1}{2^n} \openone$, and one qubit called the control ($C$) in the (pseudo)-pure state $\rho$. The system then undergoes some unitary evolution $W$ which can be efficiently implemented using the standard quantum circuit model (a set of one and two-qubit unitary gates), to yield the final state $\Omega_f$. At the end an expectation value for an observable $\mathcal{O}$ is determined on the ``(pseudo)-pure'' $C$ qubit. Repeating this procedure gives an estimate of
\begin{gather}
\braket{\mathcal{O}}_\Omega = \tr (\mathcal{O} \; \Omega_f) =\tr \left\{\mathcal{O} \left[ W \left(\rho \otimes \frac{\openone}{2^{n}} \right) W^\dagger\right] \right\}
\end{gather}
and its variance. There is no known classical algorithm for calculating this trace efficiently. The DQC1 model is not as powerful as a model with $p$ pure qubits (sometimes called DQCp) and cannot be used to simulate an arbitrary quantum process.

\begin{figure}[t]
\resizebox{8 cm}{3.56 cm}{\includegraphics{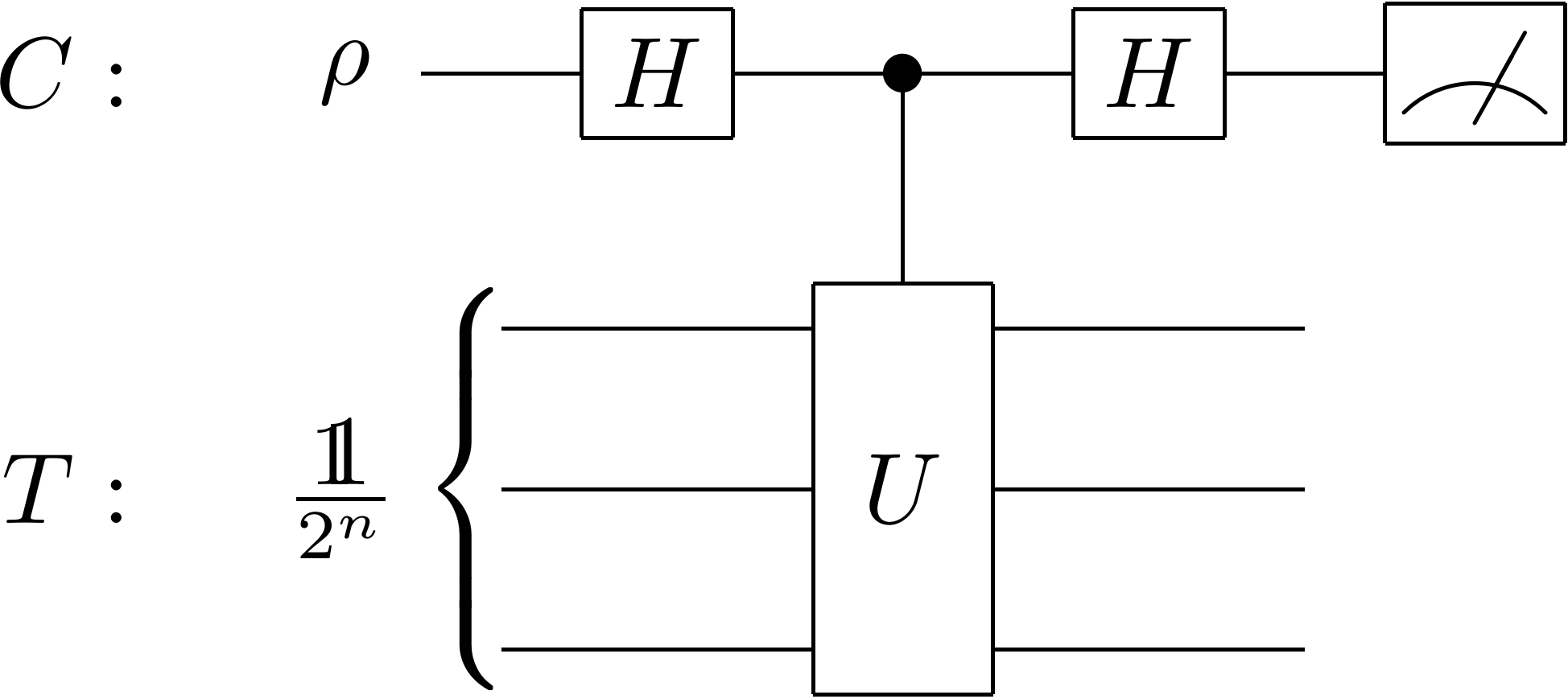}}
\caption{\label{fig:DQC1}\emph{DQC1.} The Deterministic quantum computation with one qubit (DQC1) algorithm for estimating the trace of an $n$-qubit unitary. The control qubit (top) is used to implement a controlled-unitary transformation on the target (fully- mixed) register (bottom). This circuit provides an exponential speedup over any known classical algorithm despite the fact that the input is highly mixed. The state just before the measurement is discordant, $D(T|C)>0$ but separable $E(T:C)=0$ for the T-C bipartition.}
\end{figure}

The standard DQC1 model yields the normalized trace of an $n$-qubit unitary matrix $\frac{1}{2^n} \tr(U)$ using a completely-mixed state of $n$ $T$-qubits, and a single $C$ qubit in the state $\rho = \frac{1-\alpha} {2} \openone + \alpha \ket{0}\bra{0}$. We now limit our discussion to $\alpha=1$, but most works deal with all values $1 \ge \alpha > 0$. The trace estimation algorithm is described in the following fashion \cite{arXiv:quant-ph/0110029}: The initial total system is in a product state $\Omega = \rho \otimes \frac{1}{2^n} \openone$. A Hadamard is applied to the $C$ qubit, followed by a controlled-unitary operation between the $C$ and $T$ qubits, see Fig.~\ref{fig:DQC1}. The state of the system is
\begin{align}
\Omega_f=\frac{1}{2}\left(\openone \otimes \frac{\openone}{2^n}
 +\ket{0}\bra{1} \otimes \frac{U^\dagger}{2^n} + \ket{1}\bra{0} \otimes \frac{U}{2^n}\right).\label{dqcf}
\end{align}
Tracing out the mixed $n$ qubits we get the final state of $C$
\begin{gather}
\rho_f=\frac{1}{2}\left(
\begin{array}{cc}
1&\frac{\tr(U^\dag)}{2^n}\\
\frac{\tr(U)}{2^n}&1
\end{array}\right).
\end{gather}
The expectation values of $\sigma_{x}$ and $\sigma_{y}$ are determined on $\rho$:
\begin{align}
& \tr (\sigma_x \rho_f) = \frac{\alpha}{{2^n}} {\rm Re}[\tr(U)],\nonumber\\
& \tr (\sigma_y \rho_f) = \frac{\alpha}{{2^n}} {\rm Im}[\tr(U)],
\end{align}
where Re and Im are the real and imaginary parts of the trace respectively. The last results hold provided $\alpha >0$. Note that when $\alpha$ is sufficiently small there is no possibility of the state being entangled at any point of the computation (however this requires $1/\alpha$ to scale exponentially with the problem making the algorithm inefficient).

The estimate is independent of the input size $n$ and the number of runs required for reaching a given accuracy $\epsilon$ is a quadratic function of the accuracy \cite{arXiv:quant-ph/0505213}. Since we assume that the controlled-unitary operation can be efficiently implemented using a number of one-- or two--qubit gates, which is at most a polynomial function of $n$, the number of resources required to achieve an accuracy $\epsilon$ scales at most polynomially with the number of qubits. On the other hand the best-known classical algorithm for estimating the trace of a unitary requires resources that are an exponential function of $n$. This is intuitively related to the fact that the normalized trace is the average of all $2^n$ terms, although this intuition may fail since we are only interested in \emph{estimating} the \emph{normalized} trace.

%**************************************************************

There is no guarantee that there exists no efficient (classical) algorithm for estimating the normalized trace. \cite{arXiv:quant-ph/0505213} argue why it is unlikely that a classical algorithm can efficiently solve this problem. There are a number of equivalent problems which seem to have no efficient classical solutions and can be used as a technological motivation for implementing DQC1. A popular example is the estimation of Jones polynomials in knot theory (for a more complete list and an overview of these protocols see \cite{arXiv:0807.4490} and references therein). More generally DQC1 represents its own complexity class which is believed to be more powerful than the one associated with classical computing. Its place in the hierarchy of complexity classes as well as a number of problems associated with this class are discussed in \cite{arXiv:1109.5549}. \cite{arXiv:quant-ph/0611157} show that standard methods for classically simulating quantum processes are inefficient for simulating the trace estimation algorithm.

%**************************************************************
\subsubsection{Correlations in DQC1}

To study the entanglement in the algorithm we start by expanding the unitary $U=\sum_j e^{i\phi_j} \ket{u_j}\bra{u_j}$, where $\{\ket{u_j}\}$ is the eigenbasis of $U$ and $e^{i\phi_j}$ are the eigenvalues. The final state in Eq.~\eqref{dqcf} is
\begin{gather}\label{udecomp}
\Omega_f= \frac{1}{2^{n+1}}\sum_j \ket{a_j} \bra{a_j}\otimes\ket{u_j}\bra{u_j},
\end{gather}
where $\ket{a_j} =\ket{0} + e^{i\phi_j} \ket{1}.$
It is easy to see that there is no entanglement between $C$ and $T$ \cite{PhysRevLett.92.177906}. For a more general bipartite cut, the situation is more difficult. \cite{arXiv:quant-ph/0505213} use a measure based on the Peres-Horodecki partial-transpose criteria and calculate the entanglement for any bipartition. They find that for any other bipartite cut there is some entanglement, however it is bounded from above by a constant. Thus for large $n$, there is vanishingly-small (genuinely $n$-partite) entanglement with respect to the maximal entanglement possible. Since entanglement is usually present in all but one bipartition, these results suggest that the speedup may be related to the distribution of entanglement rather than the amount of entanglement.

In a seminal paper, \cite{arXiv:0709.0548} discuss the role of discord in DQC1. At the output of the computation $C$ has some discord with respect to $T$. The main difficulty in calculating the discord is finding the optimal measurement on $C$. Different unitary operations, in general, give different bases for making the optimal measurement. For a typical unitary, chosen according to the Haar measure, the measurement basis plays a minor role in calculating the discord, and it is possible to estimate the discord using any measurement on the $x-y$ plane. The $x$--basis measurement is chosen to calculate discord for a typical random unitary. For any unitary the entropy of the system does not change and is given by $S(\Omega)= n$. For a typical unitary the trace is very small $|\tr(U)|\ll 1$ and the local entropy is $S(\rho_f) \approx 1$. The conditional entropy for large $n$ with a measurement on the $x$ basis is given by $S(T|\{\Pi_x\}) \approx n+1-\log(e)$, so the discord is $D(T|C)=2-\log(e)$, a constant fraction of the maximum possible ($D_{\rm max}=1$). Therefore discord scales like the efficiency. This is the first quantitative evidence that quantum correlations other than entanglement play a part in the speedup associated with a quantum algorithm.

The work above was followed by an attempt to quantify the correlations between $C$ and $T$ qubits using other forms of quantum correlations \cite{arXiv:0811.4003}, namely MID and locally-noneffective unitary operations (LNU). The latter is a measure of correlations based on the disturbance of a state due to a local unitary operation which does not change the marginals \cite{EPL.75.1}. LNU is ineffective in quantifying the quantum correlations in DQC1. In general LNU behaves differently from discord, and most notably vanishing discord does not imply vanishing LNU. On the other hand, they find that MID is a good measure of quantum correlations in the trace-estimation algorithm. For a typical unitary the expression reduces to $M=1$ which is the maximum value for MID. However we can see that this numerical result is slightly ambiguous due to the degeneracy of $\tau_f$ (the final state of the register). Using the fact that the discord vanishes for measurements on $T$ rather than $C$ (with the optimal measurement basis being the eigenbasis of $U$), MID can be made to match discord to a good approximation.

\cite{arXiv:1004.0190} analyze the role of discord in the speedup of DQC1 using geometric discord. Rather than calculating the correlations for a random (typical) unitary, they find the class of unitary matrices which do not produce discord in DQC1. These unitary operations are of the form $U=e^{i\phi} V$ with $V^2=\openone$, where $\{V\}$ is a set of Hermitian unitary operations. Since $V$ has $\pm 1$ eigenvalues, $\ket{a_j}=\ket{\pm}$ in Eq.~\eqref{udecomp}, which is then clearly classical. They suggest that an efficient classical algorithm for calculating the trace of these unitary operations cannot be found. However this is a simpler problem than the more general one and it may be possible that an efficient algorithm exists.

We note that the fact that discord vanishes at the end of the algorithm does not necessarily diminish the role of discord in quantum algorithms. A quantum circuit implementation of the DQC1 algorithm may generate discord at some points along the circuit and these may vanish at some later point. As an example we take a unitary composed of a phase gate $u_\phi$ on one qubit of $T$ followed by a hermitian unitary operation $U_H$ on $T$ followed by a final reverse phase gate so $U=u_{\phi} V u_\phi^\dag$ with $V^2=\openone$. After the phase gate the state has nonvanishing discord but discord vanishes at the end of the operation. This reminds us that discussions of the role of correlations in quantum algorithms only make sense when referred to specified gate sets.

%**************************************************************
\subsubsection{Purification of DQC1}\label{purdqc1}

\cite{arXiv:1006.2460} analyze the purification of the DQC1 system. Purifying the mixed target state, $T$, puts it in an entangled state with the environment $E$ so that the state $\ket{\Psi}_{ECT}$ is pure. Using their monogamy relation for pure states, Eq.~\eqref{fanchineq}, the authors show how entanglement and discord get redistributed in the system. An alternative way of finding the entanglement between $C$ and $TE$ is to note that the entropy of $C$ gives the entanglement, i.e. $E_F(C:TE)=S(C)$. They use their result to suggest that the power behind the DQC1 model is derived from the power to redistribute entanglement in the purified system. Note that at the end of the process $S(C)$ is only dependent on $\tr(U)$.

The idea of purification has one further implication which has so far (to the best of our knowledge) not been discussed. Purification of an $n$-qubit system requires at most $2n$ qubits. One may use the standard methods for simulating pure quantum computation on this purified $2n$-qubit system. Thus the computation only has an exponential advantage over a classical algorithm if entanglement in the purified system remains unbounded.

%**************************************************************
\subsubsection{Experiments}

Of the experimental implementations of DQC1 two are of particular interest, since they involve an explicit measurement of the correlations between $C$ and $T$ at the end of the computation. In an optical implementation of DQC1, \cite{arXiv:0807.0668} use two initially-pure qubits. They add noise (independently) to both qubits and follow this by a rotation with phase $\phi$. They calculate the trace of the rotation matrix for various values of $\phi$ and various degrees of mixedness for $C$. Then they calculate discord and entanglement after performing complete tomography on the two-qubit state and find agreement with the theoretical predictions.

In an NMR implementation of DQC1, \cite{arXiv:1105.2262} detect discord using the witness given in \cite{arXiv:1004.0190}, see Sec.~\ref{svd}, both before and after the controlled-unitary operation. They estimate the trace of a three-qubit unitary operation which is useful for the calculation of the Jones polynomial. Before the computation discord is zero, while after the computation it is nonzero for all values of $\alpha$ (the initial purity of $C$).

%**************************************************************
\subsubsection{Other partitions}

So far discord in DQC1 has only been calculated for the $C:T$ bipartition. The importance of this particular cut lies in the view that the unknown unitary $U$ acts on $T$ while the information about $\tr(U)$ is encoded in $C$. However entanglement exists in other bipartitions and it would be interesting to investigate how discord fares along those cuts and, perhaps more importantly, how multipartite quantum correlations behave in DQC1.

%**************************************************************
\subsection{Metrology}
%**************************************************************

Apart from its role in speeding up computation, correlations play a role in improving the precision of some measurements. Again, for pure states the necessity of entanglement is well understood since the optimal strategies always involve entanglement \cite{Science.306.1330}. Other forms of correlations only come into play in the presence of noise. This is closer to the realistic scenario but far from the optimal scenario. The results regarding discord in quantum-metrology protocols with noisy states are quite surprising and especially relevant in the light of experimental constraints. In this respect, it is most appropriate to compare the quantum strategy to other strategies with the same amount of noise.

Let us describe quantum metrology using the following paradigm \cite{PhysRevLett.96.010401}. A qubit undergoes unitary evolution $\ket{0} \to \frac{1}{\sqrt{2}} (\ket{0}+\ket{1}) \to \frac{1}{\sqrt{2}} (\ket{0} + e^{i\phi}\ket{1})$ with some unknown phase $\phi$, see Fig.~\ref{fig:metro}a. We would like to get the best estimate of this phase using a large but limited number $N$ of initial (probe) states\footnote{To be more precise, we should take into account all resources used, i.e. space (number of qubits) and time (number of gates). However these reduce to only the number of initial qubits for this example.}. We can estimate the phase by measuring the output state and the precision increases with the number of states as ${1}/{\sqrt{N}}$. This limit comes from the central limit theorem and is usually referred to as the shot noise or standard quantum limit. However, this limit can be broken using quantum effects such as entanglement, giving a new limit to the precision which increases to $1/N$. This is the Heisenberg limit, an improvement of $\sqrt{N}$ over the shot-noise-limited classical strategies.

%**************************************************************
\subsubsection{Quantum enhancement}

\begin{figure}[t]
\begin{center}
\resizebox{8 cm}{10.48 cm}{\includegraphics{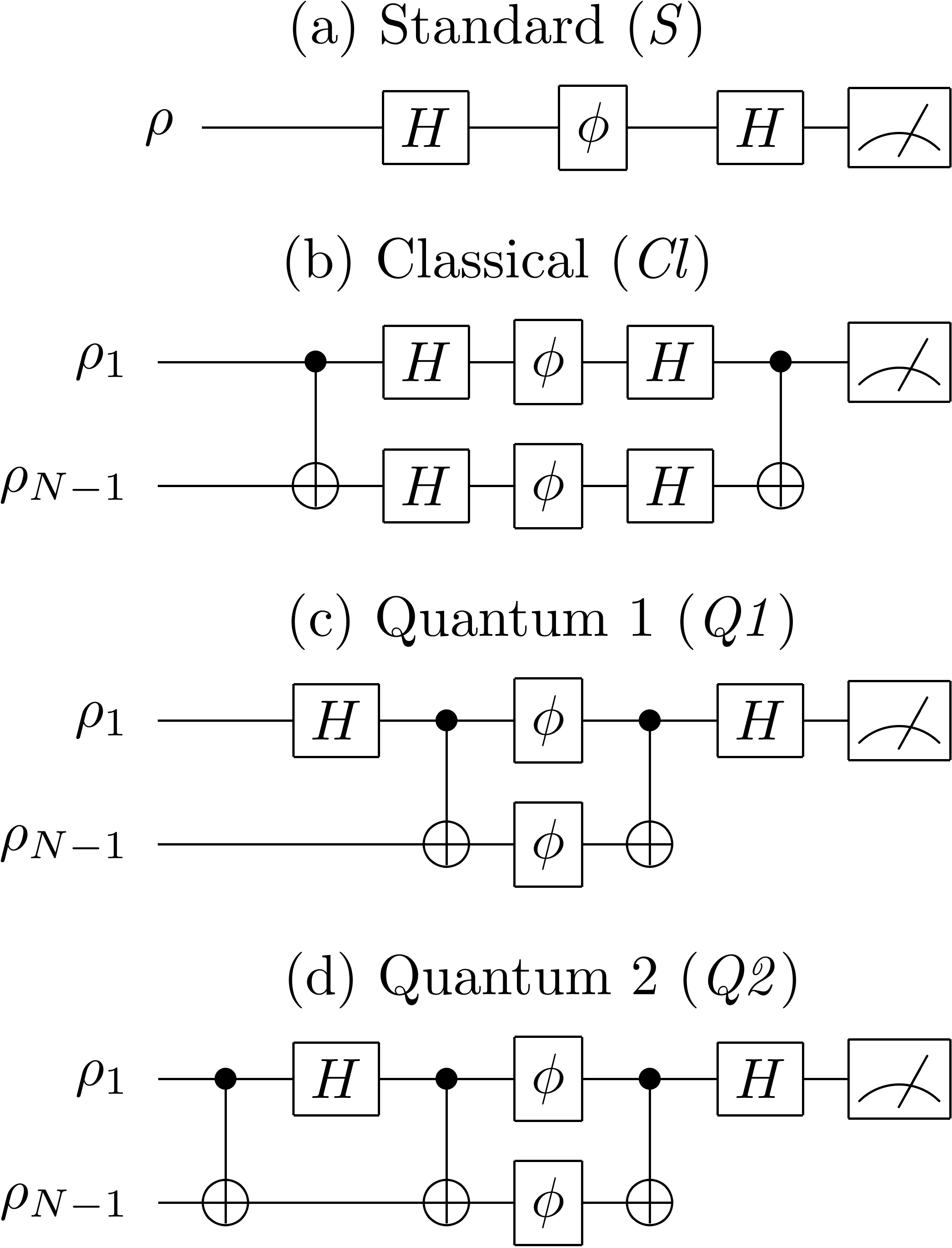}}
\end{center}
\caption{\label{fig:metro} \emph{Metrology with noisy states.} Four protocols for estimating the phase $\phi$. All input qubits are of the same form (Eq.~\eqref{eq:metrologyin} of the main text). (a) The standard protocol $S$ where the probes are independent. (b) The classically-correlated protocol $Cl$, for which the first c-\textsc{not} gate puts the qubits in a classically correlated state. (c, d) The quantum protocols $Q1$ and $Q2$, for which the Hadamard on the control qubit followed by a c-\textsc{not} quantum correlates the probe states. The quantum protocol $Q2$ is the most efficient, giving quadratic enhancement over $S$ even past the point where entanglement vanishes. It is also the only protocol which remains significantly discordant in the high-noise limit. This figure is reproduced from \cite{arXiv:1003.1174}.}
\end{figure}

\cite{arXiv:1003.1174} compare four different strategies for estimating $\phi$. Rather than assuming pure states at the input, mixed states of the form
\begin{gather}
\label{eq:metrologyin}
\rho = \frac{1+p}{2} \ket{0} \bra{0} + \frac{1-p}{2} \ket{1} \bra{1}
\end{gather}
are assumed for every qubit. They compare and analyze the correlations and precision as functions of the (mixedness) parameter $p$ for $0\le p \le 1$. $\rho$ is pure when $p=1$. Of the four strategies, (see Fig.~\ref{fig:metro}), two are classical, $S$ and $Cl$, and two are quantum, $Q1$ and $Q2$. In every case, the input is an (uncorrelated) product state, for which each qubit is assigned the same value for the mixedness parameter above. In the standard strategy $S$ the input is sent through a Hadamard gate first, then through the unitary where the phase is acquired. For the $Cl$ strategy, they apply c-\textsc{not} gates between the first and every other qubit with the first qubit as the control. This generates classically-correlated states. For strategy $Q1$, the Hadamard gate comes before the c-\textsc{not} which produces quantum-correlated probes, and entanglement is finite for $p>1/N$. In strategy $Q2$, the states are first classically correlated via c-\textsc{not} gates, and then quantum correlated via a Hadamard followed by another set of c-\textsc{not} gates. This also produces states that are entangled when $p > 1/N$. This strategy is used in a recent experiment \cite{PhysRevA.82.022330}.

%**************************************************************

Comparing the precision of each strategy, the quantum strategy $Q2$ is found to scale $1/\sqrt{N}$ better than the strategy $S$ for all $p$. The precision for the classical strategies is approximately the same and scales like $1/(p\sqrt{N})$ for large $N$ and small $p$. However the main point lies in the comparison of correlations in all strategies. The classical correlations of $Cl$ do not seem to have a significant effect, while the quantum correlations are associated with a quantum advantage. The strategy $Q2$ has more quantum correlations for noisy states and fares much better than $Q1$. However for low-noise states, both $Q1$ and $Q2$ have a very large discord and both give similar results for the precision. The advantage of $\sqrt{N}$ of $Q2$ over $S$ is maintained even after the loss of all entanglement.

%**************************************************************
\subsection{The role of correlations in other algorithms}
%**************************************************************

While the results for DQC1 and metrology indicate that correlations other than entanglement are at least partly responsible for the quantum advantage in some schemes, their exact role as well as the role of entanglement is not fully understood. The Gottesman-Knill theorem \cite{Book.Nielsen.Chuang} tells us that some algorithms involving highly-entangled states can be efficiently simulated classically. So even where entanglement is necessary for some advantage, it is not sufficient. We next review results regarding the relation between some types of correlations and some quantum scenarios.

%**************************************************************
\subsubsection{One-way quantum computation}

\cite{PhysRevA.84.022324} analyze the correlations in a noisy implementation of one-way (measurement-based) quantum computation due to \cite{arXiv:0910.1116}. This model involves a sequence of single-qubit measurements on a graph state and feed-forward. The initial state of the computation should be highly entangled, as this is clearly the main resource for this model, although a large amount of entanglement does not guarantee that a resource state is useful for achieving algorithmic speedups \cite{PhysRevLett.102.190501}. However, certain kinds of noise that reduce entanglement are sometimes less disruptive than other kinds of noise that do not affect entanglement strongly. The same is true for quantum discord. The effects of two kinds of noise on the performance of some algorithms are analyzed. These are compared to their effects on entanglement and discord. The noise models are in the form of a phase flip, where with some probability a $\sigma_z$ operation is applied, or white noise, where with some probability $\sigma_x,\sigma_y$ or $\sigma_z$ is applied. Entanglement is strongly affected by white noise and is less sensitive to phase flips, while discord is only slightly more sensitive to phase flips than to white noise.

A simple one-way algorithm for remote state preparation is described as follows: Starting with a two-qubit graph state $\ket{G}=\frac{1}{2}(\ket{00}+\ket{01}+\ket{10}-\ket{11})$ the challenge is to prepare the second qubit in the state $\ket{\psi} = \cos (\phi/2) \ket{0} -i \sin (\phi/2) \ket{1}$. This is accomplished by making an orthogonal projective measurement on the first qubit in the eigenbasis $\ket{M_\pm}=1/\sqrt{2}(\ket{0}\pm e^{-i\phi}\ket{1})$. The result gives the required state up to an application of a Pauli $\sigma_x$ depending on the measurement result. \cite{PhysRevA.84.022324} study this protocol with the application of noise applied to the first qubit. The fidelity of the outcome is, for some parameters, better for less entangled or discordant states. When entanglement vanishes, the protocol still gives better fidelity than a random state (the worst case scenario). As is often the case, discord does not vanish up to the point where the noise is maximal and fidelity drops to its lowest value of $\frac{1}{2}$. For this example, a different measure of quantum correlations Eq.~\eqref{mepot}, called minimum entanglement potential \cite{PhysRevLett.106.220403}, is directly proportional to the fidelity.

They also investigate the implementation of a general rotation and a c-\textsc{not} gate. This protocol involves more qubits in the cluster state. They apply noise to all qubits that are measured in the protocol. The different kinds of noise have the same effect on fidelity despite their very different effects on different types of correlations. Both examples imply that correlations are not the best indicator for efficiency in these protocols.

Measurement-based quantum computation provides an interesting platform for studying stronger-than-classical correlations. \cite{arXiv:0805.1002} use the following description of measurement-based quantum computer to study the computational power of correlations: A measurement-based quantum computer consists of ``two components, a correlated multipartite resource and a classical control computer''. They note that the control computer has less computational power than a universal classical computer. Next they show that various types of correlations in the multipartite state can be used to increase the computational power of the classical control. For example, a single GHZ state is not enough to promote the control computer to a universal classical computer, however a polynomial supply of these three-qubit states is sufficient. It would be interesting to study the role of discord in this model.

%**************************************************************
\subsubsection{Algorithms with highly-mixed states}

Implementation of the Deutsch-Jozsa and Grover algorithms using highly-mixed (NMR-type) states are discussed in \cite{arXiv:0906.3656}. Implementing these in the case where entanglement vanishes requires large resources. The number of runs required for a good estimate of the outcome is exponentially large if we require no entanglement. However, for a single run of the Deutsch-Jozsa and Simon algorithms, a quantum computer reveals more information about the function than a classical one even in the limit of vanishing entanglement \cite{arXiv:quant-ph/030618}. This again indicates that entanglement does not account fully for the quantum advantage although it still plays a role in getting a polynomial or an exponential advantage in many algorithms.

The reason behind the quantum advantage is presently unknown. Neither discord nor entanglement fully explain all phenomena. Another option is distinguishability \cite{arXiv:0906.3656}. These features are interrelated but not synonymous and it seems that all three of them and potentially other concepts play a role in the advantage of quantum algorithms over classical ones.

%**************************************************************
%**************************************************************
\section{Interpretation of quantum correlations} \label{SEC_PHYS_INTERPRETATION}
%**************************************************************
%**************************************************************

Some measures of correlations are motivated by the possibility to interpret them as the ``quantumness" or ``classicality" of a system, see Sec.~\ref{SEC_MEASURES}. The foremost of these are the quantum discord \cite{arXiv:quant-ph/0105072}, classical correlations \cite{JPhysA.34.6899} and MID \cite{PhysRevA.77.022301}. Discord captures the idea of superposition in a system: in terms of quantum discord, a system is classical if and only if it can be written as an eigenbasis of pointer states which are orthogonal, Eq.~\eqref{CLASSICAL_QUANTUM_STATES}. Thus if some pointer states must be described as superpositions of others, the state is discordant. Correlations and superposition play a large role in various distinctly-quantum phenomena. Most prominent of these are: the role of correlations in measurement and decoherence \cite{arXiv:quant-ph/0011039, arXiv:quant-ph/0105127, arXiv:0911.4307, arXiv:1012.4903, arXiv:1110.1664}; and the role of correlations in thermodynamics regarding the difference between local and nonlocal operations\cite{arXiv:quant-ph/0112074, arXiv:quant-ph/0202123, arXiv:0803.4067, arXiv:1002.4913, arXiv:1105.4920}.

We refer the reader to Sec.~\ref{SEC_QI}, devoted to quantum information, for descriptions of tasks such as state merging \cite{arXiv:1008.3205, arXiv:1008.4135}, entanglement activation \cite{PhysRevLett.106.220403} and dense coding, whose performance is linked to discord and gives it operational meaning.

%**************************************************************
\subsection{Einselection}\label{Einselection}
%**************************************************************

Zurek's original motivation for defining quantum discord was an information-theoretic approach to decoherence mechanisms such as \emph{environmentally-induced superselection} (einselection) \cite{arXiv:quant-ph/0011039,arXiv:quant-ph/0105127}. Discord is related to the information loss due to a quantum measurement process \cite{arXiv:quant-ph/0105072}. \emph{The most-classical} basis is the one which minimizes discord \cite{arXiv:quant-ph/0011039}, and vanishing quantum discord is a sign of classicality. The classical correlations between a measurement device and the system can also be used as a measure of how much the measurement disturbs the system if we also take into account the loss of coherence (decay of entanglement) with a purifying environment \cite{PhysRevA.84.052309, arXiv:1110.1664}.

Einselection (see \cite{arXiv:quant-ph/0105127} and references therein) is introduced to explain how systems made up of quantum parts can ``become'' classical after interaction with the environment. The notion of classicality here is the inability to observe superpositions (of quantum states) in the classical world. Einselection is the process whereby the environment picks out a preferred measurement basis, effectively imposing a superselection rule which forbids quantum superposition of pointers of measuring devices. During a measurement the environment interacts with the measuring apparatus leading to decoherence in the preferred pointer basis (for example that of a live or dead cat) and this is an effective loss of information. If einselection is effective, the system-apparatus state $\rho_{SA}$ is decohered into a new state with vanishing discord \cite{arXiv:quant-ph/0011039} $D(S|A)=0$. The pointer basis is the basis which minimizes discord (to zero). It corresponds to the {\it superselection sectors} of the apparatus. At this point the measurement apparatus can only be observed in one of the pointer states. The value of discord can generally be used to measure the efficiency of einselection \cite{arXiv:quant-ph/0105072} and classically-correlated states are seen as an indicator of a superselection rule. They calculate discord over orthogonal measurements since only those are of interest for the ideal einselection process.

Following decoherence, the records of the measurement are stored in some classical memory of the environment by being cloned into multiple copies which can be accounted for by the {\it redundancy ratio} \cite{arXiv:quant-ph/0011039}. The observer reads out the results of a measurement by collecting information from the environment (for example by interacting with photons). However although he usually only sees a fragment of the environment, this fragment contains enough information to identify the result of the measurement and allows other observers to {\it objectively} record the same result. This process is known as {\it quantum darwinism} \cite{arXiv:0911.4307}. It is intimately related to einselection, which is successful if the fragment of the environment contains the result of the measurement but no other information about the system. \cite{arXiv:0911.4307} calculate the discord between the system and the fragment of the environment in the pointer basis. This is related to the information about the initial coherence between the pointer states that is recorded in the given fragment. In the case of perfect decoherence in the chosen pointer basis the discord is zero. For a good decoherence mechanism the size of the fragment and the initial state of the environment play only a minor role. Note that since information is only {\it effectively} lost, the whole of the environment still contains information about superposition.

%**************************************************************
\subsection{Maxwell's demon}\label{maxwell}
%**************************************************************

The amount of extractable work from a quantum version of Szilard's engine \cite{ZeitsPhysA.53.840} depends on how we implement the engine. Maxwell's demons attempt to break the second law of thermodynamics by extracting the maximal amount of work from an engine regardless of the entropy. An information-theoretic approach to taming Maxwell's demon has far-reaching implications in different branches of physics such as general relativity and quantum mechanics \cite{RevModPhys.81.1}.

For a $d$-dimensional system in a state $\rho$ in contact with a heat reservoir of temperature $T$, the amount of work that a demon can extract from the system in one run is $k_BT\log(d)$ where $k_B$ is Boltzmann's constant in the relevant units. For simplicity we set $k_BT=1$ and measure the work in bits. The scheme to perform this work is to measure the state of the system and use this information to extract $\log(d)$ bits of work from the now known pure state. The demon can repeat this process until the bath temperature goes to zero, effectively creating a perpetual motion machine of the second kind (extracting work from a heat bath without any loss).

Quantum exorcism reminds us that the demons must keep some record of their measurement results. Erasing this record would cost them some work, enough to balance the books. The amount of work, given by Landauer's principle, is proportional to the entropy of the measurement device. The total work extracted from the system after erasure is given by
\begin{gather}
W_{\rm demon}=\log(d)-S(\{p_a\}),
\end{gather}
where $\{p_a\}$ are the probabilities of the results of demon's measurement. This entropy is minimized when the measurement is made in the eigenbasis of the system giving an entropy $S(\{p_a\})=S(\rho)$. To perform the maximal amount of work the demon should know the state of the system and be able to perform the optimal measurement. However if our demon is unable to perform some operations, in particular if he is restricted to local measurements, he can perform less work. These less powerful local beings are sometimes called local \emph{goblins} \cite{arXiv:1002.4913}. The difference in the amount of work performed by demons and goblins is a measure of correlations. If we allow the goblins to communicate classically, the difference in work is a measure of nonclassical correlations. Following \cite{arXiv:1105.4920} we call this measure {\it demon discord}. Various scenarios for communication and local knowledge give a different demon discord. Regardless of the physical picture the demon discord involves some (possibly complex) measurement strategy $\mathcal{M}$ which the goblins implement. The demon discord is given by
\begin{gather}
D_{\rm demon}=S(\{p_a\})-S(\rho),
 \end{gather}
where $\{p_a\}$ are the probabilities of the measurement outcomes for strategy $\mathcal{M}$ and state $\rho$. We assume that the goblins can communicate when they erase their records \cite{arXiv:1105.4920}. The demon discord depends on the measurement strategy $\mathcal{M}$, and different approaches give different results. The two main paradigms used to describe the demon discord scenario are the demon approach described above, and the closed local operations and classical communication (CLOCC) approach described in Sec.~\ref{SEC_DEFICIT}. In the latter the demon discord is \emph{work deficit}, see also Sec.~\ref{SEC_POVM_DEFICIT}. Here we follow the demon approach.

\cite{arXiv:quant-ph/0202123} describes the following scenario: A measurement apparatus $A$ is correlated with the system $B$. We would like to extract work from the system-apparatus state by using a classical strategy. First we make an orthogonal projective measurement $\{\Pi_a\}$ on the apparatus, use the resulting state to extract work $\log(d_A)-S(\{p_a\})$ from $A$, then we use the measurement result to update the state of $B$. Finally, we use this state to extract $\log(d_B)-S(B|\{\Pi_a\})$ bits of work from $B$. The total amount of work extracted using this classical strategy is
\begin{gather}
W_{\rm goblin}=\log(d_{AB})-S(\{p_a\})-S(B|\{\Pi_a\}).
\end{gather}
Comparing with the optimal quantum strategy, we get the one-way work deficit which is the same as the thermal discord $D_{\rm demon}= \tilde \Delta^\to = \tilde D_{\thm}$. The thermal discord is also used by \cite{arXiv:quant-ph/0306023} to quantify the entropic cost associated with resetting synchronized clocks. We note that the assumption that the classical strategy involves orthogonal projective measurement can be generalized to POVMs, see Sec.~\ref{SEC_POVM_DEMONS}.

In another scenario, \cite{arXiv:1002.4913} describe {\it nonlocal} versus {\it local} strategies with one-way communication. Here the goblins $A$ and $B$ only have local information about their own states, but $B$'s knowledge of his state increases once he knows the outcome of $A$'s measurement. $A$'s best strategy is to get the maximum amount of work from her system by making a measurement in the eigenbasis of her local state. The work extracted is $\log(d_A)-S(\rho_A)$. After $B$ gets $A$'s measurement result, he extracts $\log(d_B)-S(B|\{\Pi^{\rm Eig}_a\})$ bits of work from his system. Comparing with the nonlocal demon, the resulting demon discord is the discord measured in the local eigenbasis of $A$ $D_{\rm demon}=D(B|\{\Pi^{\rm Eig}_a\})$, given in Sec.~\ref{SEC_MD_DISCORDS}.

The work extracted using a Szilard engine is strongly related to the work required to erase information. \cite{PhysRevA.72.032317} use a similar idea to define quantum, classical and total correlations through the amount of work required to delete each. The scheme for removing correlations is the application of a random local unitary from some given set. This does not change the correlations of the system until the relevant party ``forgets'' which unitary they applied. Forgetting is directly related to work through the erasure principle. The mutual information is the minimum amount of work required to remove all correlations using this process, giving it an operational interpretation. \cite{arXiv:1002.0314} examine the work due to correlations in closed system with energy conservation, which corresponds to only allowing global unitary operations that change the mutual information and not the eigenvalues of the density operator. \cite{arXiv:0902.0735} show that any multipartite state is unitarily connected to a classically-correlated state, but not necessarily to a product state. The implication is that in general the lowest amount of mutual information due to global unitary operations is not zero. \cite{arXiv:1110.2371} show that the states with maximum mutual information, for a fixed spectrum, are the generalized Bell-diagonal states (which have maximally-mixed reduced states). The states with minimum mutual information, again for a fixed spectrum, turn out to be classical states, however in a nontrival manner, see \cite{arXiv:1112.3372}.

\cite{arXiv:0803.4067} consider the role of correlations in a quantum photo-Carnot engine. They consider various scenarios where correlated two-level atoms are the quantum heat reservoir. They compare this efficiency to the efficiency where the atoms are not correlated. The correlated (quantum) reservoir is found to be more efficient, and they relate the improvement to the total correlations given by mutual information.

%**************************************************************
\subsection{Superselection}
%**************************************************************

In principle, it is possible to engineer in a laboratory an arbitrary quantum operation, and therefore a state modeled by an arbitrary vector in a Hilbert space. States of many natural systems, however, do not explore the whole Hilbert space. Superselection rules constrain physically-admissible states and operations in quantum theory. They arise from fundamental restrictions such as conservation laws or relativistic invariance \cite{PhysRev.88.101, PhysRevD.1.3267, JMathPhys.15.2198}, as well as for more pragmatic reasons such as the lack of suitable reference frames or detailed knowledge about underlying interactions \cite{PhysRev.155.1428, RevModPhys.79.555}.

In its traditional form, a superselection rule (SSR) is specified by a Hermitian operator, $N$, commuting with all observables of the theory, and the requirement that no observed states of the theory are nontrivial superpositions of the eigenstates of $N$ belonging to different eigenvalues. For example, a particle-number SSR forbids coherent superpositions of states with different number of particles, i.e. all states and operations have to commute with the particle-number operator. Accordingly, the most general density matrix under this SSR is block-diagonal in the basis of total particle-number states. For example, an entangled state $\ket{\Psi} = \frac{1}{\sqrt{2}} (\ket{12} + \ket{21})$ contains three particles in total and can be prepared via joint operations of $A$ and $B$, whereas state $\frac{1}{\sqrt{2}} (\ket{12} + \ket{34})$ is forbidden.

If $A$ and $B$ are restricted to local operations respecting SSR and classical communication (SSR-LOCC), all they can prepare are fully-classically-correlated states of Eq.~\eqref{FULLY_CLASSICAL_QUANTUM_STATES}, as now they have to commute with \emph{local} particle-number operators \cite{PhysRevLett.91.010404}. The converse statement does not hold, e.g., state $\ket{+} \otimes \ket{+}$ with $\ket{+} = \frac{1}{\sqrt{2}}(\ket{0} + \ket{1})$ is classical but it is not compliant with local SSR. Therefore, the set of states that can be prepared via SSR-LOCC is different from the set of separable states. This is quantified by so-called superselection-induced variance, see \cite{PhysRevA.70.042310,PhysRevLett.92.087904}. Finally, only states that cannot be prepared via SSR-LOCC are useful reference frames that allow violation of a Bell inequality \cite{NewJPhys.13.043027}.

The permissible types of correlations that $A$ and $B$ can generate by joint operations are also altered under SSR. We now show that under particle-number SSR, there are no classical-quantum states: we only have fully-classical or fully-quantum (with discord both sides) states. A general CQ state is of the form $\chi = \sum_a p_a \Pi_a \otimes \rho_{B|a}$. Consider this state rotated to the particle-number basis of $A$:
\begin{align}
\chi' \equiv & U_A \chi U_A^\dagger = \sum_n p_n \Pi_n \otimes \rho_{B|n} \\
 = & \sum_{n,m,n',m'} \chi'_{nm;n'm'} \ket{nm} \bra{n' m'},
\end{align}
where $U_A \ket{a} = \ket{n}$, due to the block-diagonal form we have $n=n'$, and $\ket{m}$ is the particle-number basis for $B$. Since total number of particles must be fixed, $n+m = n' + m'$, it follows that $m$ has to be equal to $m'$ and therefore $\chi'$ is fully classical. Any such state remains fully classical under local unitary operations and therefore $\chi$ is fully classical too. Furthermore, since the state of $B$ satisfies the SSR locally, it follows that the only way to write down a fully-classical state under the SSR is to write it using local particle-number bases, i.e., a set of $\{\Pi_a\}$ has to be equivalent to the set of $\{\Pi_n\}$.

We finish this subsection by noting that in studies of superselection it is important to keep fixed the number of modes accessible to $A$ and $B$. Namely, the SSR is effectively lifted if one allows adding locally new modes. For example, all quantum operations of a $d$-level system can be realized on two modes $\ket{k,d-k+1}$, with $k=1,\dots, d$ under particle-number SSR \cite{PhysRevLett.91.010404}.

%**************************************************************
\subsection{Nonlocality without entanglement}
%**************************************************************

\cite{arXiv:quant-ph/9804053} coin the term {\it quantum nonlocality without entanglement} with the following example: $A$ and $B$ share a system initially prepared in one of nine orthogonal states
\begin{align}\label{NLWE}
&\ket{0\pm1}\otimes\ket{2} && \ket{1\pm2}\otimes\ket{0}\nonumber \\
&\ket{2}\otimes\ket{1\pm2}&&\ket{0}\otimes\ket{0\pm1}\\
&\ket{1}\otimes\ket{1}.\nonumber
\end{align}
$A$ and $B$ need to discover which state they are given. However, as it turns out, it is impossible to deterministically distinguish these states using LOCC, even though the states are orthogonal product states in the $AB$ space.

The name {\it nonlocality without entanglement} suggests some relation with discord. \cite{arXiv:quant-ph/0203007} first study this in terms of quantum deficit. \cite{arXiv:1002.4913} explore this idea and find that the discord approach, and any other approach based on the density matrix, is not effective in identifying such nonlocality. For example, an equal mixture of these nine states gives a completely-mixed state, a classical state.

However, following \cite{QuantInfProc.9.711} if we assume that a third party $C$ is handing out these states to $A$ and $B$ with a record, then the total state is
\begin{gather}
\rho_{ABC}=\frac{1}{9}\sum_{c=1}^9 \ketbra{\alpha_c}{\alpha_c} \otimes \ketbra{\beta_c}{\beta_c} \otimes \ketbra{c}{c},
\end{gather}
where $\ket{\alpha_c\beta_c}$ is a state from Eq.~\eqref{NLWE} and $\{\ket{c}\}$ forms an orthonormal basis in $C$. The last state clearly has finite discord: $D(C|A)>0$, $D(C|B)>0$, but $D(C|AB)=0$ and $D(AB|C)=0$. On the other hand, consider the equal mixture of four Bell states, which also has a fully-mixed density operator and is not distinguishable by LOCC \cite{PhysRevLett.87.277902}. In this case, when we bring in $C$ as the classical flag, just like above, all quantum correlations are vanishing: $D(C|A)=0$, $D(C|B)=0$, $D(C|AB)=0$, and $D(AB|C)=0$. Other versions of this approach give similar results where either distinguishable entangled states give discord or nondistinguishable entangled states give no discord (depending on which bipartition is studied). These kinds of methods may however be valid if we restrict our discussion to sets of product states as in \cite{arXiv:1009.2571}.

Going back to $D(A|B)$, the other extreme is a discordant state mixing two orthogonal entangled states, say $\frac{1}{\sqrt{2}}(\ket{00}+\ket{11})$ and $\frac{1}{\sqrt{2}}(\ket{01}+\ket{10})$. These states could be locally distinguished but give a discordant (even entangled) density matrix.

Local distinguishability is not something that can be easily deduced from the density matrix. While there is a relationship between local distinguishability of an ensemble and quantumness \cite{QuantInfProc.9.711}, the density matrix of the ensemble does not capture this simply, especially in the case of orthogonal states.

%**************************************************************
%**************************************************************
\section{Dynamics of correlations}\label{dynamics}
%**************************************************************
%**************************************************************

In this section we explore a variety of studies of discord in different dynamical systems. In Sec.~\ref{decoherenceanddiscord}, we summarize various studies looking at various (generic) features of the evolution of discord for multipartite systems (typically two qubits) subject to different types of decoherence processes. These features include robustness to sudden death, sudden change behavior, dependence on the initial state, and freezing of correlations for finite periods. Comparative studies of discord and other correlations measures such as entanglement or quantum mutual information can reveal similar or highly-dissimilar behavior. One application is identifying whether discord can serve as a relevant indicator of quantum correlations, or the onset of particular dynamical changes, for example: classicalization of quantum walks \cite{PhysRevA.81.062123, PhysRevA.83.064302}, mutual synchronization of dissipative quantum harmonic oscillators \cite{arXiv:1105.4129}, and efficient energy transfer in the Fenna-Matthews-Olson protein photosynthetic complex \cite{arXiv:0912.5112}. Next we proceed to the general theory of open quantum systems. In Sec.~\ref{lazystates} we outline how the concept of lazy states leads to a simple test for the presence of nonclassical system-environment correlations in terms of the rate of change of the entropy for the system. Then in Sec.~\ref{NCPdyn}, we review the connections between the classes of initial system-environment correlations, and the possibility for describing the open-system dynamics using completely-positive maps. Finally, in Sec.~\ref{relativityanddiscord} we describe studies of the degradation of entanglement shared between an inertial party and a party undergoing constant acceleration, arising from the Unruh effect in the accelerated frame.

%**************************************************************
\subsection{Decoherence, dephasing, and dissipation}\label{decoherenceanddiscord}
%**************************************************************

A number of works investigate the dynamics of discord in open quantum systems by comparing the evolution of different types of correlations in specific models, typically two qubits coupled to two local baths or one common bath. Several factors can affect the evolution, namely, the initial state for the system and environment, the types of system-environment interaction, and the structure of the reservoir.

One important distinction is between dissipative and nondissipative decoherence. The former describes processes of spontaneous emission for which energy is lost from the system. In the latter case, the system-environment interaction is described by a quantum-nondemolition Hamiltonian which commutes with the system Hamiltonian, and decoherence (such as dephasing) occurs without transfer of energy.

Another distinction is the form of the spectral function which describes the coupling of the system to reservoir modes of different frequencies. In the Markovian (memoryless or white-noise) limit the spectrum is flat. However, when the spectral density changes significantly for frequencies close to the characteristic system frequency, the reservoir acquires finite temporal correlations and non-Markovian evolution results. Non-Markovianity typically leads to phenomena such as oscillations, revivals or sudden birth, as coherence lost to the environment returns to the system.

In all models considered, one should note that different measures of discord can sometimes record different behavior, as is the case for the examples given in \cite{arXiv:1004.5281, arXiv:1104.4043}. Also higher-dimensional models might exhibit very different features from those for pairs of qubits -- an area where few studies have been reported so far: particular exceptions include works which address the question of when local decoherence channels can increase as well as decrease quantum correlations, discussed in Sec.~\ref{LOGQC}.

%**************************************************************
\subsubsection{No death for discord}

A question about the robustness of discord is inspired by studies of entanglement sudden death (ESD) for two qubits having no direct interaction but subject to a process of spontaneous emission. ESD (see \cite{arXiv:0910.1396} for a brief review) is said to occur when the initial entanglement, as quantified by the concurrence or the entanglement of formation, falls to and remains at zero after a finite period of evolution for some choices of the initial state. Does discord present similar behavior? In the first study addressing this question, \cite{arXiv:0905.3376} compare the evolution of concurrence and discord for two qubits, each subject to independent Markovian decoherence (dephasing, depolarizing and amplitude damping). Looking at initial states such as Werner states and partially-entangled pure states, the authors find no sudden death of discord even when ESD does occur; the discord decays exponentially and vanishes asymptotically in all cases.

An intuitive explanation for this outcome is given by a simple example: mixing a state having finite discord with the identity, as corresponds to a simple process of depolarization, can never make the discord vanish other than when the identity itself is reached \cite{arXiv:0908.3157}. In contrast entangled states become separable when sufficiently mixed with the maximally-mixed states. More generally, sudden death of discord might not be expected to occur even in more complicated models on the account of the set of classical states having zero measure. (This is in contrast to the case of separable states which have finite volume in the full set of states, leading to common occurrence of ESD.) In fact, a large number of other studies looking at the dynamics of discord for specific models, discussed below, have also failed to find sudden death phenomena.

\cite{arXiv:0809.1746} examines the fundamental difference in the behavior of discord and entanglement from a different angle. ESD is shown to result from the manner in which the total system is partitioned: that is to say, multipartite entanglement between all interacting components does not vanish when the bipartite entanglement of the reduced state for the qubit subsystem becomes zero. Discord captures all classes of nonclassical correlations, not just nonseparability, and consequently does not exhibit sudden death.

\cite{arXiv:0908.3157} give a formal proof that interaction with any (local or common) Markovian bath can never lead to a sudden and permanent vanishing of discord (unless the infinite-time limit is reached). The proof uses the nullity condition Eq.~\eqref{eq:CommuteMarginalTest}, and argues that the occurrence of sudden death would imply an infinite set of linearly-independent equations which can never be satisfied. Note however that although sudden death is not possible in a strict sense, discord can be exponentially suppressed for finite periods, and it can vanish at discrete times and periodically \cite{arXiv:0911.1096}, e.g. in dissipative atom-cavity systems in the dispersive limit \cite{arXiv:1101.5429,arXiv:1101.4983}. Furthermore, discord need not decay to zero in the asymptotic limit \cite{arXiv:0912.1468}, since an environment sometimes preserves certain types of correlations. \cite{arXiv:1007.1492} give an example of this by considering two atoms coupled to a common-dissipative cavity mode: the evolution is strongly dependent on the initial correlations, and when the initial state includes a contribution from the subradient state the discord tends to a finite value. The choice of initial state can have other effects too: a model for which two atoms subject to independent and collective spontaneous emission, as well as the dipole-dipole interaction is explored in \cite{arXiv:1111.2646}, and it is shown that the speed of decay of several types of quantum correlations can be simultaneously and strongly enhanced by local unitary transformations of the initial state.

For completeness we point out that our main conclusion so far, that sudden death of discord cannot occur for systems subject to Markovian decoherence, also holds for models where the Markovianity assumption is relaxed or explicit non-Markovian assumptions are made. \cite{arXiv:0910.5711} take a model of two noninteracting qubits subject to independent noise channels, and consider the evolution and transfer of classical and quantum correlations across different partitions of the system (intra-system, system-environment, and environment-environment). Many quantitative differences are present for the case of dissipative (amplitude damping) versus nondissipative (phase damping, bit flip, bit-phase flip, and phase flip) decoherence. \cite{arXiv:0911.1845, arXiv:0911.1096} compare the evolution of entanglement and discord for two noninteracting qubits subject to dissipative decoherence induced by reservoirs having a Lorentzian spectral distribution. \cite{arXiv:0911.1845} assume independent reservoirs, while \cite{arXiv:0911.1096} consider the cases of independent reservoirs and a common reservoir. Again discord exhibits a combination of asymptotic decay and discrete points of disappearance and revival, while ESD is sometimes observed across the same parameter range. A common reservoir leads to especially complex behavior with many sudden changes in the discord evolution \cite{arXiv:0911.1096}. \cite{arXiv:0912.1468} look at a pair of double quantum dots having two excess electrons which interact via tunneling and the Coulomb interaction. They use two independent or a common reservoir(s) with an Ohmic spectral density to incorporate phonon-induced decoherence and investigate the effects of changing temperature. \cite{arXiv:1009.5710} show that discord and entanglement have revivals even when the environment is classical.

The lack of discord sudden death raises the question of whether it is possible to establish a hierarchy of correlations according to robustness under decoherence (suitably interpreted as direct quantitative comparisons of different measures is not meaningful in general). Studies along these lines reach mixed conclusions. \cite{arXiv:1005.1043} examine a model involving two-mode Gaussian states which are coupled to a common or two independent Ohmic bath(s). They compare the persistence of intensity correlations below the shot-noise limit, entanglement, and discord for initial squeezed-thermal states. \cite{arXiv:1006.3943} examine a system of three uncoupled qubits, for which the qubit level spacings are subject to stochastic time-dependent perturbations (Ornstein--Uhlenbeck noise), and an initial state which is chosen to be a mixture of the identity and the W state. Again discord is immune to sudden death, with Bell nonlocality, bipartite and tripartite entanglements all demonstrating sudden death for some choices of parameters. Non-Markovianity delays occurrences of sudden death, or slows the decay of correlations. Discord, entanglement and Bell-nonlocality (quantified in terms of violation of the Clauser-Horne-Shimony-Holt (CHSH) inequality) are also compared in \cite{JPhysB.44.125501} for an atom-cavity system, for which two two-level atoms couple to a single cavity mode, subject to dissipation. The cases of identical and nonidentical detunings for each atom are compared. It is found that, depending on the choice of initial state and detuning, discord is readily induced, whereas entanglement and nonlocality are often not. Bell nonlocality is certainly the most fragile of the three types of quantum correlations across all cases considered. \cite{arXiv:1006.1856} compare concurrence, Bell nonlocality, teleportation fidelity and discord for two qubits subject to independent or collective decoherence in both dissipative and nondissipative models. \cite{arXiv:1106.0289} look at ESD using the Koashi-Winter formula in Eq.~\eqref{koashiwinter}, which has discord on one side and entanglement on the other side. To use the Koashi-Winter formula they must restrict their study to the unitary dynamics of a tripartite pure state and study the ESD of a bipartition. They find that if $E_F(A:B)$ goes to zero then $D(A|C)=S(A|B)$ and $D(B|C)=S(B|A)$, where $C$ is the purification of $AB$ (at all times).

ESD is an artefact of the rise in entropy of the system and the finite volume occupied by separable states. In some respects it represents a weak point in using entanglement measures as a sign of quantum correlations. The lack of this somewhat-artificial phenomenon makes discord a better indicator of quantum correlations than entanglement in many situations.

%**************************************************************
\subsubsection{Frozen discord}

Studies of nondissipative decoherence for two qubits in Bell-diagonal states yield important insights into the persistence of nonclassical correlations for this family of states. These states are parametrized by three parameters in the Bloch representation: $\frac{1}{4}(\openone \otimes \openone + \sum_i c_i \sigma_i \otimes \sigma_i)$ (discussed previously in Sec.~\ref{sec:evalD42qubits} and Sec.~\ref{sec:GDanalytic}). In \cite{arXiv:0905.3396}, the authors look at a model for which each qubit is subject to independent phase damping/phase flip, bit-flip, and bit-phase flip channels. The different channels yield equivalent behavior, and the three types of evolution are determined by the initial state: (a) the amount of classical correlations remains constant while discord decays monotonically; (b) the amount of classical correlations decays monotonically until it freezes at a transition point, discord decays monotonically with an increase in the decay rate at the same transition point; (c) both classical and quantum correlations decay monotonically throughout. \cite{arXiv:1001.5441} present a very similar freezing phenomenon to (b), also under the nondissipative-independent-Markovian reservoirs assumption. In this case, for special choices of the initial state, a transition is observed from a period of classical decoherence (for which the discord remains constant while the amount of classical correlations decays), to a period of quantum decoherence (for which the discord decays and the amount of classical correlations remains constant). (ESD occurs at a different point in the evolution.) Before the transition the distance, (as captured by $D_R$), between the system state and the closest classical state remains constant, while the amount of correlations of that classical state decays. At the transition point, the closest classical state becomes constant, and the system state approaches it asymptotically.

\begin{figure}[t]
\resizebox{8 cm}{8.22 cm}{\includegraphics{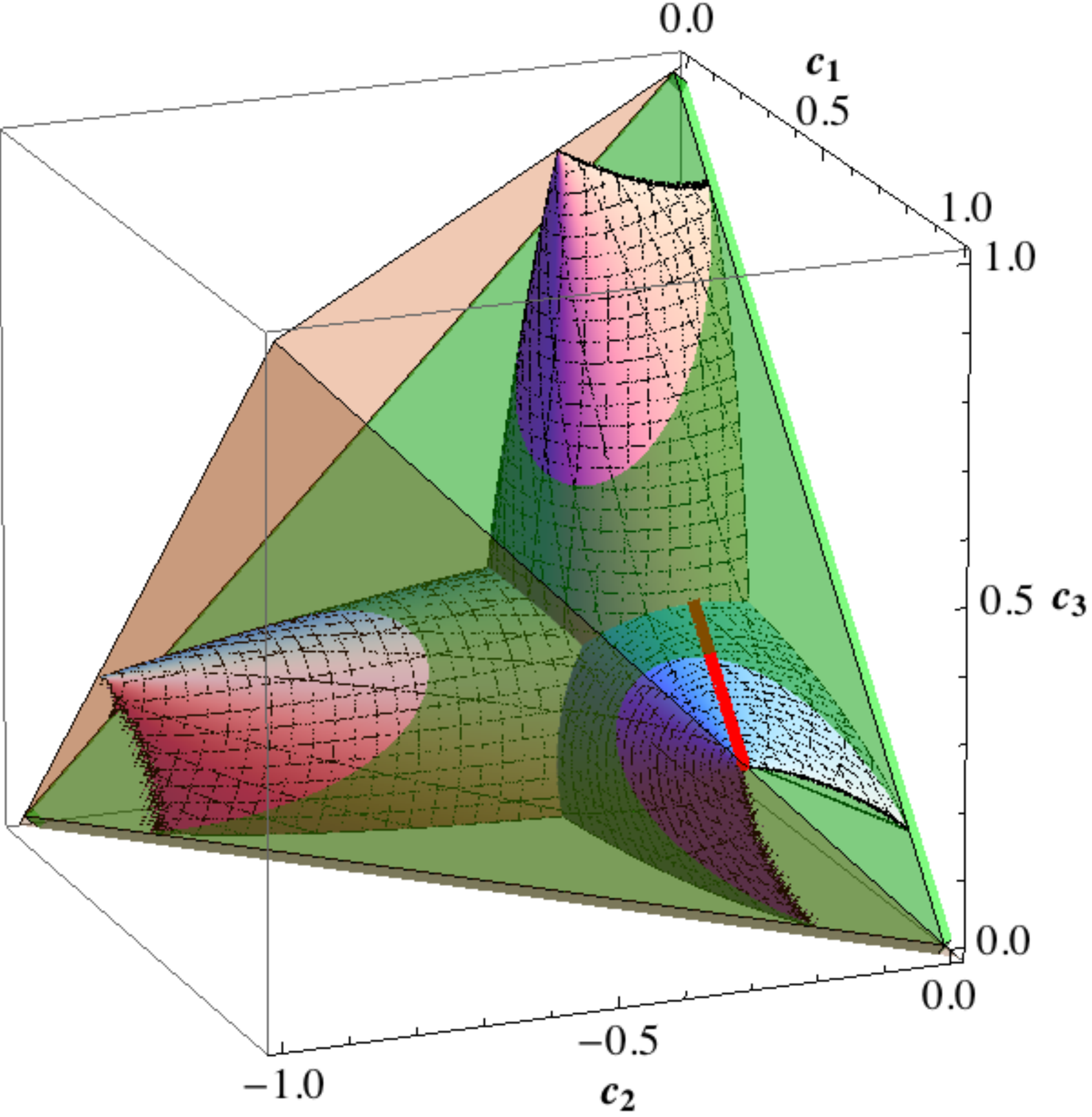}}
\caption{\label{fig:frozen}(Color online.) \emph{Freezing of correlations.} A geometric view of the phenomenon of frozen discord, for two qubits evolving within the space of Bell-diagonal states due to nondissipative decoherence --- typified by (Markovian) phase-flip channels acting independently on the qubits. The representation of the space of Bell-diagonal states is as explained earlier in Fig.~\ref{FIG_DG}, with one octant shown. Here, the Cartesian axes are parametrized by the entries $c_i$ of the (diagonalized) correlation tensor for the Bell-diagonal states in the Bloch representation. Entanglement of formation, quantum mutual information and classical correlations increase with greater distance from the origin. In contrast, discord grows with distance from each axis, and the level surfaces of discord define intersecting tubes centered around the axes. The straight-line trajectory shows a typical evolution under decoherence, for which $c_3$ is constant and $c_1,c_2$ decay exponentially while maintaining a constant ratio. The trajectory lies initially in a surface of constant discord, crosses the intersection of the tubes in correspondence to the transition from classical to quantum decoherence, and terminates at the $c_3$ axis. This figure is reproduced from \cite{arXiv:1006.2775}.}
\end{figure}

\cite{arXiv:1006.2775} provide a complete picture for the frozen-discord phenomenon. They explore the geometry of the Bell-diagonal states in terms of level surfaces for discord as illustrated in Fig.~\ref{fig:frozen}. In this geometric picture, the diagonal entries of the correlation tensor $c_1,\; c_2, \; {\rm and} \; c_3$ define a three-dimensional Cartesian coordinate system. The (physical) Bell-diagonal states define a tetrahedron in this space, with the separable subset defining an octahedron within, centered on the origin (illustrated in Fig.~\ref{FIG_DG} of Sec.~\ref{sec:GDanalytic}). Entanglement of formation, quantum mutual information and classical correlations increase with distance away from the origin; however, the surfaces of constant discord define intersecting tubes centered on the Cartesian axes. The effect of the phase-flip channel in the model of \cite{arXiv:1001.5441} is to define straight-line trajectories through the Bell-diagonal tetrahedron, starting towards the faces and terminating at a Cartesian axis. Each trajectory follows a discord level surface until the point at which the tubes along different axes intersect; this intersection corresponds to the change from classical to quantum decoherence.

%**************************************************************

The phenomenon of frozen correlations appears in a number of other models which also assume nondissipative decoherence and initial Bell-diagonal states. The model of \cite{arXiv:1001.5441} has been extended to local non-Markovian dephasing noise by \cite{arXiv:1006.1805}. For the non-Markovian case, the study finds similar occurrences of frozen discord to the Markovian case, as well as transitions between classical and quantum decoherence. However, for the non-Markovian model there are typically multiple transitions and complicated (damped) oscillatory behavior due to memory effects. Similar phenomena have been seen in the model of \cite{arXiv:1005.4204}, which assumes a nondissipative coupling to a common-Ohmic environment; more specifically, certain choices of initial-state and reservoir parameters lead to a critical time at which there is a sudden change in the evolution of the classical and nonclassical correlations. Discord is amplified or preserved up to the critical time. \cite{arXiv:1010.3789} look at a quantum chaos model with two qubits, with one qubit coupled to a quantum-kicked top which induces strong dephasing effects. In a chaotic regime and large-spin limit of the quantum kicked top, there are similar dynamical features as are found for the Markovian models discussed above, i.e. a period of classical decoherence followed by a period of quantum decoherence. The regular regime for the quantum-kicked top has some similarities to the non-Markovian models previously mentioned, although the non-Markovian effects have a different origin. Lastly, \cite{arXiv:1107.3939, arXiv:1112.2050} investigate the connections between the discontinuous behavior of discord in dynamics to similar behavior in quantum phase transitions, see Sec.~\ref{mbp}. They find that this common behavior in both types of study comes from the same mathematics. It would be interesting to give physical arguments that unify the two.

In conclusion, freezing phenomena have been found to be a robust feature of a family of two-qubit models subject to nondissipative decoherence. Different measures of discord record only minor differences \cite{arXiv:1010.3521, PhysRevA.82.052122} (see \cite{arXiv:1104.4043} for a different conclusion). These dynamical features seem to be specific to the nondissipative case, and are not found for the model of dissipative decoherence discussed in \cite{arXiv:1006.1856} for example. An experimental demonstration of the dynamics under phase-damping is reported by \cite{arXiv:0911.2848}, who used a parametric down-conversion source and simulated a noise channel on one of the qubits using birefringent quartz plates of variable length. \cite{arXiv:1104.3885} demonstrate sudden-change behavior using NMR, and it is found to be robust in the presence of an additional energy relaxation process. \cite{arXiv:1107.3428} recently reviewed these and many other aspects of quantum discord. Finally we note some theoretical extensions: A study of a qubit-qutrit system subject to independent and common Markovian dephasing noise is presented in \cite{ PhysLettA.375.4166}. Negativity, classical correlations, discord and geometric discord are compared for two families of initial states and a range of sudden transition, freezing and amplification behavior is observed. \cite{PhysRevA.83.022321, arXiv:1112.4318} explore extensions to a larger family of initial two-qubit X-states in Eq.~\eqref{xstate}, for which, in the Bloch representation, the Bloch vectors are aligned in the $z$ direction and the correlation tensor is diagonal.

%**************************************************************
\subsection{Local operations generating quantum correlations}\label{LOGQC}
%**************************************************************

Recently some authors have pointed out that (non-unitary) local operations can generate quantum correlations. \cite{arXiv:1105.5548, arXiv:1105.5551} show that the action of certain local channels can enhance quantum correlations, including in continuous-variable systems \cite{arXiv:1108.2143}. An intuitive example of such a process is seen by taking a CC state $\chi_{ab} = p_{00} \ket{00}\bra{00} + p_{11} \ket{11}\bra{11}$ and acting on the side of $A$ with the channel that takes $\ket{0} \mapsto \ket{0}$ and $\ket{1} \mapsto \ket{+}$. This yields the QC state $\chi_{Ab} = p_{00} \ket{00}\bra{00} + p_{11} \ket{+1}\bra{+1}$.

\cite{arXiv:1106.2028} prove that a classical state of two qubits is preserved if and only if the local channel acting is either unital, that is to say it maps the maximally-mixed state to itself: $\mathcal{B}_{\rm unital}(\openone)=\openone$, or a semi-classical channel: $\mathcal{B}_{\rm sc}(\rho) = \sum_a p_a \Pi_a \rho \; \Pi_a$. They go on to prove that any distance-based measure of quantum correlations for two qubits is decreasing under the action of unital and semi-classical channels. \cite{arXiv:1110.5109} give a similar result in terms of mixing channels, a channel that increases the entropy for all inputs: $S[\mathcal{B}_{\rm mixing}(\rho)] \ge S(\rho) \quad \forall \; \rho.$ They show that for qubits unital channels are mixing channels. However, by an explicit construction, they show that a mixing channel on higher-dimensional systems can in fact create quantum correlations.

\cite{arXiv:1112.3141, arXiv:1112.5700} give the necessary and sufficient condition for preserving a classical state. They show that the classicality of a state is preserved if and only if a channel preserves vanishing commutators: i.e. $[\mathcal{B}(\rho),\mathcal{B}(\sigma)]=0$ for all $[\rho,\sigma]=0$.

Lastly, we should remark on the interpretation of the original quantum discord due to \cite{arXiv:1012.4903} given in Eq.~\eqref{Dis_PE}, that is discord as measured by $B$ is equal to the minimal partial entanglement, ${P_E(\tilde \rho_{A|B:E})} = E_D(\tilde \rho_{AB:E}) - E_D(\tilde \rho_{B:E})$. This leads to the conclusion that local operations made by $A$ cannot increase the discord as measured by $B$. This is due to the fact that the local operation can only lower the positive quantity in $P_E(\tilde \rho_{A|B:E})$. A local operation of $A$ can increase $D(B|A)$ but not $D(A|B)$.

%**************************************************************
\subsection{Lazy states and decoherence}\label{lazystates}
%**************************************************************

\cite{arXiv:1004.5405} propose a simple test to detect the presence of nonclassical system-environment correlations: If the time derivative of the entropy of the reduced state of the system is nonzero at $t=\tau$, then the system-environment state is quantum correlated at that time. More precisely they show that the time derivative at $t=\tau$ is vanishing if and only if $[\rho_\Sy \otimes \openone_\E, \rho_{\SE}]=0$. This is the same condition as the one given in Eq.~\eqref{eq:CommuteMarginalTest} for nullity of discord. States satisfying the condition are called lazy states. The final result of the paper is thus rather interesting:
\begin{gather}
\left| \frac{d}{dt} S(\rho_\Sy) \right| \leq \|H_{\rm int}\|
\left\| \left[ (\log(\rho_\Sy) \otimes \openone_\E), \rho_{\SE} \right] \right\|_1,
\end{gather}
where $\| x_i \|_1 = \sum_i |x_i|$ and both sides are evaluated at $t=\tau$. The time derivative of the entropy of the system is bounded by the interaction Hamiltonian, and a function of the state of the system, as well as the state of the system plus the environment. This shows how quantum correlations between the system and the environment provide bounds on the entropy rate of the system.

We know that the set of lazy states is of zero measure, and for Markovian dynamics the system never becomes lazy \cite{arXiv:0908.3157}. Therefore the rate of entropy change never vanishes. On the other hand, recently \cite{arXiv:1109.0602} report that ``almost all states are pretty lazy" for sufficiently large environments, i.e. the entropy rate is rather low. All of these results suggest that quantum correlations play a nontrivial role for real open systems, as the lack of laziness is a necessary feature of decoherence.

%**************************************************************
\subsubsection{Trace distance quantum-correlations witness}

More recently, a measurement-based witness is put forth in \cite{arXiv:1108.2699}. The initial state of system is subjected to a measurement in its own basis: $\rho_\Sy = \sum_s \Pi_s \rho_\Sy \Pi_s$. However, the total system-environment state under this measurement is invariant if and only if discord is zero, i.e. $D(\E|\Sy)=0 \Leftrightarrow \rho_{\SE}=\sum_s \Pi_s \otimes \openone_\E \rho_{\SE} \Pi_s \otimes \openone_\E$. Next, the measured system-environment state is allowed to evolve and the final state of system is determined. In a different experiment the system is not measured in its basis, but rather simply allowed to evolve. If the final states of the system are different for the two trials then one can conclude that the initial system-environment correlations are quantum.

%**************************************************************
\subsection{Complete positivity}\label{NCPdyn}
%**************************************************************

The dynamics of an open quantum system can be thought of as a contraction of the unitary dynamics of the system with its environment: $\mathcal{B}(\rho_\Sy) \equiv \tr_\E(U \rho_\Sy \otimes \rho_\E U^\dag)$, where $\mathcal{B}(\cdot)=\sum_{ee'} \braket{e| U\sqrt{\rho_\E} |e'} (\cdot) \braket{e' |\sqrt{\rho_\E} U^\dag |e}$ is a completely-positive dynamical map. The usual assumption that the dynamics of the system is described by a completely-positive map requires that the system and the environment are initially uncorrelated. When this assumption is relaxed the situation is much more complicated. There is a rich history of investigations into the initial correlations between a system and its environment and the positivity of dynamics of the system \cite{arXiv:0910.5568}.

\cite{arXiv:quant-ph/0703022} show that when the system is classically correlated to its environment $\rho_{\SE}=\sum_s p_s \Pi_s \otimes \rho_{\E|s}$, for any combined unitary evolution of that state, the open dynamics of the system can be described by a completely-positive dynamical map. The proof goes as the following. Using the linearity of the dynamics we have
\begin{align}
\mathcal{B}\left(\rho_\Sy\right) =& \tr_\E \left[ U \left(\sum_s p_s \Pi_s \otimes \rho_{\E|s} \right) U^\dag \right] \nonumber \\ =& \sum_s p_s \tr_\E \left( U \Pi_s \otimes \rho_{\E|s} U^\dag \right).
\end{align}
We may consider the action of completely-positive maps
\begin{align}
\mathcal{B}_s (\cdot) \equiv& \tr_\E \left( U \Pi_s \otimes \rho_{\E|s} U^\dag \right) \nonumber\\
\equiv& \sum_{ee'} \braket{e| U\sqrt{\rho_{\E|s}} |e'} (\cdot) \braket{e' |\sqrt{\rho_{\E|s}} U^\dag |e}
\end{align}
acting on pure states $\Pi_s$ of the system: $\mathcal{B} \left( \rho_\Sy \right) =\sum_s p_s \mathcal{B}_s \Pi_s$. Next we use the idempotent and orthonormality of projections $\Pi_s$ to rewrite the action of the map as $\mathcal{B}(\rho_\Sy) =\left( \sum_s \mathcal{B}_s \Pi_s \right) \sum_{s'} p_{s'} \Pi_{s'}$, with $\rho_\Sy = \sum_{s'} p_{s'} \Pi_{s'}$. Finally we have $\mathcal{B}=\sum_s \mathcal{B}_s \Pi_s$.

The intuition here is that the dynamical map acts on the probability vector $p_s$, which defines the full state of the system. Therefore the dynamics looks very much like a classical stochastic map\footnote{However, this dynamical map can change the basis of the probability vector, differentiating it from a classical stochastic map.}. \cite{arXiv:0808.0175} present a method for describing the dynamics (for a class of states) for which nullity of discord is necessary and sufficient for completely-positivity. However, dynamical maps for initially correlated system-environment states are not the same as maps derived from quantum process tomography \cite{PhysRevA.76.042113}. When dealing with quantum process tomography one needs deal with any initial correlations carefully, including classical \cite{arXiv:1012.1402}.

Mathematically the connection between different classes of initial system-environment correlations and the properties of the resulting dynamical maps can be spelled out by breaking up the dynamical map as $\mathcal{B}=\mathcal{T}_{\E}\circ \mathcal{U}\circ \mathcal{A}$, where $\mathcal{T}_{\E}$ is the trace with respect to the environment and $\mathcal{A}$ is termed an assignment map. The assignment map \cite{PhysRevLett.73.1060, arXiv:0910.5568} takes the system state to a correlated state of the system-environment. Since $\mathcal{T}_\E$ and $\mathcal{U}$ are completely positive, the positivity of the dynamical map depends on the positivity of the assignment map. Pechukas proves that an assignment is linear, positive, and consistent ($\mathcal{T}_\E \circ \mathcal{A} = \mathcal{I}$) if and only if there are no initial correlations. If we give up the consistency requirement however, we can define the following assignment, $\mathcal{A}:\rho_\Sy \mapsto \sum_{s}{\tr}(\rho_\Sy \Pi_{s}) \Pi_{s} \otimes \rho_{\E|s}$ where $\rho_\Sy$ is an arbitrary state for the system, $\left\{ \Pi_{s} \right\}$ is a (fixed) set of orthonormal projectors for the system, and the $\rho_{\E|s}$ are operators for the environment. Comparing with Eq.~\eqref{eq:BlockDiagDecomp}, we see that this assignment leads to classical states with respect to the system. $\mathcal{A}$ here is linear and positive by construction. Combining it with $\mathcal{T}_\E \circ \mathcal{U}$, the dynamical map $\mathcal{B}$ is linear and completely positive too \cite{arXiv:quant-ph/0703022}. However, a mathematical property which fails for this choice of assignment map is consistency, namely $\tr_{\E}\left[ \mathcal{A}\left( \rho_\Sy \right) \right] \neq \rho_{\Sy}$ in general. It is interesting to note that the positivity of a general assignment map is related to the no-broadcasting theorem \cite{arXiv:0910.5568} discussed in Sec.~\ref{NLBC}.

%**************************************************************
\subsection{Relativity and cosmology}\label{relativityanddiscord}
%**************************************************************

The effects of thermal noise on correlations are extended to relativistic systems via the Unruh effect (see \cite{arXiv:quant-ph/0212023, ProgTheoPhysSupp.88.1} and references therein). This effect is due to the horizon experienced by an accelerating observer, as he does not have access to any information from beyond the horizon. In quantum information language, any degree of freedom beyond this horizon is correspondingly traced out. This leads to a thermal effect even in the vacuum state associated with an inertial observer (the Minkowski vacuum). The simplest models only discuss free fields and neglect the properties of the detector.

In this type of model, (see for example \cite{PhysRevA.82.042332}), an inertial observer $A$ shares an initially entangled qubit with an accelerating $R$ (for Rob, relativistic Bob) following a world line
\begin{align}
t_R(\tau)=&a^{-1}\sinh (a\tau) \quad {\rm and} \quad z_R(\tau)=a^{-1}\cosh (a\tau)
\end{align}
with acceleration $a$ and proper time $\tau$. $R$ now experiences a different space-time than $A$ known as Rindler space-time, which can be related to the Minkowski space-time by an appropriate transformation. However this transformation produces two causally-disconnected regions known as the left and right Rindler wedges. $R$'s trajectory is confined to the right wedge and he physically experiences an event horizon.

The initial entangled state prepared by $A$ (in Minkowski space) is a two-mode state $\ket{\psi}_M = \frac{1}{\sqrt{2}} \left(\ket{00} + \ket{11}\right)$. However, $A$'s (Minkowski) Hilbert space is not the one $R$ experiences. In Rindler space-time the Fock states $\ket{0}$ and $\ket{1}$ are expanded as an entangled state with modes in both Rindler wedges (only the relevant modes are considered). $R$ is limited to his wedge and thus experiences a mixed state. The technical details depend on the type of fields associated with the states. We look at two distinct cases: Klein-Gordon and Dirac fields. All these results can also be applied to black-hole horizons.

%**************************************************************
\subsubsection{Klein-Gordon fields}

For Klein-Gordon fields the effective (mode-modulated) Minkowski vacuum is
\begin{gather}
\ket{0}=\frac{1}{\cosh r}\sum_{n=0}^\infty(\tanh r)^n\ket{n}_R\ket{n}_L,
\end{gather}
and the first occupied state is
\begin{gather}
\ket{1}=\frac{1}{\cosh r}\sum_{n=0}^\infty(\tanh r)^n\sqrt{n+1}\ket{n+1}_R\ket{n}_L,
\end{gather}
where the subscripts $R$ and $L$ refer to the right and left wedges respectively and $\tanh r=\exp(-\pi|k|c/a)$, where $k$ is the wavenumber for the mode.

Now $A$ has access to the same initial qubit state while $R$ only has access to his Rindler-wedge infinite-dimensional state. It is easy to see that the initial pure state $\ket{\psi}_M$ is now mixed and entanglement is somewhat degraded.

\cite{arXiv:0905.3301} compares the degradation of entanglement to that of discord and MID. In a similar fashion to the dynamical cases above, discord (as measured by $A$) does not vanish even at the infinite-acceleration limit where entanglement vanishes. The same is true for MID which stays very close to the maximal value of $M=1$.

\cite{arXiv:1003.4477} consider the case of entangled detectors coupled to a scalar field which acts as a thermal bath. To achieve a more realistic scenario, the detector is switched on for a finite time. They compare discord, symmetric discord, classical correlations, and total correlations. The results show that the decay of quantum correlations is not described by a smooth function. It is worth mentioning that in this protocol some correlations are lost even in the absence of acceleration due to dark counts.

%**************************************************************
\subsubsection{Dirac fields}

Dirac particles are limited in the occupation number for each mode, and furthermore they are restricted by superselection rules \cite{Coll.Streater}. However antiparticles give an added dimension and the Minkowski vacuum is an entangled state of positive-frequency modes in the right wedge and negative-frequency modes in the left wedge. These differences from the Klein-Gordon case give qualitatively different results, one of which is the degradation of entanglement to a constant nonvanishing value in the infinite acceleration case for Dirac particles.

\cite{arXiv:0912.4129} study the effect of acceleration on quantum correlations in Dirac fields. Classical and quantum correlations all decay to some constant positive value at the limit of infinite acceleration. Looking at correlations between other partitions ($A$ and the left wedge, or the left and right wedge) the correlations increase as a function of acceleration. See \cite{arXiv:1108.5553} for further discussions."

The parity superselection rule also gives rise to different methods for classifying separability. \cite{arXiv:0705.1103} define separability criteria based on physically meaningful methods for representing the density matrix as a sum of states following the superselection rule, see also \cite{PhysRevLett.92.087904,PhysRevA.70.042310}. \cite{PC.WessmanTerno} propose an extension of this work to relativistic fermions and measures of discord, giving more meaningful measures of entanglement and discord degradation due to the Unruh effect.

%**************************************************************
\subsubsection{Cosmology}

Vacuum entanglement is another field studied in the context of relativistic quantum field theory. \cite{arXiv:quant-ph/0008006, arXiv:quant-ph/0310058} show that a pair of initially-correlated two-level systems may become entangled after an appropriate interaction with the vacuum state. Since the vacuum entanglement decays rapidly with the distance the amount of entanglement generated using this process decays very quickly as a function of the distance between the two-level systems.

\cite{PhysRevD.79.044027} use vacuum entanglement to study the entanglement of an exponentially-expanding (de Sitter) vacuum. Using the same method \cite{arXiv:1105.5212} study the classical and quantum correlations (defined through symmetric discord and orthogonal projectors) between two-level systems interacting with the vacuum of scalar fields. They calculate the correlations in both the Minkowski and de Sitter space-time and show that they decay less rapidly than entanglement. While entanglement and the capacity for violating CHSH inequalities vanish past the Hubble horizon, other correlations remain positive indefinitely. The presence or lack of quantum correlations in various types of fields beyond the Hubble horizon gives an insight into the nature of the quantum fluctuations in the early universe.

%**************************************************************
%**************************************************************
\section{Many-body physics}\label{mbp}
%**************************************************************
%**************************************************************

One of the major applications of quantum discord has been in the field of many-body physics. Many-body physics attempts to understand the physics of a large number of quantum particles interacting with each other. Correlations in such systems play an important role because the macroscopic physics does not simply come from microscopic degrees of freedom \cite{Science.177.4047}. Quantitative treatments of entanglement in such systems are fruitful \cite{RevModPhys.80.517, RevModPhys.82.277}. We give a brief review of several many-body systems and the studies of quantum discord within them. A great deal of the work on discord and many-body physics is on quantum phase transitions (QPTs), as discord identifies the critical points. Next we look at the dynamics of discord in spin-chains followed by the effects on discord of system temperature.

%**************************************************************
\subsection{Quantum phase transition}
%**************************************************************

The ground state of a quantum system can be in different quantum phases at zero temperature. When an external parameter of the Hamiltonian, called the \emph{control parameter}, is varied the phase of the system can change. This phenomenon is known as \emph{quantum phase transition} (QPT). The quantum phases are different from classical phases because they are strictly determined by the properties of the ground state. The transition from one quantum phase to another occurs at a \emph{critical point} characterized by certain nonanalytic behavior in the ground-state energy \cite{Book.Sachev, PhysRevA.74.052335}.

In statistical mechanics QPTs are well studied \cite{Book.Sachev}, and recent studies of QPTs utilize entanglement \cite{RevModPhys.80.517} and area laws \cite{RevModPhys.82.277}. Correlations play an important role in critical systems, for instance entanglement generally follows an area law in noncritical systems. That is to say, in a spin system the entanglement between two spins is inversely proportional to the distance between the two spins. In critical systems the area law is broken, which means that the entanglement is genuinely multipartite and spread across the whole system. It is known that the nonanalyticity of energy is manifested in terms of entanglement between the nearest and the next-nearest neighbors at a critical point \cite{PhysRevA.66.032110, Nature.416.608}.

QPTs separate different phases of matter, which are governed by the external control parameter \cite{Book.Wen}. Over the last 20 years, it has been realized that not all phases of matter are accounted for by the symmetry of the ground state. Different patterns in correlations describe different phases, e.g. high-temperature superconductors and fractional quantum Hall liquids.

Several authors study quantum discord in critical systems in the thermodynamic limit $N \to \infty$. The most-studied systems are 1-d systems, and there discord can indicate the critical points of QPTs. The nonanalyticity of the energy at the critical point can manifest itself in discontinuous behavior of the derivatives of the energy. If the $n$th derivative shows nonanalytic behavior then it is an $n$th order QPT. This has led researchers to examine the behavior of different correlations near the critical point, especially their analyticity properties as revealed by differentiation. The evidence that quantum discord plays an important role in QPTs comes from showing that it leads to a scaling law, which is not the case for entanglement.

%**************************************************************
\subsubsection{1--d lattice}

A 1-d spin chain has a general Hamiltonian of the form
\begin{gather}\label{ChainHam}
H = \sum_i \left( J_x \sigma_x^i \sigma_x^{i+1} + J_y \sigma_y^i \sigma_y^{i+1} + J_z \sigma_z^i \sigma_z^{i+1} + h \sigma_z^i \right),
\end{gather}
where the last term is the external magnetic field. This Hamiltonian has parity and translational symmetry, which is also enjoyed by its ground state. In general, this Hamiltonian is not exactly diagonalizable. However, there are well-known special cases. The reduced states of two qubits coming from such a chain typically have the form of an X-state, given in Eq.~\eqref{xstate}, for which analytic expressions for quantum discord and classical correlations are known. We begin by discussing the role of correlations in some of the special cases.

One of the first studies on quantum discord in a spin chain is given by \cite{arXiv:0809.1723} who analyzed QPT in a transverse Ising chain. The Ising case is obtained from Eq.~\eqref{ChainHam} by setting $J_x <0$, $J_y = J_z = 0$, and $h=-1$. At $J_x=0$ all spins are in $z$-eigenstates and as $J_x \to -\infty$ all spins are $x$-eigenstates, with the critical point at $J_x = -1$. The author finds that at the critical point entanglement between nearest neighbors is not maximal, but entanglement between next-nearest neighbors is maximal. Quantum discord is not maximal in both of these cases. The author also considers a 1-d antiferromagnetic $XXZ$ spin model by setting the parameters in Eq.~\eqref{ChainHam} to $J_x = J_y = 1$, $h=0$, and letting $J_z$ vary. The critical point in this model is at $J_z = 1$, at which the concurrence is maximal, classical correlations are minimal with a discontinuity, and discord is maximal (with a discontinuity) for nearest neighbors. In a different study, \cite{arXiv:0905.1347} considers an Ising model and shows that the derivatives of discord display the characteristic (logarithmic) divergence of the critical Ising model. He also considers a symmetric $XXZ$ model, and Lipkin-Meshkov-Glick (LMG) model, $J_x=J_y=-J/2$, $J_z=-\Delta J/2$, and $h=0$. Quantum and classical correlations of the nearest neighbors are shown to be discontinuous at $\Delta = \pm 1$. For the LMG model, QPT occurs at $\Delta=1$. Quantum and classical correlations are equal and finite for $h<1$ and vanish for $h>1$. The derivative is discontinuous at $\Delta=1$. A special case of the LMG model analyzed with mutual information is presented in \cite{arXiv:1111.5225} (also see the references therein for analyses with other correlations measures). \cite{arXiv:1002.3906} study the XXZ model, the LMG model, and the anisotropic $XY$ spin chain with $J_x=-J(1+\gamma)$, $J_y=-J(1-\gamma)$, $J_z=0$ and $h=-1$. They show that quantum discord is present between neighbors farther than the next-nearest, while entanglement may be absent for these neighbors.

\cite{arXiv:1006.3332, arXiv:1104.1105} study the behavior of various macroscopic quantities and entanglement in the $XXZ$ model at the critical point as temperature goes above absolute zero. They show that none of these quantities can pick out the critical points, while quantum discord is shown to be discontinuous at the critical point at finite temperature. These first studies set a benchmark for quantum correlations in many-body physics, and specifically QPTs. Other similar studies are molecular magnets described by a symmetric spin trimer and a tetramer in \cite{arXiv:1012.0650}, a 1-d lattice with Dzyaloshinskii-Moriya interaction \cite{arXiv:1012.2788}, long-range correlations in \cite{arXiv:1012.5926, arXiv:1110.6381}, and finite-temperature QPTs with three-spin interactions \cite{PhysRevA.83.052323}.

%**************************************************************
\subsubsection{Global discord}

Based on the diagrammatic approach of \cite{arXiv:1104.1520} described in Sec.~\ref{SEC_DIAGRAM_UNIFICATION}, \cite{arXiv:1105.2548} derive a multipartite version of quantum discord, called global discord. This is followed up by construction of a multipartite quantum-correlations witness by \cite{arXiv:1108.4001}. Both tools are put to use in studying the Ashkin-Teller model \cite{arXiv:1105.2548, arXiv:1108.4001}. The multipartite discord picks up the QPT points as before. \cite{arXiv:1201.3252} study global discord, along with other measures of quantum correlations, in the Ising model. They find that MID is not a good indicator for the critical point of the QPT, while global discord scales linearly with the number of qubits.

%**************************************************************
\subsubsection{Factorization}

Usually the ground state of the Hamiltonian is entangled, however for specific values of $h$ the ground state is completely factorized \cite{PhysicaA.112.235}. This is the point where all correlations vanish, quantum and classical. Finite discord then indicates the departure from the factorization point. \cite{arXiv:1012.4270, arXiv:1112.0361} utilized quantum discord in an $XY$ model to study symmetry breaking, leading to states that are not of the X-state form. They show that quantum discord has exponential scaling near the point, and based on that argue that discord has scaling behavior. The factorization point is $h \approx 0.7$. Their study also hints that the factorization phenomenon gives rise to nontrivial correlations when the ground state is on either side of the factorization-critical point. Lastly, to compute the optimized discord in non-X-states, \cite{arXiv:1112.3280} make use of tomographically-complete POVMs introduced in \cite{PhysRevA.70.052321}. Interestingly, four-element POVMs are determined (in many cases) by the values of $J_x$, $J_y$, and $J_z$ in the Hamiltonian in Eq.~\eqref{ChainHam}. Also see \cite{arXiv:1105.0027}, who analyze discord near the factorization point.

%**************************************************************
\subsubsection{Topological phase transition}

\cite{arXiv:0912.3874} explore a 2-d Castelnovo-Chamon model. The Hamiltonian reads
\begin{gather}
H=-\lambda_0 \sum_p B_p - \lambda_1 \sum_s A_s + \lambda_1 \sum_s e^{-\beta \sum_{i \in s} \sigma^z_i},
\end{gather}
where $\lambda_{0,1} > 1$, $A_s = \prod_{i\in s} \sigma^x_i$, and $B_p = \prod_{i \in p} \sigma^z_i$. $A_s$ is a star operator with vertex $s$ and $B_p$ is a plaquette operator acting on the four spins on the edges. The system has toric boundary conditions. Such systems have a phase transition from a topological phase to a magnetic phase. They find that quantum correlations between any two sites are always zero and are distributed globally. However, the mutual information between any two neighbors is able to pick up the critical point by discontinuous behavior. Therefore, it is the classical correlations that detect the transition. Also entanglement between any site and the rest picks up the critical point. Since this is a bipartite pure state, discord is the same as entanglement. In \cite{arXiv:1009.2846} a 1-d cluster-like system with
\begin{gather}
H=-J \sum_i (S_i+ B \sigma^z_i),
\end{gather}
$J>0$ and $S_i=\sigma^x_{i-1} \sigma^x_{i} \sigma^x_{i+1}$ is analyzed. Such systems also admit a topological phase transition. In contrast to the 2-d case, here quantum discord is finite though suppressed. However, correlations measured by the mutual information and quantum discord, near the critical point, show reversed power-law decay as a function of the distance. They comment that it is the richness of the 2-d topology over the 1-d topology that leads to vanishing discord between neighbors in the former and finite discord in the latter.

%**************************************************************
\subsubsection{Scaling}

While discord plays an important role in quantifying the quantum correlations, often quantum mutual information is sufficient to indicate critical behavior and is much easier to compute. However the scaling of quantum correlations is important from a conceptual point of view. The scaling of correlations is perhaps more interesting than the analytic behavior studied above. By scaling we mean how discord (or other correlations measures) fare as a function of distance and the number of sites considered. Many of the studies above show that discord scales linearly with the number of qubits, while it scales logarithmically with distance, i.e. it decays exponentially. Additionally, researchers find that discord scales differently to entanglement. It would be fruitful to compare the scaling of mutual information with the scaling of discord.

%**************************************************************
\subsection{Time and temperature}
%**************************************************************

%**************************************************************
\subsubsection{Discord and temperature}

Similar to the studies of dynamics of discord in Sec.~\ref{dynamics}, several studies of discord (in comparison with entanglement) as a function of temperature have been carried out. In \cite{arXiv:0911.3903}, the behavior of discord is analyzed as a function of temperature $T$ in two-qubit $XXZ$ ($J_x = J_y =J$ and $J_z \ne 0$ in Eq.~\eqref{ChainHam}) and $XXX$ ($J_x = J_y =J_z = J$ in Eq.~\eqref{ChainHam}) models. In the former entanglement is always zero, while discord is zero for low $T$, but it increases with $T$ before asymptotically dying as $T \to \infty$. In the latter, entanglement sudden death is observed as a function of $T$, while discord does not have sudden death. They also observe regrowth of quantum discord, i.e. it decreases as $T$ increases and it begins to grow again. Entanglement sudden death is contrasted with a lack of such an effect for discord as a function of $T$ in a two-qubit $XXZ$ system with Dzyaloshinskii-Moriya interaction in \cite{arXiv:1002.0176}. A similar study of discord in an $XX$ model ($J_x = J_y$ in Eq.~\eqref{ChainHam}) with two qubits is presented in \cite{arXiv:1007.3169}, with different strengths of external magnetic field ($h$ in Eq.~\eqref{ChainHam} is $h_1$ and $h_2$ for the two qubits respectively); discord, classical correlations, and entanglement are considered as functions of $T$ and magnetic-field strength. All of these works involve the discord of X-states, given in Eq.~\eqref{xstate}. \cite{arXiv:1104.1525} study three qubits in a $XXZ$ model with the Dzyaloshinskii-Moriya interaction. They find the discord and classical correlations to be discontinuous. Lastly, a comparison between different measures of discord is carried out in \cite{arXiv:1105.2866}, and discord is analyzed in \cite{arXiv:1104.1276} based on the available experimental data for both antiferromagnetic and ferromagnetic interactions for $Cu(NO_3)_2 \cdot 2.5H_2O$ molecules, hydrated and anhydrous copper acetates, and ferromagnetic binuclear copper acetate complex $[Cu_2L(OAc)] \cdot 6H_2O$.

%**************************************************************
\subsubsection{Dynamics in chains}

\cite{arXiv:1011.5309} link the dynamics of entanglement, quantum discord, and zero-way work deficit. They study an infinite $XY$ model with $J_x=\frac{\lambda}{2}(1+\gamma)$, $J_y=-\frac{\lambda}{2}(1-\gamma)$, and $J_z=0$ in Eq.~\eqref{ChainHam}. The dynamics is generated by applying the external transverse field at time $t=0$, i.e. $h(t<0) = 0$ and $h(t>0) = h$). They find that quantum entanglement suffers from sudden death and revival, but discord and work deficit do not. However, the revival of entanglement is shown to be related to the behavior of quantum discord and work deficit. They show that if for some fixed time the entanglement in two nearest neighbors vanishes around the critical field $h = h_c$, it revives for $h > h_c$ if
\begin{gather}
\frac{\partial D}{\partial h}{\big |}_{\sim h_c} > 0.
\end{gather}

In a study of propagation of correlations in a $XXZ$ ($J_x = J_y=-2 J$, $J_z=0$, and $h=-2h$ in Eq.~\eqref{ChainHam}) spin chain, \cite{arXiv:1105.5548} show that discord is better transported when compared to entanglement across 50 spins. The dynamics of total, quantum, and classical correlations in two qubits coming from the ground state of a transverse Ising model ($J_x=-J$, $J_y=J_z=0$, and $h=-1$ in Eq.~\eqref{ChainHam}) subjected to dephasing and decohering Markovian channels are analyzed in \cite{arXiv:1107.3939, arXiv:1112.2050}. For the bit-flip and the amplitude-damping channels at large times, only classical correlations survive. In the latter case, quantum correlations are higher than classical initially, and therefore there is a crossing point. For the phase-flip channel there are two crossing points for quantum and classical correlations, and the classical correlations are frozen \cite{arXiv:1006.1805, arXiv:0911.2848}. They analyze the behavior of these crossing points, their derivatives, and the difference in crossing points (for the phase-flip case) as a function of $J$. They find singular behavior in the derivatives at the critical value of $J=J_c=1$. Recently \cite{arXiv:1111.4879} study the dynamics of atoms in a BEC in terms of quantum discord and mutual information.

%**************************************************************
\subsubsection{Ergodicity}

\cite{arXiv:1112.1856} study ergodicity of quantum correlations versus entanglement in a similar system as above. The specifics of this $XY$ model are $J_x=\frac{\lambda}{2}(1+\gamma)$, $J_y=-\frac{\lambda}{2}(1-\gamma)$, and $J_z=0$ in Eq.~\eqref{ChainHam}), and $h(t<0) = 0$ and $h(t>0) = h$). An ergodic physical quantity has the property that its time average is the same as its ensemble average. They compare ergodicity of the concurrence and negativity versus discord and deficit. They find that entanglement measures are always ergodic, while discord and deficit are not for specific values of the transverse field.

%**************************************************************
%**************************************************************
\section{Conclusions}
%**************************************************************

Quantum discord encapsulates the idea that two equivalent ways of looking at correlations in classical information theory give different results when generalized to quantum information theory. In quantum physics, we can have classical correlations, but we also have correlations that exceed them. This excess is called quantum discord and is a more general concept than quantum entanglement (in the sense that all entangled states are also discordant, but not vice versa). We have shown that discord features in a number of different areas, not only in quantum-information theory and quantum computation, but also in many-body physics, thermodynamics and open-systems dynamics.

Despite intense and exciting research over the last 10 years, there are still a number of challenging outstanding problems that drive much effort in the field. We believe that among the most exciting are the following.

Can any quantum computation be efficiently performed with just discordant states and without any entanglement? This has been an outstanding question for at least 20 years and, although most results indicate that the answer is in the negative, we still do not have any formal proofs either way. The intuition supporting the negative conclusion is based on the fact that a classically-correlated state (i.e. the one that is useless for quantum information processing) can by local operations be converted into a discordant one. Since this is easy to do, so the argument goes, discord should not give us any additional power. We know that this argument is not entirely convincing since such operations might still be hard to simulate classically. The jury is therefore still out on this one.

Are there any useful information protocols involving just discord and not entanglement? This question is similar to the previous one and it is clear that protocols such as super-dense coding and teleportation require quantum entanglement, but these are by no means the only useful ways of processing quantum information. Cryptographic protocols such as BB84 are based on discord and require no entanglement (though they could be said to be not as secure as the entanglement-based quantum cryptography). Remote state preparation can be shown to require discord, but we still await convincing applications of this and related protocols. 

Is discord a useful order parameter in many-body physics? We are still searching for phase transitions that cannot be characterized in any other way than by using discord. Perhaps some forms of topological phase transitions will require discord of many systems, since being topological implies that no local operation can perturb the system out of the ordered phase. Here again, all evidence points to a battle between discord and entanglement.

What is responsible for the quantum to classical transition (if such a transition exists in the first place)? This question is intimately related to the process of quantum measurement and the hope is that discord might shed further light on this intricate and deep problem.

No matter what the answers end up being to these questions, it is certain that understanding quantum correlations is a subject that will preoccupy the minds of physicists, mathematicians and computer scientists for a long time to come. And, who knows: studying the nature of correlations in the world around us might even help us catch a glimpse of the theory that comes to supersede quantum physics.

%**************************************************************
%**************************************************************
\acknowledgements
First of all we regret any omissions in this review. We have tried to be fair and all-inclusive in citing any work on measures of quantum correlations. This is a thriving field and many papers will appear (on the arXiv and otherwise) even as this manuscript is being concluded. We chose a deadline for papers to be included in this review: Specifically, any paper appearing in print or on the arXiv before the end of 2011 is included, provided the content is fitting of the review.

We acknowledge financial support by the National Research Foundation and the Ministry of Education of Singapore. KM and VV thank the John Templeton Foundation for support. AB acknowledges the hospitality of the Centre for Quantum Technologies, the University of Oxford and funding from EQuS. TP is supported by NTU start-up grant. We are grateful to many friends who have read and commented on this manuscript: G. Adesso, L. Amico, D. Browne, {\v C}. Brukner, Q. Chen, B. Daki\'c, A. Datta, U. Devi, F. Fanchini, A. G\'orecka, M. Gu, the Horodecki family, M. Lang, S. Luo, J. Oppenheim, P. Perinotti, M. Piani, A. Rajagopal, C. Rodr\'iguez-Rosario, D. Terno, S. Vinjanampathy, A. Winter, S. Yu, C. Zhang, and W. Zurek.

\bibliography{rmp_SF}

\end{document}